\documentclass[a4paper,fleqn,usenatbib,useAMS]{mnras}

\usepackage[T1]{fontenc}
\usepackage{ae,aecompl}

\usepackage{graphicx}	
\usepackage{amsmath}	
\usepackage{amssymb}	
\usepackage{float}
\usepackage{subcaption}
\usepackage{mathrsfs}
\usepackage{booktabs}
\usepackage{multirow}
\usepackage{threeparttable}

\newcommand\vece{{\bmath e}}

\newcommand\veck{{\bmath k}}

\newcommand\vecr{{\bmath r}}

\newcommand\vecu{{\bmath u}}
\newcommand\vecv{{\bmath v}}

\newcommand\dd{\mathrm{d}}
\newcommand\DD{\mathrm{D}}
\newcommand\ee{\mathrm{e}}
\newcommand\ii{\mathrm{i}}

\newcommand\p{\upartial}

\newcommand{\dbar}{\dd\hspace*{-0.18em}\bar{}\hspace*{0.14em}}

\title[Non-linear spiral waves in accretion discs]
{Non-linear spiral waves in accretion discs}

\author[Joshua J. Brown and Gordon I. Ogilvie]
{Joshua J. Brown\thanks{E-mail: jb2228@cam.ac.uk} and Gordon I. Ogilvie\thanks{E-mail: gio10@cam.ac.uk}
\\
Department of Applied Mathematics and Theoretical Physics,
University of Cambridge, Centre for Mathematical Sciences,\\
Wilberforce Road, Cambridge CB3 0WA, UK
}

\pubyear{2024}

\begin{document}
\label{firstpage}
\pagerange{\pageref{firstpage}--\pageref{lastpage}}
\maketitle

\begin{abstract}
We derive a simple, accurate, non-linear, global equation governing spiral density waves in thin, non-self-gravitating, inviscid accretion discs. These discs may have any slowly varying surface density or temperature profile. For specific `self-similar' disc profiles, solutions to our equation match (novel) \emph{smooth} non-linear exact spiral solutions derived via a separate method, which highlight that non-linear spiral waves need not shock. Indeed, at low amplitudes, we find that dispersion can overcome wave steepening, and may prevent the inner spiral wakes excited by low mass planets (below roughly $1\%$ of a thermal mass) embedded in protoplanetary discs from shocking. At high amplitudes, we find a simple universal description of non-linear spiral waves with shocks, as well as caps on the possible amplitude and wave action flux of non-linear spirals both with and without shocks, depending on how many arms they have. We further find that highly non-linear spirals are far more loosely wound than their linear counterparts. These developments shed light on why two-armed spirals are prevalent across a range of astrophysical systems which don't necessarily possess an intrinsic twofold symmetry, and why they appear surprisingly loosely wound in observations. These results are supported by very high-resolution numerical simulations.
\end{abstract}




\section{Introduction}

Steadily rotating density waves in gaseous accretion discs are naturally sheared by the rapid differential orbital flow into spirals. As a result, spiral waves are observed or inferred ubiquitously across a wide range of astrophysical contexts. In recent years, high-resolution telescopes and the advent of extreme adaptive optics have facilitated the direct imaging of protoplanetary discs, revealing large-scale spiral structures in the discs MWC 758 \citep{benisty_asymmetric_2015}, SAO 206462 \citep{2016A&A...595A.113S,2025A&A...699L..10M}, HD100453 \citep{rosotti_spiral_2020}, TW Hya \citep{2019ApJ...884L..56T}, UX Tau \citep{2020A&A...639L...1M} and AB Aur \citep{2024Natur.633...58S} to name a few. Kinks in velocity channel maps have also been attributed to spiral shock waves generated by disc-embedded planets \citep{2018ApJ...860L..13P}.

Numerical and theoretical studies representing a variety of astrophysical systems frequently exhibit spiral waves, which often play important direct and indirect roles in shaping and regulating their host systems. Such systems include planets embedded in protoplanetary discs, reviewed by \citet{2012ARA&A..50..211K} and \citet{2023ASPC..534..685P}, which excite trailing spiral waves that drive angular momentum exchange between the planet and the disc (leading to orbital evolution of the planet), as well as the analogous situation involving black holes or stars embedded in active galactic nucleus (AGN) discs \citep{secunda_orbital_2019}. External perturbers such as binary companions outside or inside circumstellar discs \citep{1979MNRAS.186..799L,1994MNRAS.268...13S}, stellar flybys and captures \citep{1994ApJ...424..292O,2023MNRAS.521.3500S,2025MNRAS.tmp.1506P} and the host star of a circumplanetary disc system \citep{rivier_circum-planetary_2012,2016ApJ...832..193Z} also excite large-scale, dynamically important spiral waves. Inward-propagating acoustic spirals excited by galactic bars have been inferred to play a key role in the formation of nuclear rings near galactic centres \citep{2024MNRAS.528.5742S}. Spirals may be excited by shadowing due to misaligned inner discs \citep{2025ApJ...980..259Z}, or acoustic interaction with a misaligned inner disc \citep{2025MNRAS.542.1430R}. Spiral waves also occur as direct and indirect outcomes of instabilities, for example the gravitational instability (GI) \citep{bethune_spiral_2021}, magneto-rotational instability (MRI) \citep{heinemann_excitation_2009} and via vortices generated via the Rossby wave instability (RWI) \citep{1999ApJ...513..805L,2010ApJ...725..146P}.

The early theory of spiral density waves was developed to try to explain spiral structure in galaxies. A linear theory was derived by \citet{1964ApJ...140..646L}, and a related theory was developed for density waves observed in Saturn's rings, reviewed by \citet{1984prin.conf..513S}. Linear waves in gaseous discs excited at Lindblad resonances (e.g. by satellites or companions \citep{1979ApJ...233..857G,1987Icar...69..157M}) were also understood to form a coherent spiral pattern via a linear superposition of modes \citep{ogilvie_wake_2002}; it was later realised that this pattern possesses multiple arms due to the action of dispersion \citep{miranda_multiple_2019}.

Various non-linear theories have been developed to describe different aspects of non-linear spiral waves in a range of settings. With density waves in galaxies and Saturn's rings in mind, \citet{1985ApJ...291..356S} derived within a Lagrangian framework a non-linear equation for tightly wound density waves in self-gravitating, pressure-less discs, which describes their shape as well as propagation over short radial distances. \citet{yuan_resonantly_1994} explored smooth forced solutions to this equation in the vicinity of a Lindblad resonance, including an extension to model pressure. Many further contributions in this context are summarised by \citet{2016ARA&A..54..667S}.

Early simulations of close binaries in gaseous (non-self-gravitating) discs showed prominent two-armed spirals \citep{1986MNRAS.219...75S}, motivating a theory of spiral wave-driven accretion developed by \citet{spruit_stationary_1987} and \citet{1987MNRAS.229..517S}. \citet{spruit_stationary_1987} found exact global non-linear `self-similar' solutions for spiral waves containing shocks in discs with very specific profiles, and a closely related family of solutions was later found by \citet{hennebelle_spiral-driven_2016}. With the same systems in mind, \citet{larson_non-linear_1990} derived a weakly non-linear wave equation for tightly wound, low-amplitude waves, and solved for smooth waves as well as waves with (very weak) shocks.

\citet{goodman_planetary_2001} developed an approximate weakly non-linear model (in analogy with one dimensional non-linear gas dynamics) for the evolution of spiral waves excited by planets in the shearing sheet model of astrophysical discs (e.g. \citet{2017MNRAS.472.1432L}). \citet{rafikov_nonlinear_2002} extended this study to a global disc model, applicable to weakly non-linear highly acoustic waves\footnote{Dispersion, which leads to the formation of multiple spiral arms, is neglected in this highly acoustic limit.} (with radial wavenumber $k_r$ much larger than the inverse scale-height $1/H$), which highlighted the tendency of planet-driven waves to shock. The shock-driven deposition of angular momentum by these waves modifies the angular momentum profile of the disc itself, leading to secular disc evolution including accretion and gap-opening \citep{goldreich_disk-satellite_1980,1984ApJ...285..818P}.

\citet{heinemann_weakly_2012} derived a weakly non-linear theory of spiral waves in the shearing sheet, which describes the evolution and shocking of waves excited in turbulent accretion discs. They used a Lagrangian coordinate system which further naturally followed wave crests, and included dispersion (which they showed to be unable to subdue the combined effects of shear amplification and wave steepening even at low amplitudes) via an antiderivative operator. Similarly, \citet{fromang_properties_2007} derived an exact equation for non-linear axisymmetric density waves in the shearing sheet (which also used a Lagrangian description), finding finite-amplitude smooth solutions. Their wave equation has much in common with the global spiral wave equation (\ref{MasterEqn}) which we derive in section \ref{GDW}.

Removed from the specific context of accretion discs, sophisticated and general mathematical descriptions of non-linear waves have also seen important advancements since the earliest developments in spiral density wave theory. These include Whitham's description of non-linear dispersive modulated wave trains and their evolution \citep{whitham_non-linear_1965,whitham_linear_1974}, as well as the generalised Lagrangian mean (GLM) description of waves on a mean flow \citep{andrews_exact_1978,andrews_wave-action_1978}, a good introduction to which is given in \citet[chapter 10]{buhler_waves_2014}.

Each of the aforementioned perspectives on non-linear waves and spiral waves in accretion discs contributed importantly to the development of our global non-linear model (equation (\ref{MasterEqn}) below) and its solutions presented in this paper.

The remainder of the paper is structured as follows. In section \ref{S7} we find exact smooth spiral wave solutions to the 2D Euler equations for specific `self-similar' disc profiles. In section \ref{GDW} we derive the global non-linear equation for spiral waves, supported by calculations in appendices \ref{appx2} and \ref{appxb1}. In section \ref{SFPA} we explore some (semi-)analytic solutions to the wave equation, and in section \ref{PDSW} we present simulations of low and high mass planet-driven spiral waves. We discuss our findings in section \ref{discussion}, and draw our conclusions in section \ref{summary}.

\section{Steady self-similar smooth spirals}\label{S7}

One key objective of this paper is to demonstrate and emphasise that steady non-linear spiral waves without shocks not only exist, but they represent an important possible outcome for the inner wake of low-amplitude driven spiral waves. In this section we find steady non-linear waves for which the effects of wave steepening and dispersion are in perfect balance. These steady waves have zero angular pattern speed (a necessary condition for self-similarity), and exist well within their corotation radius (which is formally at $r = \infty$). 

Much of this section follows analysis by \citet{spruit_stationary_1987} and subsequently \citet{hennebelle_spiral-driven_2016}, who studied large amplitude self-similar non-linear spiral waves possessing multiple shocks in their profile, finding exact semi-analytical solutions to the 2D Euler equations. Spruit's waves exist in discs with specific radial profiles for density and temperature, which allow for a uniform accretion rate.

For smooth solutions to be possible, a different radial density profile is necessary, which permits the radial angular momentum flux (or wave action flux) of the waves to remain constant. \citet{hennebelle_spiral-driven_2016} considered this profile, but only searched for shocking solutions (though it's possible these authors were aware of the smooth counterparts). Whilst the spiral waves found below require a specific background state to remain self-similar, it will become clear (namely through analysis presented in section \ref{GDW}) how steady non-linear spirals behave in discs with more general profiles.

\subsection{Governing equations}\label{GE}

For simplicity, we adopt an approximate 2D model for the flow in the disc. Whilst 2D models have been connected in a precise manner to the linear behaviour of a `2D mode' within adiabatic discs \citep{lubow_wave_1993, brown_horseshoes_2024}, this connection doesn't extend to the non-linear case. Indeed, as \citet{fridman_possibility_1996} point out, a non-linear 2D model is applicable when both the disc is thin and the characteristic timescales of the processes studied are long. Even then, the resulting 2D equations differ somewhat from those traditionally adopted. Nevertheless, 2D models offer a valuable simplicity over a 3D description, whilst capturing much of the important physics believed to be at play.

Therefore, as is traditional, we adopt the steady 2D Euler equations for an adiabatic flow about a star or massive object with potential $\Phi(r) = -\frac{G M_\star}{r}$:
\begin{subequations}\label{EE}
\begin{align}
    &\nabla\cdot\left(\Sigma \vecu \vecu + P \mathbf{I}\right) = -\Sigma\nabla\Phi,\label{ME}\\
    &\nabla \cdot\left(\Sigma \vecu\right) = 0,\\
    &\vecu \cdot\nabla\left(P\Sigma^{-\gamma}\right)=0.
\end{align}
\end{subequations}
Here, $\Sigma$ represents the surface density of the disc, $P$ the vertically integrated pressure, and $\gamma$ the constant effective adiabatic index in this vertically integrated system. Note the absence of forcing in the system (\ref{EE}): the waves which appear below are freely propagating.

Taking $\vecr \times$(\ref{ME}) for $\vecr = r\vece_r$, and exploiting the symmetry of the momentum flux density tensor, we obtain the angular momentum equation, which is particularly important for us:
\begin{equation}\label{AME}
    \nabla\cdot\left(\vecr \times \left(\Sigma \vecu \vecu + P \mathbf{I}\right)\right) = {\bmath{0}}.
\end{equation}
In order for our waves to maintain a self-similar profile, they must maintain a constant level of non-linearity. We'll see that only a particular choice of background disc profile allows this to be the case: equation (\ref{AME}), which describes the conservation of angular momentum flux (which for these waves is intimately linked to wave action flux, a connection we demonstrate explicitly in appendix \ref{appx1}), determines the relationship between the radial evolution of the wave's level of non-linearity and the radial profile of the disc.

The final equation which we'll make (brief) use of is the energy equation (which may be derived from the system (\ref{EE})):
\begin{equation}\label{ENE}
    \nabla\cdot\left(\Sigma \vecu \left(\frac{1}{2}\left|\vecu\right|^2 + \frac{\gamma}{\gamma - 1}\frac{P}{\Sigma} + \Phi\right)\right) = 0.
\end{equation}

We write the radial and azimuthal components of the flow velocity as
\begin{equation}
    \vecu = u \vece_r + \left(r \Omega + v\right)\vece_\theta,
\end{equation}
for
\begin{equation}
    \Omega(r) = \sqrt{\frac{G M_\star}{r^3}}.
\end{equation}
We non-dimensionalise the equations by setting:
\begin{subequations}
    \begin{align}
        r &= r_p \tilde{r},\\
        (u,v) &= r_p\Omega(r_p)(\tilde{u},\tilde{v}),\\
        \Sigma &= \Sigma_p\tilde{\Sigma},\\
        P &= [r_p\Omega(r_p)]^2\Sigma_p\tilde{P},
    \end{align}
\end{subequations}
for reference radius and surface density $r_p$ and $\Sigma_p$ (we reserve the subscript '$0$' notation (e.g. $\Sigma_0(r)$) for quantities from a background or reference disc). The Euler equations (\ref{EE}) become
\begin{subequations}\label{EEND}
    \begin{align}
        \tilde{\Sigma}\left[\left(\tilde{r}^{-3/2}\!+\frac{\tilde{v}}{\tilde{r}}\right)\frac{\p \tilde{u}}{\p \theta} + \tilde{u}\frac{\p\tilde{u}}{\p\tilde{r}} - \frac{\tilde{v}^2}{\tilde{r}} - 2\tilde{r}^{-3/2}\tilde{v}\right] &= -\frac{\p\tilde{P}}{\p\tilde{r}},\\
        \tilde{\Sigma}\left[\left(\tilde{r}^{-3/2}\!+\frac{\tilde{v}}{\tilde{r}}\right)\frac{\p\tilde{v}}{\p \theta} + \tilde{u}\frac{\p\tilde{v}}{\p\tilde{r}} + \frac{\tilde{r}^{-3/2}\tilde{u}}{2} + \frac{\tilde{u}\tilde{v}}{\tilde{r}}\right] &= -\frac{1}{\tilde{r}}\frac{\p\tilde{P}}{\p \theta},\!\\
        \frac{\p}{\p \tilde{r}}\left[\tilde{r}\tilde{\Sigma}\tilde{u}\right] + \frac{\p}{\p \theta}\left[\tilde{\Sigma}\left(\tilde{v} + \tilde{r}^{-1/2}\right)\right] &= 0,\\
        \tilde{u}\frac{\p}{\p \tilde{r}}\left[\tilde{P}\tilde{\Sigma}^{-\gamma}\right] + \frac{\tilde{v} + \tilde{r}^{-1/2}}{\tilde{r}}\frac{\p}{\p \theta}\left[\tilde{P}\tilde{\Sigma}^{-\gamma}\right] &= 0,
    \end{align}
\end{subequations}
and the angular momentum equation (\ref{AME}) becomes
\begin{multline}\label{AMEND}
    \frac{\p}{\p \tilde{r}}\left[\tilde{r}^2\tilde{\Sigma}\tilde{u}\left(\tilde{v}+\tilde{r}^{-1/2}\right)\right] +\\ \frac{\p}{\p \theta}\left[\tilde{r}\tilde{\Sigma}\left(\tilde{v} + \tilde{r}^{-1/2}\right)^2 + \tilde{r}\tilde{P}\right] = 0.
\end{multline}

Following \citet{spruit_stationary_1987} (and \citet{hennebelle_spiral-driven_2016}), we introduce the azimuthal `spiral' coordinate
\begin{equation}
        \psi = \theta + \beta(\tilde{r}),
\end{equation}
so that
\begin{equation}
    \frac{\p}{\p \tilde{r}}\bigg|_\theta = \frac{\p}{\p \tilde{r}}\bigg|_\psi + \frac{\dd \beta}{\dd \tilde{r}}\frac{\p}{\p \psi}, \quad \frac{\p}{\p \theta} = \frac{\p}{\p \psi}.
\end{equation}
Lines of constant $\psi$ correspond to spirals, with which our wave crests will be aligned. Ensuring all terms in the resulting equations scale with $\tilde{r}$ in the same way, we obtain the similarity ansatz:
\begin{subequations}\label{SSA}
    \begin{align}
        &\tilde{u} = \tilde{r}^{-1/2}U(\psi),\label{SSAU}\\
        &\tilde{v} = \tilde{r}^{-1/2}V(\psi),\label{SSAV}\\
        &\tilde{\Sigma} = \tilde{r}^{-n_r}D(\psi),\\
        &\tilde{P} = \tilde{r}^{-n_r-1}P(\psi),\\
        &\beta' = \tilde{r}^{-1}b,
    \end{align}
\end{subequations}
for $b$ a constant eigenvalue of the problem determining the angle $\varphi$ between lines of constant $\psi$ (which are aligned with the spirals) and the radial direction via
\begin{equation}
    b = \tan\varphi.
\end{equation}
Lines of constant $\psi$ correspond to curves $\theta = \text{const} - b \ln \tilde{r}$, that is, the spirals are logarithmic. The pitch angle of the spirals is $\upi/2 - \varphi$, and the spirals are tightly wound for $b \gg 1$.

Upon substitution of the ansatz (\ref{SSA}), the equations of motion (\ref{EEND}) read:
\begin{subequations}\label{SSEsys}
    \begin{align}
        &WU' - \frac{1}{2}U^2 - V^2 - 2 V = (n_r+1)\frac{P}{D} - b\frac{P'}{D},\label{SSEa}\\
        &WV' + \frac{1}{2}UV +\frac{1}{2}U = -\frac{P'}{D},\label{SSEb}\\
        &(DW)' + \left(\tfrac{1}{2} - n_r\right)DU = 0,\label{SSEc}\\
        &W\left(\frac{P'}{P} - \gamma\frac{D'}{D}\right) + ((\gamma-1)n_r-1)U = 0,\label{SSEd}
    \end{align}
\end{subequations}
for $W = 1 + V + bU$. The angular momentum equation (\ref{AMEND}) becomes
\begin{equation}\label{AMESS}
(DW(1+V)+P)'=(n_r-1)DU(1+V),
\end{equation}
which may also be derived from equations (\ref{SSEb}) and (\ref{SSEc}). 

It's instructive at this stage to note the symmetry of the system (\ref{SSEsys}) under the transformation $b \to -b$, $U \to - U$, $\psi \to - \psi$ (which for smooth solutions also leaves the boundary conditions unaffected). Solutions with $b>0$ correspond to inward propagating trailing spirals, and conversely those with $b<0$ are outward propagating leading spirals.

Interestingly, \citet{spruit_stationary_1987} found that outward-propagating leading \emph{self-similar} spirals containing shocks in their profiles seemed not to exist: presumably this is because such solutions (which are assumed to possess zero net angular momentum flux) involve a wave transporting a negative flux outwards, which is unable to cancel the positive inward advected flux of angular momentum carried by the induced accretion flow.

In light of the aforementioned symmetry, we may without loss of generality restrict our analysis to the case $b>0$, which is also the case of greater physical interest, as it corresponds to trailing spirals propagating inwards, interior to their corotation radius. Whilst rare, the case $b<0$, corresponding to leading spirals propagating outwards far inside their corotation radius, is physically possible. For example \citet{2025MNRAS.542.1430R} observe such leading spirals, which were generated via an acoustic interaction with a slowly precessing misaligned inner disc.

Returning now to the angular momentum equation (\ref{AMESS}), we integrate from $0$ to $2\upi$, yielding
\begin{equation}\label{nrcond}
    (n_r-1)\int_{0}^{2\upi}DU(1+V)\dd \psi = 0.
\end{equation}
That is, either the wave possesses no net radial angular momentum flux, or $n_r = 1$. We therefore set $n_r = 1$, corresponding to the density profile $\Sigma_0 \propto r^{-1}$. In doing so we restrict ourselves to strictly non-accreting solutions\footnote{This may be seen mathematically by integrating equation (\ref{SSEc}) from $0$ to $2\upi$.}, though as we're concerned with smooth waves, this is an expected feature. It follows that the total angular momentum flux normal to the spirals is constant. That is,
\begin{equation}\label{OWA}
    DW(1+V)+P = \text{const}.
\end{equation}

Eliminating $D'$ from the energy equation (\ref{SSEd}) in favour of $U'$ and $V'$ leaves us finally with the third order system:
\begin{multline}\label{SSEfinal}
    \begin{bmatrix}
DW & 0 & b \\
0 & DW & 1 \\
b & 1 & \frac{W}{\gamma P}
\end{bmatrix}
\begin{bmatrix}
U' \\ V' \\ P'
\end{bmatrix}
=\\
\begin{bmatrix}
\frac{1}{2}DU^2 + DV^2 + 2DV + 2P \\ -\frac{1}{2}DU(1+V) \\ \big(\frac{2}{\gamma} - \frac{1}{2}\big)U
\end{bmatrix},
\end{multline}
where recall $D$ and $W$ may be written algebraically in terms of $U$, $V$ and $P$. 

It's possible to find two further independent constants of the motion, and reduce the problem yet further to a 1D ordinary differential equation. These constants arise as a consequence of energy and entropy conservation (though do not directly read as such). They may both be derived in a similar way. For example, suppose for some quantity $q = q_0\tilde{r}^\lambda Q(\psi)$ we have
\begin{equation}
    \nabla\cdot\left(\Sigma \vecu q\right) = 0.
\end{equation}
It follows, having transformed into our spiral coordinates, that
\begin{align*}
    &\left(\lambda-\tfrac{1}{2}\right)DUQ + \left(DWQ\right)' = 0\\
    \implies &\left(2\lambda-1\right)(DW)'Q + (DW)'Q + DWQ' = 0\\
    \implies &2\lambda \frac{(DW)'}{DW} + \frac{Q'}{Q} = 0\\
    \implies &(DW)^{2\lambda} Q = \text{const}.
\end{align*}
Entropy conservation yields
\begin{equation}\label{entcons}
    (DW)^{2(\gamma-2)}PD^{-\gamma} = \text{const},
\end{equation}
whilst energy conservation (\ref{ENE}) yields
\begin{equation}\label{encons}
    \left(DW\right)^{-2}\left[\frac{1}{2}\left(U^2+(1+V)^2\right) + \frac{\gamma}{\gamma-1}\frac{P}{D}-1\right] = \text{const}.
\end{equation}
Despite these algebraic relationships, we opt to solve the third-order system (\ref{SSEfinal}) numerically, as a reduced 1D differential equation making use of these conserved quantities would be very complicated. The existence of these conserved quantities is nevertheless a useful fact to bear in mind when designing a numerical method, as is discussed below.

\subsection{Boundary conditions}

We're interested in periodic solutions to the system (\ref{SSEfinal}), where the period must divide $2\upi$. That is,
\begin{equation}
    U(\psi) = U(\psi + 2\upi/m)
\end{equation}
where $m$ is the number of spiral arms present (for Spruit, the equivalent variable was $n_s$, the number of shocks). Since we have enough conserved quantities to reduce the problem to a single dimension, it immediately follows that if $U(\psi)$ is periodic, then $V(\psi)$ and $P(\psi)$ are also periodic with the same period.

We have four further constraints to enforce. Firstly, we specify
\begin{equation}\label{MCSS}
    \frac{1}{2\upi}\int_0^{2\upi} D \dd \psi = 1,
\end{equation}
so that the mean surface density matches the surface density $\Sigma_0$ of a reference state. In practice, it's easiest to enforce this condition by fixing the constant in equation (\ref{OWA}) to be an arbitrary value, before later rescaling $D \to \lambda D$, $P \to \lambda P$, choosing $\lambda$ such that the condition (\ref{MCSS}) is satisfied. Note that this operation leaves all other flow variables in the non-linear system (namely $U$, $V$, $W$ and the sound speed, proportional to $P/D$) unchanged. Secondly, we must specify the amplitude of the wave, which we enforce by selecting a target value for the radial angular momentum flux $\mathcal{B}$ (which we show is equal to the wave action flux in appendix \ref{appx1}). This flux is given by
\begin{equation}\label{AMFint}
    \mathcal{B} = \int_0^{2\upi} DUV \dd \psi,
\end{equation}
which is equal to the integral appearing in equation (\ref{nrcond}) because the mass flux vanishes.

Thirdly, we should enforce a target value for the disc's average (non-dimensional) sound speed, $C$, which we define via
\begin{equation}\label{Cdef}
    C^2 = \frac{1}{2\upi}\int_0^{2\upi}\frac{\gamma P}{D} \dd \psi.
\end{equation}
The non-dimensional value $C$ measures the physical sound speed in comparison to the orbital velocity $r\Omega$, and so is related to the average aspect ratio of the disc, $h_0$, via $C = \sqrt{\gamma}h_0$. These first three constraints also fix the three constants of integration in equations (\ref{OWA}), (\ref{entcons}) and (\ref{encons}).

Finally, the system (\ref{SSEfinal}) has a translational symmetry in the coordinate $\psi$, requiring a final constraint to specify a reference phase. We therefore have four constraints which each require a degree of freedom to enforce, and four degrees of freedom: $U(0)$, $V(0)$, $P(0)$ and $b$, which may be varied until the numerical solution possesses the desired statistics. 

\begin{figure*}\centering
  \includegraphics[width=0.495\linewidth]{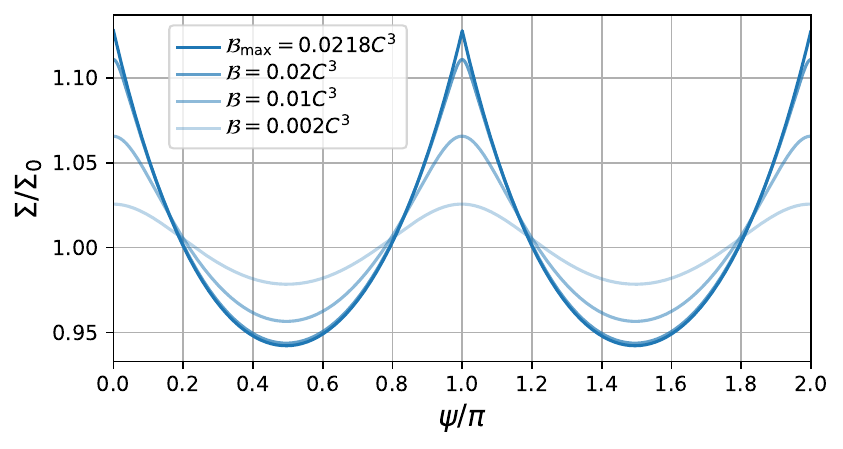}
  \includegraphics[width=0.495\linewidth]{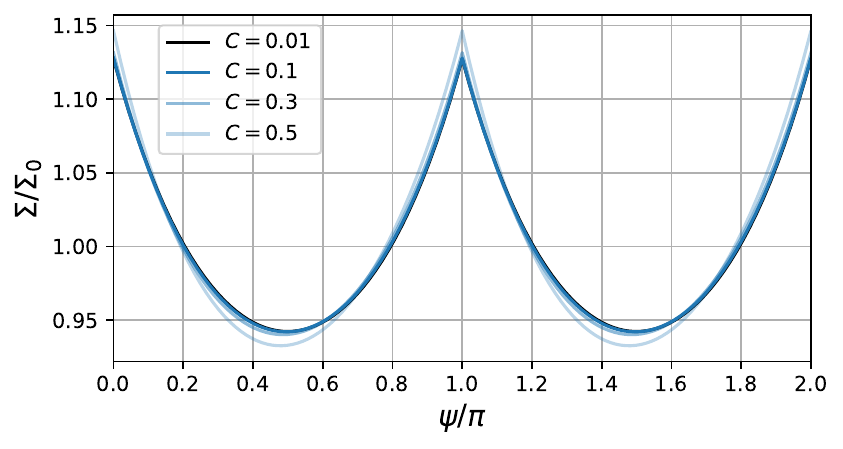}
  \includegraphics[width=0.495\linewidth]{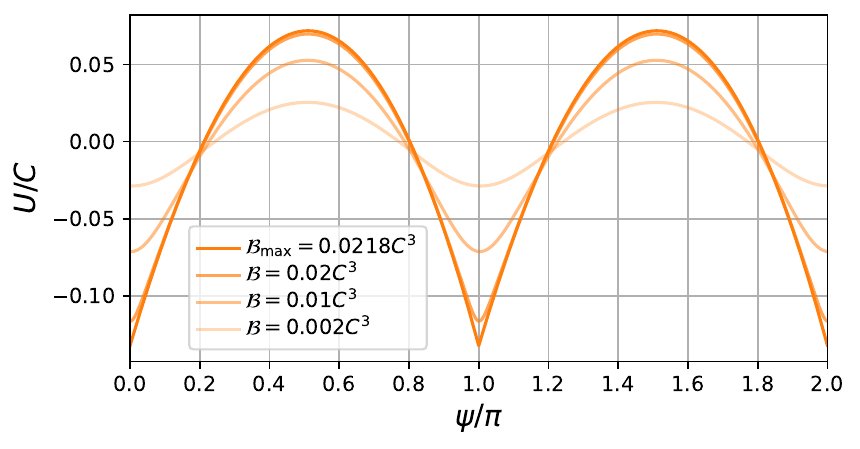}
  \includegraphics[width=0.495\linewidth]{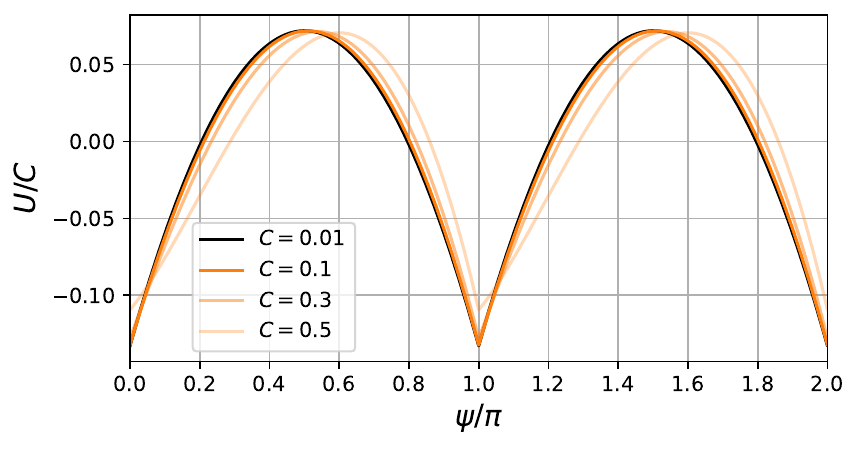}
  \includegraphics[width=0.495\linewidth]{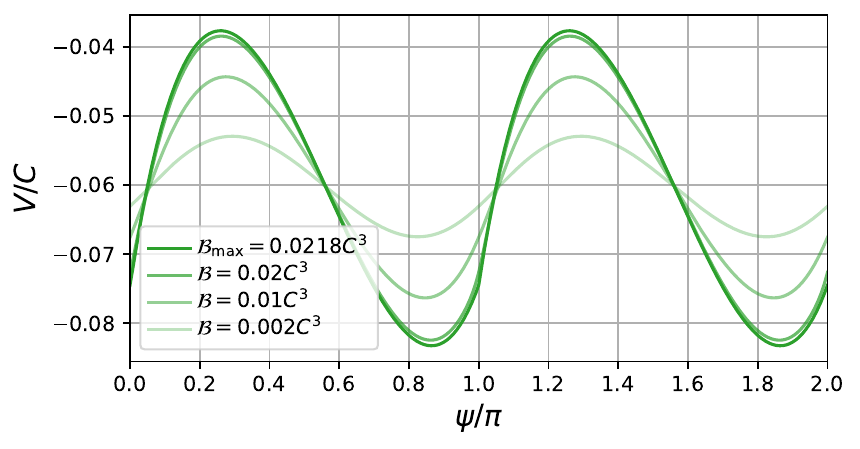}
  \includegraphics[width=0.495\linewidth]{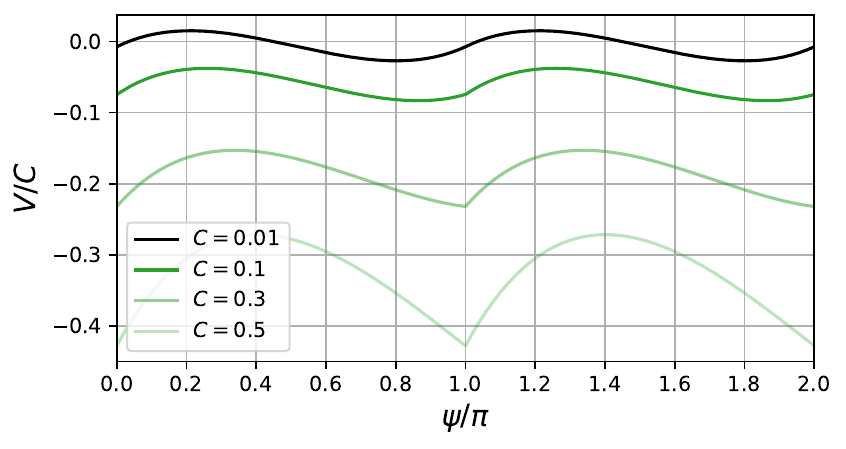}
\vspace{-1em}
\caption{Left: azimuthal cross-sections of surface density, radial velocity and azimuthal velocity perturbation in non-linear self-similar spiral solutions to the system (\ref{SSEsys}) with $C = 0.1$. The background disc has $\Sigma_0\propto r^{-1}$. Lighter colours correspond to waves of lower amplitude, with a lower radial wave action flux. $\mathcal{B}_{\text{max}}$ is the largest possible radial wave action flux of a 2-armed smooth steady spiral wave in the chosen reference disc. The functions $U(\psi)$ and $V(\psi)$ are defined in equations (\ref{SSAU}) and (\ref{SSAV}). Right: comparison of `cuspy' waves with $\mathcal{B} = \mathcal{B}_\text{max}$ for discs with different aspect ratios. The surface density profiles are remarkably similar across the range of aspect ratios, especially for the thin discs. In all cases $\gamma = 5/3$.}
  \label{SSSprofiles}
\end{figure*}

\subsection{Linear problem}\label{linprob}

We can gain some useful insight and intuition from the linearised problem (as well as a comparison point for numerical solutions), so present its solution below.

We linearise about the uniform solution, letting:
\begin{subequations}
\begin{align}
    &U = \delta U,\\ 
    &V = \sqrt{1-\frac{2}{\gamma}C^2} - 1 + \delta V,\\
    &P = \frac{C^2}{\gamma} + \delta P,\\
    &D=1 + \delta D,
\end{align}
\end{subequations}
where $C^2$ is the (non-dimensional) squared adiabatic sound speed relative to $r\Omega$, which may be treated as a variable parameter. (Incidentally, for convenience we've normalised the density in advance, choosing the constant in equation (\ref{OWA}) to be $1-C^2/\gamma$). We set the perturbed variables proportional to $\ee^{\ii m\psi}$.

Upon linearising the governing equations (\ref{OWA}) and (\ref{SSEfinal}), and substituting the above ansatz, we find the dispersion relation
\begin{equation}\label{DR}
m^2\left(1-\frac{C^2(1+b^2)}{1-\frac{2}{\gamma}C^2}\right) = 1 + \frac{2-\gamma}{\gamma}\frac{C^2}{\left(1-\frac{2}{\gamma}C^2\right)^2}.
\end{equation}
(Details of this calculation have been omitted for brevity). Recall $m$ is the number of spirals and $\varphi = \tan^{-1}\!b$ the angle made by the spirals with the radial direction. Reassuringly, in the limit of a thin disc $C \ll 1$, we see that (\ref{DR}) becomes the familiar dispersion relation for steady density waves\footnote{Specifically, from its definition we have $b = K_r/m$, with $K_r = rk_r$ the non-dimensional radial WKB wavenumber (and $k_r$ the true wavenumber). Employing the approximation $C \ll 1$, equation (\ref{DR}) becomes $m^2 - C^2(m^2 + K_r^2) = 1$. Reintroducing dimensions yields the perhaps more familiar time-independent dispersion relation $m^2(\Omega - \Omega_p)^2 = \Omega^2 + c^2\left|\veck\right|^2$, for $\Omega_p = 0$ the angular pattern speed.}.

Notice that solutions with $m = 1$ require a disc with $C\geqslant 1$, a value outside the realm of our interest (and one which would cast serious doubts over the validity of our 2D flow assumption).

The full linear solution is (neglecting the uniform mode which corresponds to a small modification to $C$):
\begin{subequations}
    \begin{align}
        \delta U = & \, 2A\left[1-\left(1+\frac{2}{\gamma}\right)C^2\right]\cos(m \psi),\\
        \delta V = & \, A\bigg[2 bC^2\cos(m \psi) - \frac{1 + (\frac{2}{\gamma}-1)C^2}{m}\sin(m \psi)\bigg],\\
        \delta P = &\, A\bigg[\frac{4C^2\alpha}{\gamma m}\sin(m \psi)-2 bC^2\alpha\cos(m \psi)\bigg],\\
        \delta R = &\, A\bigg[\frac{2(1-C^2)}{m\alpha}\sin(m \psi)-2b\alpha\cos(m \psi)\bigg],
    \end{align}
\end{subequations}
for arbitrary small amplitude $A$ and $\alpha = \sqrt{1-\frac{2}{\gamma}C^2}$. We find the linear solution is in good agreement with low amplitude numerical solutions of the full non-linear system (\ref{SSEfinal}).

\subsection{Non-linear numerical solutions}

\begin{figure}\centering
  \includegraphics[width=0.99\linewidth]{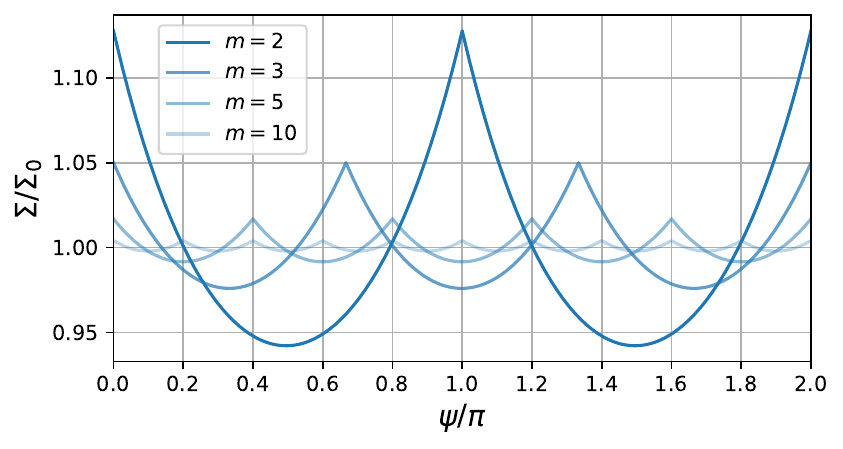}
  \vspace{-1em}
\caption{Comparison of density profiles for `cuspy' waves with $\mathcal{B} = \mathcal{B}_\text{max}$, $C = 0.1$, $\gamma = 5/3$ and various numbers of spiral arms. The background disc has $\Sigma_0\propto r^{-1}$. The two-armed spiral solution has a far higher amplitude than all others.}
  \label{SSSnscomp}
\end{figure}

\begin{figure}\centering
  \includegraphics[width=0.99\linewidth]{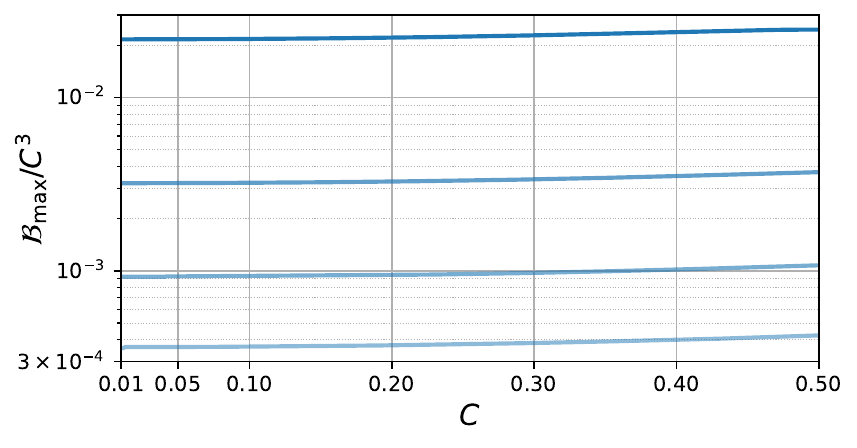}
  \vspace{-1em}
\caption{A comparison of the maximum possible wave action flux for self-similar smooth spiral waves with $2$ (darkest) to $5$ (lightest) arms; again we've taken $\gamma = 5/3$. We note the asymptotic scaling of $\mathcal{B}_\text{max}$ with $m^{-4}$, which we demonstrate later in section \ref{SFSW}.}
  \label{RWAcomp}
\end{figure}

\begin{figure}\centering
  \includegraphics[width=0.99\linewidth]{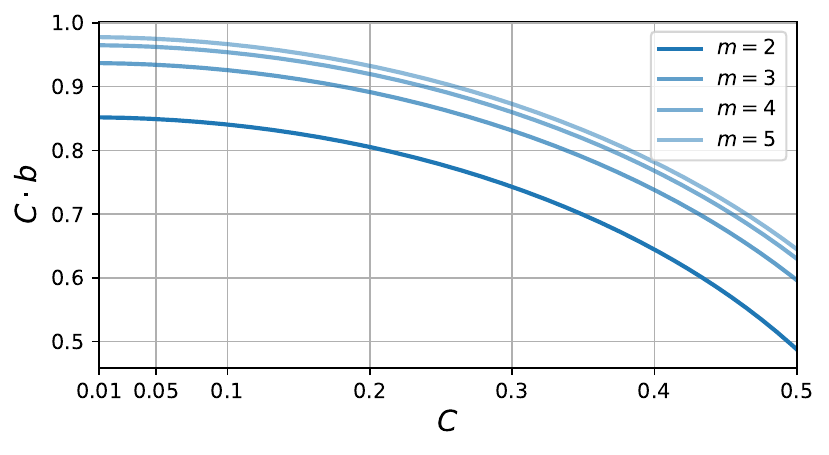}
  \includegraphics[width=0.99\linewidth]{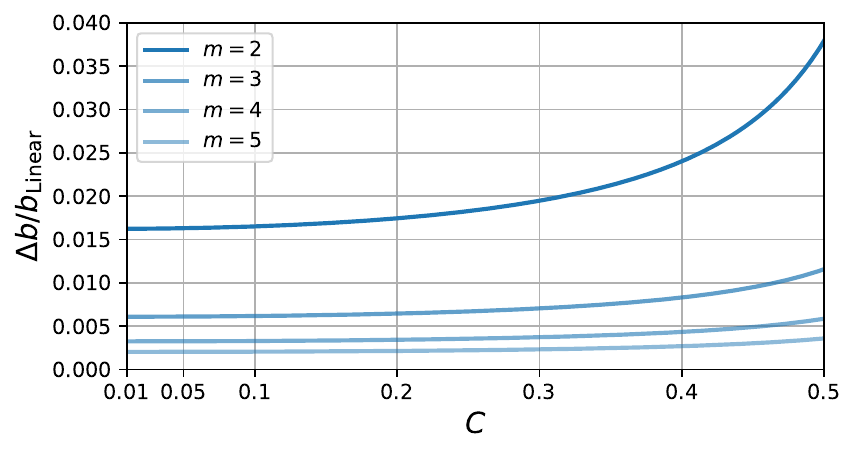}
  \vspace{-1em}
\caption{Top: spiral angle proxy $b = \tan\varphi$ for cuspy smooth spirals of maximal amplitude with various numbers of arms. Tightly wound spirals have $b \gg 1$. $C$ (defined in equation (\ref{Cdef})) measures the aspect ratio of the disc. Bottom: comparison of the winding angle for linear waves (from the dispersion relation (\ref{DR})) with that of the non-linear cuspy waves. Specifically, $\Delta b = b_{\text{linear}} - b_{\text{cuspy}}$. In both graphs $\gamma = 5/3$.}
  \label{Banglecomp}
\end{figure}

\begin{figure*}\centering
  \includegraphics[width=0.865\linewidth]{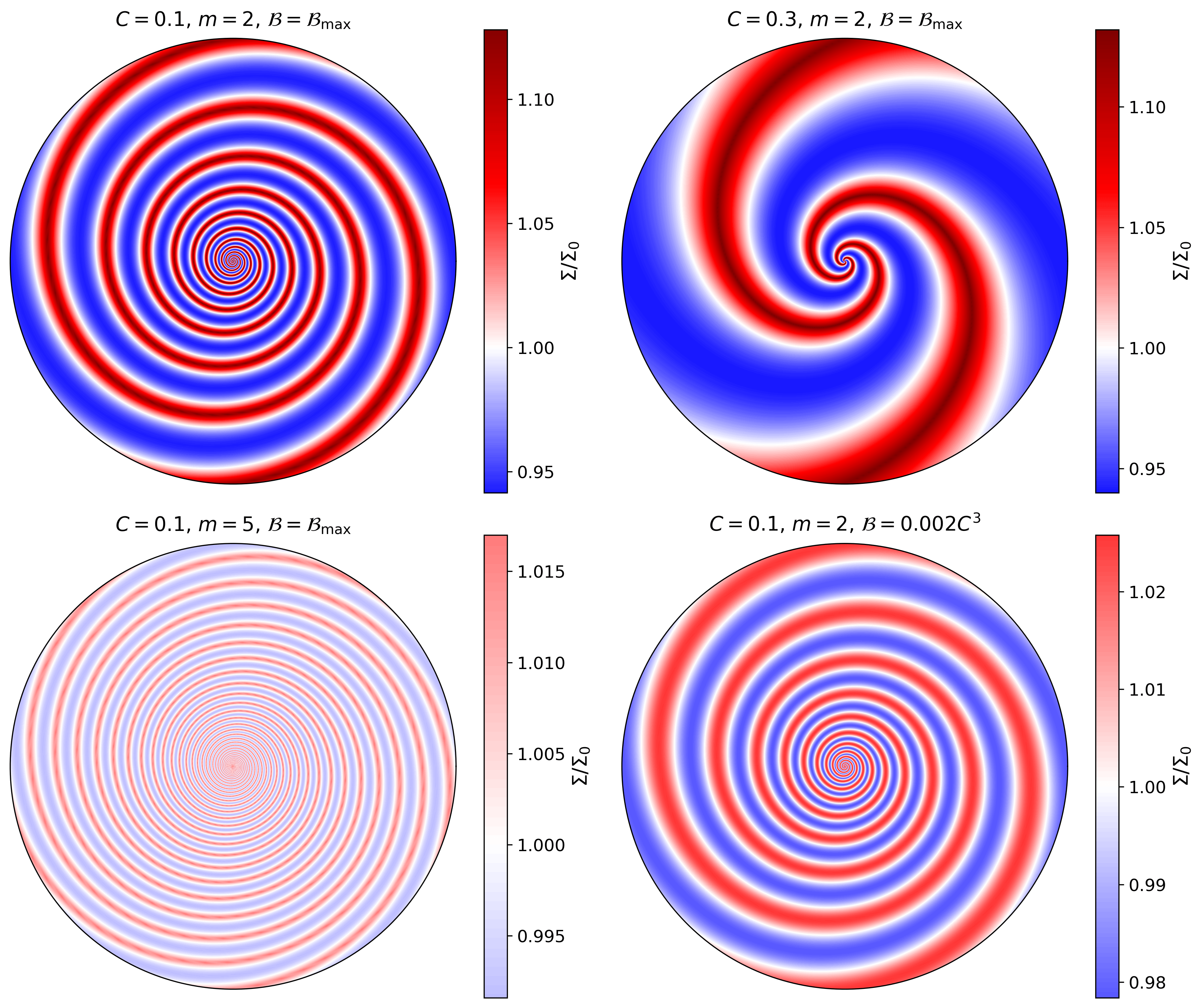}
  \vspace{-1em}
  \caption{Density plots of smooth non-linear spiral solutions to the system (\ref{SSEsys}). The background disc has $\Sigma_0\propto r^{-1}$ and $\gamma = 5/3$. Three plots show `cuspy' waves, which carry a maximal amount of radial wave action flux ($\mathcal{B}_{\text{max}}$) whilst remaining smooth (that is, containing no shocks in their profile). The bottom right plot shows a lower-amplitude two-armed wave with a more sinusoidal azimuthal shape. As well as its lower amplitude when compared with the two-armed cuspy wave in the top left plot, its positive and negative density perturbations are roughly equal in magnitude.}
  \label{SSSpirals}
\end{figure*}

We solved numerically the equations (\ref{OWA}) and (\ref{SSEfinal}), exploring a range of possible wave amplitudes, disc aspect ratios and numbers of spiral arms.

Figure \ref{SSSprofiles} shows azimuthal profiles for two-armed spiral waves of various amplitudes, and `cuspy' waves in discs of various aspect ratios. The `cuspy' waves carry a maximal radial wave action flux possible for smooth, steady self-similar spiral waves. The cusp of the waves lies precisely at a sonic point, where the flow velocity normal to the spiral wave crests is equal in magnitude to the sound speed. A comparison of cuspy waves with different numbers of spiral arms is shown in figure \ref{SSSnscomp}. Figure \ref{RWAcomp} shows the variation of the maximal radial wave action flux for discs with different aspect ratios and spiral waves with different numbers of arms. Figure \ref{Banglecomp} shows the variation of the spiral winding angle with disc aspect ratio, and compares values for non-linear `cuspy' waves with the linear predictions from the dispersion relation (\ref{DR}). Finally, figure \ref{SSSpirals} shows global surface density plots of these spiral waves in a few different scenarios.

Of particular interest to us is that our disc can support a finite wave action flux whilst remaining smooth. For example, a two-armed spiral in a self-similar disc with $C = 0.1$ permits a (non-dimensional) radial wave action flux of up to $0.0218 C^3$ whilst remaining smooth. For comparison, the same flux is carried by the inner spiral wave generated by a planet of mass ratio $q = M_p/M_\star = 0.24 h^3$.\footnote{To see this, note that in the local approximation, a planet in an (isothermal) disc excites a wave with angular momentum flux $\mathcal{B}_{\text{planet}} = 0.37 (G M_p)^2 \Sigma_p r_p \Omega_p/c_p^3 = 0.37 \left(q/h^3\right)^2 c_p^3 \Sigma_p r_p/\Omega_p$ \citep{dangelo_three-dimensional_2010,zhu_planet-disk_2012,brown_horseshoes_2024}. Reintroducing dimensions, $\mathcal{B}_{\text{max}} = 0.0218 c_p^3 \Sigma_p r_p/\Omega_p$, which, equating the sound speeds in the two models, is equal to the planet-driven angular momentum flux for $q = 0.24 h^3$.}

The wave action flux may be seen to scale naturally with $C^3$ from its expression when written in Lagrangian coordinates, $\mathcal{B} = \oint P \dd \xi_r$ (this expression is derived in appendix \ref{appx2}). Here, $P$ scales with $C^2$ and the radial displacement $\xi_r$ scales with $C$. This implies a leading order (in $C$) cancellation in the Eulerian integral expression for the angular momentum flux (\ref{AMFint}), which ostensibly scales as $C^2$. A quick numerical check verifies that this cancellation is indeed present.

Outstanding questions include how spiral waves in discs with different background profiles may be described, and how more general non-linear spiral wave packets containing multiple wavenumbers (akin to the waves excited by embedded planets) may be treated. It is these questions which we devote the majority of the remainder of this paper to answering.

\section{The spiral density wave equation}\label{GDW}

In this section we derive a simple, second-order, fully non-linear equation governing spiral density waves in a thin disc. This equation is valid away from the corotation radius of the waves, and gives an asymptotically exact description for thin, circular discs. The equation is formulated in a Lagrangian coordinate system, but the conversion back to standard Eulerian variables is simple. Whilst we originally set out to investigate non-linear aspects of planet-disc interactions, the generality of the non-linear model (\ref{MasterEqn}) extends considerably beyond this initial remit, with applications to steadily rotating (2D) spiral waves in a variety of contexts. Much of the derivation is comparable to the approach taken by \citet{1985ApJ...291..356S}, who studied non-linear waves excited at Lindblad resonances in pressure-less, self-gravitating discs (though they also discussed a post-hoc extension to include an isothermal pressure term).

In the following analysis we'll make use of two key approximations in addition to the adoption of 2D equations. We assume firstly that the sound speed $c$ and velocity perturbation $\vecv'$ (measuring the departure from a reference circular orbital flow) are far slower than the orbital flow in the frame corotating with the planet (or pattern, or perturber), which has angular frequency $\Omega_p$. Secondly, we assume that azimuthal and radial derivatives obey $\p_\theta \ll r\p_r \sim 1/h$. That is, the waves are tightly wound.\footnote{In most scenarios, this second condition is naturally satisfied as a consequence of our first approximation. For example, a wave propagating at the sound speed will naturally be sheared as it travels radially by the fast orbital flow into a tightly wound spiral. Note further that all the solutions in section \ref{S7} with $C \ll 1$ were tightly wound (see figure \ref{Banglecomp} in particular).}

The 2D steady Euler equations in the frame corotating with the planet (or pattern) read in polar coordinates:
\begin{subequations}\label{FULLEQS}
\begin{align}
    &u_r \p_r u_r + \frac{u_\theta}{r}\p_\theta u_r - \frac{u_\theta^2}{r} - 2\Omega_p u_\theta + \frac{\p_rP}{\Sigma} = \Omega_p^2 r - \p_r \Phi, \\
    &u_r \p_r u_\theta + \frac{u_\theta}{r}\p_\theta u_\theta + \frac{u_\theta u_r}{r} + 2\Omega_p u_r + \frac{\p_\theta P}{\Sigma r} = - \frac{\p_\theta \Phi}{r},\\
    &\frac{1}{r}\p_r\left(\Sigma r u_r\right) + \frac{1}{r}\p_\theta\left(\Sigma u_\theta\right) = 0,\label{MCglob}
\end{align}
\end{subequations}
with
\begin{equation}
    \Phi = -\frac{G M_\star}{r} + \Phi_p,
\end{equation}
where $\Phi_p$ is the gravitational potential of the planet or perturber, and $\vecu = u_r \vece_r + u_\theta \vece_\theta$ is the total flow velocity in the frame corotating with the planet.

We further assume the specific entropy to be conserved on streamlines,
\begin{equation}
    \vecu\cdot \nabla s = 0,
\end{equation}
from which we deduce $s = s_0(R_0)$, where $R_0$ is the original radial location of a fluid element displaced to $r=R$, and $s_0(r) = \text{const} + \ln\left[P_0(r)/\Sigma_0(r)^\gamma\right]$ is the specific entropy distribution of a circular background reference disc. That is,
\begin{equation}\label{EntLag}
    P = P_0(R_0)\left(\frac{\Sigma}{\Sigma_0(R_0)}\right)^\gamma,
\end{equation}
and further the sound speed $c$ may be related to the sound speed in the reference disc, $c_0 = \sqrt{\gamma P_0/\Sigma_0}$, via
\begin{equation}
    c^2 = c_0^2(R_0)\left(\frac{\Sigma}{\Sigma_0(R_0)}\right)^{\gamma-1}.
\end{equation}
Now, since the velocity perturbation is far slower than the orbital flow, it follows that the Lagrangian displacement $\bmath{\xi}$ of fluid elements away from the same circular reference flow is much shorter than the orbital radius. That is, $\left|\bmath{\xi}\right| \ll r$. As a corollary, it will be unimportant whether we evaluate $\Sigma_0$, $c_0^2$ and $P_0$ in the above equations at $R_0$ or $R$, as each of these background quantities is assumed to be slowly varying with radius. (The same is not true for $\Sigma$, $c^2$ and $P$ however, as in general these variables may vary on a shorter length-scale).

We treat mass conservation in a similar Lagrangian manner. We imagine a map transforming fluid elements from a circular reference flow to the non-linear flow which we wish to describe,
\begin{equation}\label{map}
    \bmath{R}_0 \to \bmath{R} = \bmath{R}_0 + \bmath{\xi}.
\end{equation}
Going forward, we'll use the subscript '$0$' to denote flow fields associated with the reference flow. We also define the `Lagrangian' coordinates $r_0$ and $\theta_0$ as the preimages of $r$ and $\theta$ under the map (\ref{map}). Equation (\ref{MCglob}) for mass conservation may be expressed in terms of the Jacobian of this map, $\mathbf{J} = \mathbf{I} + \nabla_0 \bmath{\xi}$, namely 
\begin{equation}\label{MCLag}
    \Sigma = \Sigma_0(R_0)/\det(\mathbf{J}).
\end{equation}
Since $\left|\bmath{\xi}\right| \ll r$, the radial displacement, $\xi_r = \bmath{\xi}\cdot \vece_{r} \approx \bmath{\xi}\cdot \vece_{r_0}$\footnote{Here $\vece_{r_0} = \bmath{R}_0/\left|\bmath{R}_0\right|$ is the Lagrangian unit radial basis vector associated with the circular reference system.}, is very well approximated by $R-R_0$. Similarly, we may self-consistently neglect terms like $\frac{1}{r}\p_\theta \xi_r$, $\frac{1}{r}\p_\theta\xi_\theta$ and $\bmath{\xi}/r$ within our approximation regime. The expression for the Jacobian then reads
\begin{equation}
    \mathbf{J} = \begin{pmatrix} 1 + \frac{\p \xi_r}{\p R_0} & 0\\ \frac{\p \xi_\theta}{\p R_0} & 1\end{pmatrix} + \ldots.
\end{equation}
It follows that the expression for $\det(\mathbf{J})$ takes the simple form
\begin{equation}
    \det(\mathbf{J}) = \frac{\Sigma_0}{\Sigma} \approx \frac{\p R}{\p R_0}.
\end{equation}

We now return to the momentum equations, and define the angular velocity $\Omega(r)$ via
\begin{equation}
    r\Omega^2 = \frac{G M_\star}{r^2} + \frac{\p_rP_0}{\Sigma_0}.
\end{equation}
We let $u_r = u$, $u_\theta = r(\Omega- \Omega_p) + v$. Approximating $u_\theta \p_\theta \approx r(\Omega- \Omega_p)\p_\theta$, and ignoring the contribution from the planet's potential, the now free equations of motion become
\begin{subequations}\label{Momfree}
\begin{align}
    &(\Omega- \Omega_p)\p_\theta u + u \p_r u - 2\Omega v + \left(\frac{\p_rP}{\Sigma} - \frac{\p_r P_0(r)}{\Sigma_0(r)}\right) = \frac{v^2}{r}, \\
    &(\Omega- \Omega_p)\p_\theta v + u \p_r v + 2 B u = - \frac{uv}{r} - \frac{1}{\Sigma r}\p_\theta P,
\end{align}
\end{subequations}
where $B(r) \equiv \Omega + \frac{r}{2}\frac{\dd \Omega}{\dd r}$ is the Oort parameter.

It will transpire that each term on the right-hand side of the above system may be self-consistently neglected within our approximation scheme, and we'll see conversely that each term on the left-hand side contributes importantly.

We define the non-dimensional radial coordinate variable (which is similar, though not identical to that chosen in \citet{rafikov_nonlinear_2002}):
\begin{equation}\label{xdefn}
    x = \int_{r_p}^r \frac{\Omega(r')-\Omega_p}{c_0(r')}\dd r'.
\end{equation}
This variable choice is particularly convenient, as the characteristic curves traced by spiral waves in the $(x,\theta)$ coordinate system are very well approximated by straight lines. The choice of the lower integration limit here is discretionary, but a convenient choice is the corotation radius of the pattern (with $\Omega(r_p) = \Omega_p$). In the case of an embedded planet, this is very close to the planet's orbital radius. If however there is no corotation radius, for example in the case $\Omega_p \leqslant 0$, then replacing $r_p$ with an arbitrary reference radius suffices.

The original location of a fluid element initially at $R_0$ expressed in this new coordinate system is
\begin{equation}
    X_0 = \int_{r_p}^{R_0} \frac{\Omega(r)-\Omega_p}{c_0(r)}\dd r.
\end{equation}
We define the $x-$displacement, which we denote $\eta$ (to avoid confusion with the previously defined $\bmath{\xi}$), via
\begin{equation}
    \eta = X - X_0.
\end{equation}
Now, because both $\Omega(r)$ and $c_0(r)$ are slowly varying with radius, it follows that
\begin{equation}
    \frac{\p X_0}{\p R_0} = \frac{\Omega(R_0) - \Omega_p}{c_0(R_0)} \approx \frac{\Omega(R) - \Omega_p}{c_0(R)} = \frac{\p X}{\p R},
\end{equation}
hence
\begin{equation}
    \frac{\p X}{\p X_0} = \frac{\p R}{\p R_0}.
\end{equation}
It follows (from equation \ref{MCLag}) that
\begin{equation}\label{sigmarel}
    \Sigma = \frac{\Sigma_0(R_0)}{1 + \p_{X_0}\eta},
\end{equation}
and further
\begin{equation}\label{xtoX0}
    \frac{1}{\Sigma}\p_{x} = \frac{1}{\Sigma_0(R_0)}\p_{X_0},
\end{equation}
where both partial derivatives are taken at constant $\theta$.

We now aim to transform the momentum equations (\ref{Momfree}) from Eulerian to Lagrangian coordinate systems, via the map
\begin{equation}
(r,\theta) \to (X_0,\theta_0).
\end{equation}
Importantly, $u$ and $v$ will remain the Eulerian velocity components\footnote{The Lagrangian components of the perturbed flow velocity are given explicitly in equation (\ref{lagpertvel}).}, that is, projections of the flow velocity onto radial and azimuthal basis vectors $\vece_r$ and $\vece_\theta$ respectively, though we shall now think of them, as well as the other flow fields, as functions of $X_0$ and $\theta_0$. 

It turns out that because $\theta - \theta_0 \approx \xi_\theta/r \ll 1$, and $\p_\theta \sim 1$, it's unimportant whether we use Eulerian angle $\theta$ versus $\theta_0$ in the following analysis. For notational simplicity we therefore drop the `$0$' subscript on $\theta_0$.

Consequently, the relation (\ref{xtoX0}) between $x$ and $X_0$ derivatives remains valid at leading order in $h$ when $\p_{X_0}$ is considered at constant $\theta_0$. Transforming the azimuthal derivative yields\footnote{It may be seen that $\p R/\p\theta\big|_{R_0} = u/(\Omega-\Omega_p)$ by considering the angle subtended between the (closed) streamline $R(\theta,R_0) \approx R_0 + \xi_r(\theta,R_0)$ and the azimuthal direction. We can also reconcile this expression to the definition $\DD\bmath{\xi}/\DD t = \vecu - r_0(\Omega(r_0) - \Omega_p)\vece_{\theta_0}$, which notably includes a covariant derivative with respect to $\theta_0$, and introduces the (undesirable) azimuthal displacement $\xi_\theta$, as follows: $\DD\bmath{\xi}/\DD t \cdot \vece_{r_0} = (\Omega(r_0)-\Omega_p)\left(\DD\xi_r/\DD \theta_0 - \xi_\theta\right) = u - (\Omega - \Omega_p)\xi_\theta$ by equation (\ref{lagpertvel}). The terms involving $\xi_\theta$ cancel, and we see it's important that $u$ remains the Eulerian radial component of the velocity.}:
\begin{equation}\label{MatDer}
    \frac{\DD}{\DD \theta}\bigg|_{X_0} = \p_\theta + \frac{u}{c_0(r)} \p_x \equiv \p_\theta + \frac{u}{\Omega - \Omega_p} \p_r,
\end{equation}
That is,
\begin{equation}\label{urel}
    \frac{\DD \eta}{\DD \theta} = \frac{u}{c_0(r)} \approx \frac{u}{c_0(R_0)}.
\end{equation}
It's possibly helpful therefore to imagine $\frac{\DD}{\DD\theta} \equiv \p_\theta\big|_{X_0}$ as a material derivative in our new coordinate system. Indeed relabelling $\theta \to t$ yields an almost exact analogy to the waves studied by \citet{fromang_properties_2007}. 

The radial equation of motion is (neglecting $v^2/r \ll \Omega v$): 
\begin{equation}\label{reomrel}
    \frac{\DD u}{\DD \theta} - \frac{2\Omega v}{\Omega - \Omega_p} + \frac{1}{c_0(r)}\left(\frac{\p_x P}{\Sigma} - \frac{\p_x P_0(r)}{\Sigma_0}\right) = 0.
\end{equation}
Now, from equation (\ref{EntLag}) for entropy conservation, we have
\begin{equation}\label{PEQN}
    P = P_0(R_0)\left(\frac{1}{1 + \p_{X_0}\eta}\right)^\gamma,
\end{equation}
Using equation (\ref{xtoX0}) to transform the derivative with respect to $x$, further estimating remaining slowly varying factors associated with the background state involving $R/R_0 \approx 1$, and substituting in equation (\ref{PEQN}), we arrive at the equation
\begin{multline}\label{reommanip}
    \frac{\DD ^2\eta}{\DD \theta^2} - \frac{2\Omega}{\Omega - \Omega_p}\frac{v}{c_0} + \frac{1}{\gamma}\p_{X_0}\left(\frac{1}{(1 + \p_{X_0}\eta)^\gamma}\right) \\ = \frac{\p_{X_0}P_0(R_0)}{c_0^2\Sigma_0}\left(1-\frac{1}{(1 + \p_{X_0}\eta)^\gamma}\right)\approx 0.
\end{multline}
Within our approximation scheme, the term on the right-hand side is small relative to the left-hand side in both the highly non-linear case $\eta = \mathcal{O}(1)$ and the weakly non-linear case $\eta \ll 1$, because
\begin{equation}
    \frac{\p_{X_0}P_0}{c_0^2\Sigma_0} = \frac{c_0}{\Omega - \Omega_p}\frac{\p_{R_0}P_0}{c_0^2\Sigma_0} \sim \frac{c}{r(\Omega - \Omega_p)} \ll 1,
\end{equation}
and further since the term scales linearly with $\p_{X_0}\eta$ for $\eta \ll 1$. The right-hand side of equation (\ref{reommanip}) is therefore small uniformly.

The azimuthal equation of motion is (discarding $uv/r \ll Bu$):
\begin{equation}
    \frac{\DD v}{\DD \theta} + 2 \frac{Bu}{\Omega - \Omega_p} = - \frac{1}{r(\Omega - \Omega_p)}\frac{\p_\theta P}{\Sigma}.
\end{equation}
Now, the term involving the azimuthal pressure gradient has size $c/r(\Omega - \Omega_p)$ relative to the left-hand side, and may be self-consistently neglected using the tight-winding approximation. We'll further discover (see equations (\ref{aziPint}) and (\ref{Iint})) that the azimuthal pressure gradient does not produce any accumulating contributions noticeable on a global scale at leading order either.

We integrate the leftover equation with respect to $\theta$ to get
\begin{equation}
    \frac{v}{c_0} + \frac{2 B}{\Omega - \Omega_p} X = \frac{2 B}{\Omega - \Omega_p} X_0,
\end{equation}
\begin{equation}\label{SAMcons}
    \implies \frac{v}{c_0} + \frac{2 B}{\Omega - \Omega_p} \eta = 0.
\end{equation}
That is, fluid elements conserve their specific angular momentum at leading order.

Finally, we define the slowly varying dimensionless `effective squared epicyclic frequency'
\begin{equation}
    \tilde{\kappa}_0^2(R_0) = \frac{4\Omega(R_0) B(R_0)}{(\Omega(R_0) - \Omega_p)^2},
\end{equation}
which compares the epicyclic frequency $\kappa = \sqrt{4\Omega B}$ to the orbital frequency in the corotating frame, $\Omega - \Omega_p$.

Combining equation (\ref{SAMcons}) with the radial equation of motion yields the non-linear wave equation
\begin{equation}\label{GNLDWEOM}
    \eta_{\theta \theta} + \tilde{\kappa}_0^2\eta + \p_{X_0}\left[\frac{1}{\gamma}\left(\frac{1}{\left(1+\eta_{X_0}\right)^\gamma}-1\right)\right] = 0,
\end{equation}
where we've denoted derivatives $\frac{\DD \eta}{\DD \theta}$ with a subscript $\eta_\theta$, and similarly for spatial derivatives $\p_{X_0}\eta = \eta_{X_0}$.

For a Keplerian disc, note that $\tilde{\kappa}_0 \to 1$ as $R_0 \to 0$, and conversely $\tilde{\kappa}_0 \to 0$ as $R_0 \to \infty$. The term $\tilde{\kappa}_0^2 \eta$ provides a dispersive inertial contribution to an otherwise acoustic wave. Spiral waves with a given wavelength are effectively more dispersive in regions of disc where $\tilde{\kappa}_0^2$ is larger. In this sense, the dispersion of wave-packets is appreciable in the inner disc (inside the planet or pattern's corotation radius), but relatively weak in the outer disc. Remarkably, setting $\gamma = 1$ and taking the inner disc limit of equation (\ref{GNLDWEOM}), we recover the wave equation derived by \citet{fromang_properties_2007} governing non-linear \emph{axisymmetric} density waves in the shearing sheet.

\subsection{Wave action conservation and generalisation to a complete global description}

Equation (\ref{GNLDWEOM}) provides an accurate description of non-linear spiral waves over a short radial patch of the disc. What's not explicitly accounted for is how the wave evolves over very long length-scales $\sim r$. Small contributions from terms neglected in our approximation scheme may accumulate to give meaningful contributions over long length-scales. For example, one item of information not derivable from the wave equation (\ref{GNLDWEOM}) is the expression for the conserved radial wave action flux of density waves (even after linearisation).

In the appendix \ref{appx2}, we demonstrate using the Generalised Lagrangian Mean (GLM) theory of \citet{andrews_wave-action_1978} that smooth tightly wound spiral waves carry a conserved radial wave action flux\footnote{Reassuringly, we recover from this expression the linear wave action flux (e.g. \citet[equation (7)]{rafikov_nonlinear_2002}) in the small amplitude case.}:
\begin{equation}\label{WAint}
    \mathcal{B} = - \frac{R_0 \Sigma_0 c_0^3}{\Omega_0 - \Omega_p}\int_{0}^{2\upi}\frac{\eta_\theta}{\gamma}\left(\frac{1}{(1+\eta_{X_0})^\gamma} - 1\right)\dd \theta.
\end{equation}
Whilst this wave action integral is formulated in terms of Lagrangian coordinates, we obtain excellent numerical agreement when comparing to the equivalent Eulerian angular momentum flux expression (\ref{AMFint}) for the self-similar waves of section \ref{S7}, and refer the reader to section \ref{ctss} for further discussion.

A complete description of spiral waves should include both the dominant local dynamics described by equation (\ref{GNLDWEOM}), as well as the large-scale evolution of a wave in radius and azimuth, capturing how its amplitude and phase vary due to variations in the medium within which it propagates, as well as the influence of smaller terms perhaps neglected in the leading wave equation, which may nevertheless contribute a meaningful net effect when allowed to act over long length-scales.

This large-scale evolution is described by the non-linear wave action equation,\footnote{as well as the `conservation of waves' condition (\ref{CW})} which, for a general non-linear wave, is given by \citet[equation (2.15)]{andrews_wave-action_1978}. A model which captures the local wave dynamics and also reproduces this wave action equation for spiral waves in thin discs would therefore offer a complete global description.

Now, we have the freedom to introduce a slowly varying factor into equation (\ref{GNLDWEOM}) (dependent on the background disc state) without modifying the leading local order dynamics. Remarkably, there is a unique choice for which the equation of motion naturally conserves the (GLM) wave action, and thus captures faithfully the global behaviour of spiral waves obeying the full equations of motion (\ref{FULLEQS}). We verify this explicitly in appendices \ref{appx} and \ref{appxb1}. This choice is:
\begin{equation}\label{MasterEqn}
    \mathcal{F}\eta_{\theta\theta} + \tilde{\kappa}_0^2\mathcal{F}\eta + \p_{X_0}\left[\frac{\mathcal{F}}{\gamma}\left(\frac{1}{(1+\eta_{X_0})^\gamma}-1\right)\right] = 0,
\end{equation}
for slowly varying
\begin{equation}\label{Fdef}
    \mathcal{F}(X_0) = \frac{R_0 \Sigma_0 c_0^3}{\Omega_0 - \Omega_p}.
\end{equation}
Equation (\ref{MasterEqn}) now offers a global description of free, tightly wound waves in a thin accretion disc in the absence of dissipative effects. The generality is considerable, as many accretion discs are thin, and many waves therein can be considered tightly wound. Whilst the derivation presented assumes smooth solutions, solutions containing shocks may be obtained by solving equation (\ref{MasterEqn}) piecewise away from shocks and imposing appropriate jump conditions (that is, the Rankine--Hugoniot relations) across discontinuities. A practical approach to obtain solutions for shock waves of very high amplitude is outlined in sections \ref{UNLW} and \ref{WAES}.

Equation (\ref{MasterEqn}) constitutes the main result of this paper: the remainder of this paper will predominantly discuss the application of the theory just derived to waves driven by embedded planets. In section \ref{SFPA} we examine the two solution regimes which naturally arise, associated with low and high amplitude forcing.

\subsection{Lagrangian formulation of spiral waves}

Equation (\ref{MasterEqn}) is the Euler--Lagrange equation associated with the Lagrangian
\begin{multline}\label{Lagrangian}
	L = \iint \mathcal{F}\Bigg[\frac{1}{2}\eta_\theta^2 -\frac{1}{2}\tilde{\kappa}_0^2 \eta^2 \\ - \frac{1}{\gamma(\gamma-1)(1+\eta_{X_0})^{\gamma-1}} - \frac{\eta_{X_0}}{\gamma}\Bigg] \, \dd X_0 \,\dd \theta.
\end{multline}
Now, reintroducing the radial displacement $\xi_r = \frac{c_0}{\Omega-\Omega_p}\eta$, and noting $\frac{c_0}{\Omega - \Omega_p}\dd X_0 = \dd R_0$, we see that the Lagrangian may be re-cast at leading order in terms of the more familiar polar coordinates $(R_0,\theta)$ as
\begin{multline*}
	L = \iint \Sigma_0\Bigg[\frac{1}{2}\left[(\Omega-\Omega_p)\frac{\DD\xi_r}{\DD \theta}\right]^2 -\frac{1}{2}\kappa^2 \xi_r^2 \\ - \frac{c_0^2}{\gamma(\gamma-1)(1+\p_{R_0}\xi_{r})^{\gamma-1}}\Bigg] R_0 \,\dd R_0 \,\dd \theta 
\end{multline*}
\begin{equation}\label{PolarLag}
	= \int \Sigma_0\left[\frac{1}{2}\left[(\Omega-\Omega_p)\frac{\DD\xi_r}{\DD \theta}\right]^2 -\frac{1}{2}\kappa^2 \xi_r^2 - \frac{P}{(\gamma-1)\Sigma}\right] \,\dd A,\!
\end{equation}
where the derivative $\frac{\DD}{\DD \theta}$ is taken at constant Lagrangian radial coordinate $R_0$, and $\kappa^2 = 4 \Omega B$ is the (true) epicyclic frequency.

Equation (\ref{PolarLag}) is now, reassuringly, more recognisable as a simplification of the Lagrangian for hydrodynamic flow\footnote{For a direct comparison, see, for example, \citet[equation 2.6]{salmon_hamiltonian_1988}. The simplification is more easily undertaken in the case $P_0 \equiv \text{const}$, making use of equations (\ref{samcons}) and (\ref{xiazi}) in particular. In general, and not coincidentally, the manipulations required are similar to those undertaken in appendix \ref{appx2}. Indeed, in order to prove that equation (\ref{MasterEqn}) is an accurate global wave equation, it suffices to show that the integrand in equation (\ref{Lagrangian}) is everywhere the leading approximation to the Lagrangian density for 2D hydrodynamic flow (under appropriate assumptions). The simplified Lagrangian may then be seen to everywhere predict the leading approximations to the wave action density and its flux, meaning that wave evolution over large scales is accurately captured.}. In practice equation (\ref{Lagrangian}) gives the useful Lagrangian formulation of tightly wound non-linear spirals, as otherwise care is needed to appropriately treat slowly varying factors associated with the background disc.

\section{Spirals with fixed pitch angle: simplification to a 1D ODE}\label{SFPA}

Perhaps the defining property of spiral waves is that, non-linear or not, their crests trace spiral curves. The pitch angle of these curves may in general vary over the global extent of the disc, or perhaps across a spiral wave-packet itself, but nevertheless at each point on the spiral we may define a pitch angle.

We can capture this property mathematically with the condition that spiral waves may locally be described as functions of the phase $\psi = k_r r + m \theta$, for $k_r$ and $m$ radial and angular wavenumbers (which may vary over long length-scales). That is, wave crests trace contours of $\psi$, which are spiral curves making an angle $\varphi = \tan^{-1}\left(\frac{k_r r}{m}\right)$ with the radial direction\footnote{A similar principle justifies the choice of the shearing coordinates by \citet{heinemann_weakly_2012} to describe non-linear waves in the shearing sheet.}.

We therefore seek local spiral wave solutions to equation (\ref{GNLDWEOM}) which are functions of the phase variable $\phi = X_0 - \mathcal{U}\theta$, for constant `effective phase speed' $\mathcal{U}$\footnote{The choice of notation here is to emphasise and preserve the connection to the travelling waves of \citet{fromang_properties_2007}; the spiral angle is equal to 90 degrees minus the pitch angle, and is determined by $\tan \varphi = -\frac{r(\Omega-\Omega_p)}{c_0\mathcal{U}}$.} (we treat the case of slowly varying $\mathcal{U}$ in section \ref{VSA} and appendix \ref{appxb2}). This allows us to directly describe spiral waves with the model derived in section \ref{GDW}, and we'll gain considerable insight in doing so.

Equation (\ref{GNLDWEOM}) becomes the 1D ordinary differential equation (studied previously by \citet{fromang_properties_2007}):
\begin{equation}\label{Spiraleqn}
    \mathcal{U}^2\eta_{\phi\phi} + \tilde{\kappa}_0^2\eta + \p_\phi\left[\frac{1}{\gamma}\left(\frac{1}{\left(1+\eta_\phi\right)^\gamma}-1\right)\right] = 0.
\end{equation}

\begin{figure}\centering
  \includegraphics[width=0.99\linewidth]{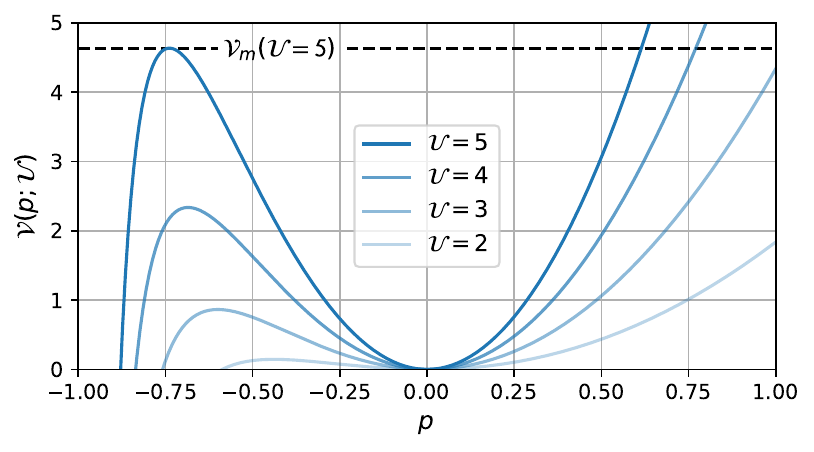}
  \vspace{-1em}
  \caption{Profile of the potential $\mathcal{V}(p\equiv \eta_\phi;\mathcal{U})$ for $\gamma = 1.4$ and various values of the effective phase speed/spiral-angle proxy $\mathcal{U}$. Higher values of $\mathcal{U}$ yield deeper wells, allowing for 'higher energy' solutions before the particle escapes the well.}
  \label{Vgraph}
\end{figure}

Equation (\ref{Spiraleqn}) conserves the energy-like quantity $\tilde{\mathcal{H}}$, obtained by multiplying (\ref{Spiraleqn}) through by $\eta_\phi$ and integrating:
\begin{equation}
    \tilde{\mathcal{H}} = \frac{1}{2}\mathcal{U}^2\eta_\phi^2 + \frac{1}{2}\tilde{\kappa}_0^2 \eta^2 + \frac{1}{\gamma(\gamma-1)}\left[\frac{1+\gamma \eta_\phi}{(1 + \eta_\phi)^\gamma} - 1\right].
\end{equation}
It's possible to interpret the dynamics as corresponding to that of a particle in a potential well. Specifically, writing $p = \eta_\phi$ and re-expressing $\eta$ in terms of $p_\phi$ via the equation of motion (\ref{Spiraleqn}), we obtain
\begin{equation}\label{ParticleAN}
    \frac{1}{2\tilde{\kappa}_0^2}\left[\mathcal{U}^2 - \frac{1}{(1+p)^{\gamma+1}}\right]^2p_\phi^2 + \mathcal{V}\left(p;\mathcal{U}\right) = \tilde{\mathcal{H}},
\end{equation}
where
\begin{equation}
    \mathcal{V}(p;\mathcal{U}) = \frac{1}{2}\mathcal{U}^2p^2 + \frac{1}{\gamma(\gamma - 1)}\left(\frac{1 + \gamma p}{\left(1 + p\right)^\gamma}-1\right)
\end{equation}
is the effective potential within which the particle sits\footnote{Equation (\ref{ParticleAN}) includes the unusual effect in which the `mass' of the imagined particle varies with position. This effect can be eliminated, and the analogy with a particle in a potential well can be made exact if $\mathcal{V}(p(\mu);\mathcal{U})$ is considered as a (transcendental) function of the canonical momentum $\mu = \mathcal{U}^2p + \frac{1}{\gamma}\left[\frac{1}{(1+p)^\gamma}-1\right]$. This gives $\tilde{\mathcal{H}} = \frac{1}{2}\tilde{\kappa}_0^2\eta^2 + \mathcal{V}(\mu;\mathcal{U})$, and Hamilton's equations are $\eta_\phi = \tilde{\mathcal{H}}_\mu = \mathcal{V}'(\mu)$, $\mu_\phi = -\tilde{\mathcal{H}}_\eta = -\tilde{\kappa}_0^2 \eta$. The exact analogy is complete when we associate $\mu$ with the position of the particle, $-\eta$ with its momentum and $1/\tilde{\kappa}_0^2$ with its mass. It's possible to write down an explicit expression for the potential $\mathcal{V}(\mu;\mathcal{U})$ in the case $\gamma = 1$, however this footnote is too small to contain it.}. Figure \ref{Vgraph} shows the profile of the potential $\mathcal{V}(p;\mathcal{U})$ for various values of $\mathcal{U}$.
Notably, the depth of the potential well formed by $\mathcal{V}(p;\mathcal{U})$ is only finite: if our imagined particle has an energy higher than this depth, it will `escape' the well. These solutions where the particle escapes the well correspond to waves with shocks in their profile, and we explore them further in section \ref{SFSW} below.

Just as all non-linear spiral solutions found in section \ref{S7} had $\tan\varphi < 1/c_0$, our spiral waves here must have $\left|\mathcal{U}\right| > 1$; indeed, the potential $\mathcal{V}$ no longer forms a well for $\left|\mathcal{U}\right| \leqslant 1$.

The local maximum of the potential $\mathcal{V}$, determining the depth of the well, may be found by differentiation and has the algebraic form:
\begin{equation}
    \mathcal{V}_m = \frac{1}{2}\mathcal{U}^2 - \frac{\gamma+1}{\gamma}\mathcal{U}^{\frac{2\gamma}{\gamma+1}} + \frac{\gamma+1}{2(\gamma-1)}\mathcal{U}^{\frac{2(\gamma-1)}{\gamma+1}} - \frac{1}{\gamma(\gamma-1)}.
\end{equation}

Smooth spiral wave solutions exist for values $\tilde{\mathcal{H}} < \mathcal{V}_m(\mathcal{U})$, and $\mathcal{V}_m(\mathcal{U})$ is an increasing function of $\left|\mathcal{U}\right| \geqslant 1$, which is apparent from the expression for its derivative
\begin{equation}
    \mathcal{V}_m'(\mathcal{U}) = \mathcal{U}\left(1- \mathcal{U}^{-2/(\gamma+1)}\right)^2.
\end{equation}
That is, more loosely wound spirals may be larger in amplitude.

Smooth spiral wave solutions to the second order differential equation (\ref{Spiraleqn}) have two constant parameters which must be set: $\tilde{\mathcal{H}}$, as well as an (arbitrary) reference value for the phase. There is a further important boundary condition which is that the waves ought to be $2\upi$-periodic in $\theta$, and this imposes a quantizing restriction on the phase speed $\mathcal{U}$, asserting that $\eta(\phi) \equiv \eta(\phi + 2\upi\mathcal{U})$. In terms of the wavelength (measured using the $X_0$ variable)
\begin{equation}
    \lambda = \int \dd \phi = \oint \frac{\dd \eta}{p},
\end{equation}
the quantisation condition yields the dispersion relation
\begin{equation}\label{Qcon}
    m \lambda = 2\upi \mathcal{U}.
\end{equation}
Just as in the case of the self-similar spirals described in section \ref{S7}, the non-linearity has no more than a $2\%$ impact\footnote{The largest non-linear discrepancy between the linear and non-linear cases is shown in figure \ref{Banglecomp}.} (for smooth spirals) on the phase speed $\mathcal{U}$ deduced from the dispersion relation (\ref{Qcon}). However, this discrepancy can be significant for large-amplitude waves with shocks, which are more loosely wound than their linear counterparts (see section \ref{UNLW}). The linear dispersion relation\footnote{which may be seen to coincide with the more familiar equation for tightly wound spirals $m^2(\Omega-\Omega_p)^2 = \kappa^2 + c^2k_r^2$,} reads
\begin{equation}
    \mathcal{U}^2 = \frac{m^2}{m^2-\tilde{\kappa}_0^2},
\end{equation}
and we see that $\tilde{\kappa}_0<1$ is necessary to allow for 1-armed spirals (a condition which is only met for radii $r>2^{2/3}r_p$ in a Keplerian disc). Smooth wave solutions of various amplitudes are discussed further in \citet{fromang_properties_2007}, though they essentially resemble the azimuthal profiles of the self-similar waves in figure \ref{SSSprofiles} (left).

\subsection{Comparison to self-similar spirals}\label{ctss}

At this point it's appropriate to compare the exact Eulerian solutions for spiral waves found in section \ref{S7} to the wave solutions of the spiral wave equation (\ref{Spiraleqn}). We expect asymptotic agreement in the limit of a thin disc with the self-similar profile $\Sigma \propto r^{-1}$, $c_s \propto r^{-1/2}$ and zero pattern frequency. Note that in this case $\mathcal{F}\equiv \text{const}$ and $\tilde{\kappa}_0 \equiv 1$, and equations (\ref{GNLDWEOM}) and (\ref{MasterEqn}) coincide.

We convert between the Lagrangian coordinates in the wave equation (\ref{Spiraleqn}) and the more familiar Eulerian coordinates (used in section \ref{S7}) by noting that $\phi = x - \eta - \mathcal{U}\theta$. At a fixed Eulerian radius (constant $x$), we therefore have
\begin{equation}\label{thetacr}
    \theta = \text{const} - \frac{\phi +\eta(\phi)}{\mathcal{U}}.
\end{equation}
Plotting $1/(1+\eta_{X_0})$ against $-(\phi +\eta(\phi))/\mathcal{U}$ therefore yields the graph of $\Sigma/\Sigma_0$ as a function of $\theta$ at constant $r$.

\begin{figure}\centering
  \includegraphics[width=0.99\linewidth]{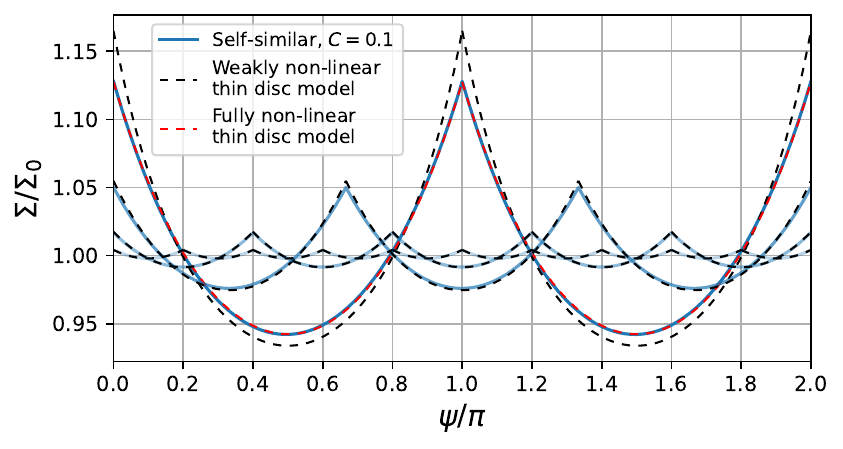}
  \vspace{-1em}
  \caption{Fully non-linear and weakly non-linear wave solutions satisfying equations (\ref{Spiraleqn}) and (\ref{NLDWNL}) respectively compared to exact solutions obtained in section \ref{S7} in the case of a `self-similar' disc with $C = 0.1$, $\Omega_p = 0$, $\Sigma_0 \propto r^{-1}$ and $T_0 \propto r^{-1}$. All solutions are plotted at constant physical radius (as opposed to constant Lagrangian coordinate $X_0$). The very small discrepancy between the two-armed fully non-linear solution and the analogous self-similar solution is due to the finite disc thickness.}
  \label{NLTDcomp}
\end{figure}

Figure \ref{NLTDcomp} compares the exact cuspy self-similar wave solutions with various $m$ shown in figure \ref{SSSnscomp} to the two-armed solution of the fully non-linear wave equation (\ref{Spiraleqn}) as well as `weakly non-linear' solutions derived in the next section.

We can also compare the angular momentum flux and wave action flux of the two classes of solutions. In the case of the two-armed cuspy spiral, equation (\ref{WAint}) gives a wave action flux $\mathcal{B} = 0.0217r_0 \Sigma_0 c_0^3/\Omega$, which differs only slightly from the Eulerian angular momentum flux value $\mathcal{B} = 0.0218r \Sigma_0 c_0^3/\Omega$ for $C = 0.1$. In fact, the y-intercept of the $m=2$ line in figure \ref{RWAcomp} is $0.0217$. This strongly suggests that the greatest angular momentum flux a steady wave may carry whilst remaining smooth is
\begin{equation}
    \mathcal{B}_\text{max} = 0.0217 r \Sigma_0 c_0^3/\Omega
\end{equation}
when $\gamma = 5/3$.

The deeper connection between the (non-linear) Eulerian and Lagrangian expressions for the angular momentum and wave action fluxes is slightly obscured by the fact that the pressure gradient contributes no radial flux in an Eulerian framework (the associated torque is entirely in the azimuthal direction). Conversely, the Lagrangian coordinate system traces flow streamlines, so that the advective flux of angular momentum naturally vanishes in this picture. Nevertheless, equations (\ref{AMFint}) and (\ref{WAint}) often describe the same quantity. For an explicit verification of their connection, we refer the reader to appendix \ref{appx1}.

\subsection{Shock formation in spiral waves}\label{SFSW}

\begin{figure*}\centering
    \begin{subfigure}{.6\textwidth}
      \includegraphics[width=0.49\linewidth]{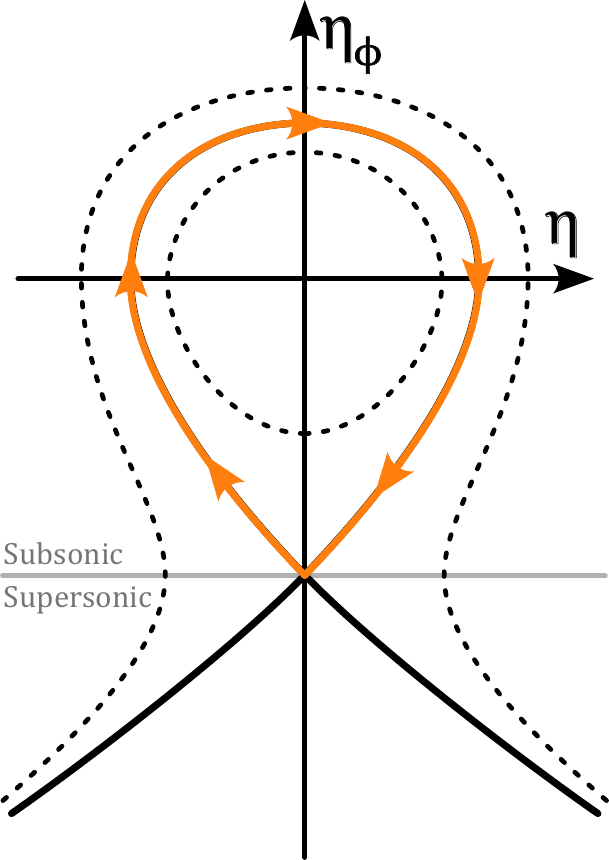}
      \includegraphics[width=0.49\linewidth]{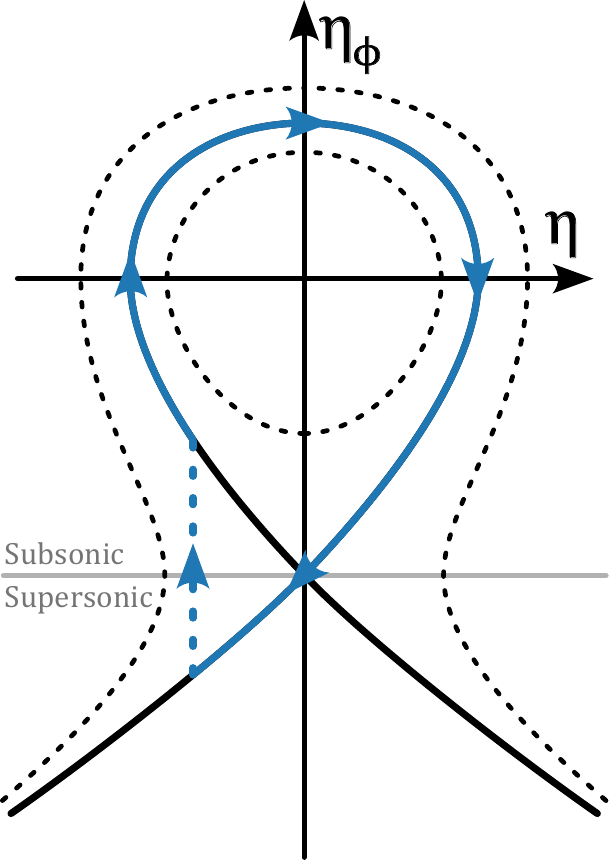}
    \end{subfigure}
    \begin{subfigure}{.39\textwidth}
      \includegraphics[width=0.99\linewidth]{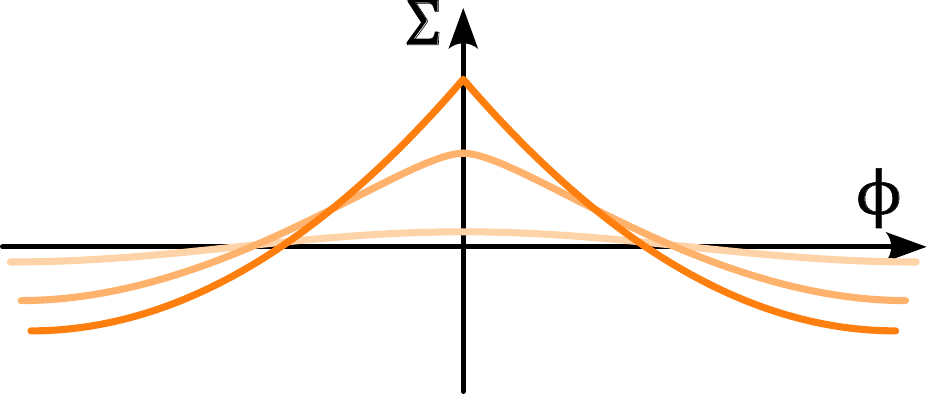}
      \includegraphics[width=0.99\linewidth]{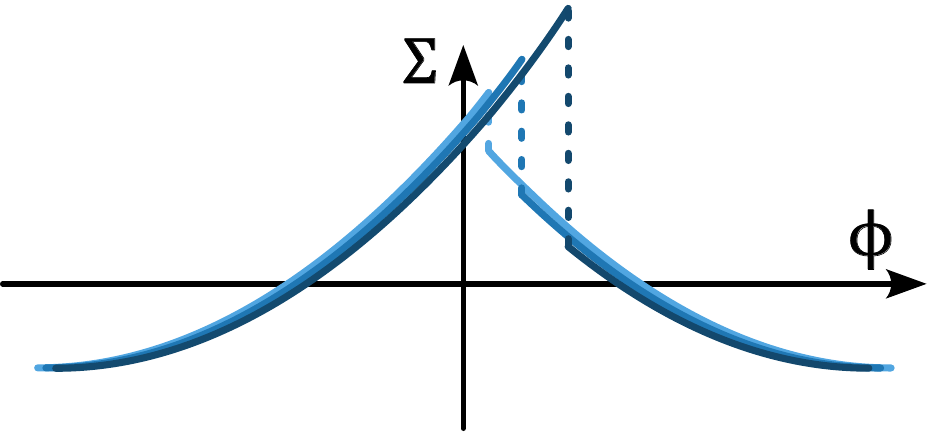}
    \end{subfigure}
\vspace{-1em}
\caption{Phase portraits of waves of varying action flux $\mathcal{B}$ satisfying equation (\ref{Spiraleqn}) alongside corresponding profiles of density perturbation. The phase portrait axes are displacement $\eta$ and canonical momentum proxy $\eta_\phi$ (related to the surface density and radial velocity via equations (\ref{sigmarel}) and (\ref{urel})). The wave action density $\mathcal{A}$ is directly proportional to the area enclosed by the curve.}
  \label{wave_break}
\end{figure*}
For simplicity and because of their elegant and illuminating solution, we now consider weakly non-linear waves (though the physical results below may be generalised to a fully non-linear description). This involves introducing the approximation $\eta_\phi \ll 1 \iff \Sigma-\Sigma_0 \ll \Sigma_0$, and keeping leading terms\footnote{there's an implicit assumption here about the scaling of each term should they all be in balance: namely $(\mathcal{U}^2-1)\sim \eta_\phi \ll 1$, $\eta \sim (\mathcal{U}^2-1)^{3/2}$, $\p_\phi \sim 1/\lambda \sim 1/\sqrt{\mathcal{U}^2-1}$, and consequently $m \sim 1/\sqrt{\mathcal{U}^2-1} \gg 1$.} in equation (\ref{Spiraleqn}), which becomes:
\begin{equation}\label{NLDWNL}
    \left(\mathcal{U}^2-1\right)\eta_{\phi\phi} + (\gamma+1)\eta_{\phi}\eta_{\phi\phi} + \tilde{\kappa}_0^2\eta = 0.
\end{equation}
This equation is very similar to that derived by \citet{larson_non-linear_1990}, though Larson worked within an Eulerian formulation (among other differences); importantly, it permits the same simple cubic solutions presented below. It has the first integral involving the Hamiltonian:
\begin{equation}
    \frac{1}{2}\left(\mathcal{U}^2-1\right)p^2 + \frac{\gamma+1}{3}p^3 + \frac{1}{2}\tilde{\kappa}_0^2\eta^2 = \tilde{\mathcal{H}},
\end{equation}
where again $p = \eta_\phi$. Dotted lines in the phase portraits in figure \ref{wave_break} show phase-space trajectories for the cases $\tilde{\mathcal{H}}>\mathcal{V}_m(\mathcal{U})$ (for which there are no smooth periodic solutions) and $\tilde{\mathcal{H}}< \mathcal{V}_m(\mathcal{U})$. The orange line traces the periodic orbit with $\tilde{\mathcal{H}} = \mathcal{V}_m(\mathcal{U})$, corresponding to the continuous `cuspy' periodic wave solution.

Smooth solutions with $\tilde{\mathcal{H}}>\mathcal{V}_m(\mathcal{U})$ are not possible, as no closed loops in phase space exist for $\tilde{\mathcal{H}}>\mathcal{V}_m(\mathcal{U})$. We gain some insight into the character of solutions containing shocks from the explicit solutions below.

It may be verified that 
\begin{equation}\label{cubicsol}
    \eta = \frac{\mathcal{U}^2-1}{2(\gamma+1)}\phi - \frac{\tilde{\kappa}_0^2}{18(\gamma+1)}\phi^3
\end{equation}
is an exact solution to the non-linear equation (\ref{NLDWNL}), with Hamiltonian $\tilde{\mathcal{H}} = \mathcal{V}_m(\mathcal{U})$, where now
\begin{equation}
    \mathcal{V}_m(\mathcal{U}) = \frac{\left(\mathcal{U}^2-1\right)^3}{6(\gamma+1)^2}.
\end{equation}
Considering the solution (\ref{cubicsol}) on the domain
\begin{equation}
    -3\frac{\sqrt{\mathcal{U}^2-1}}{\tilde{\kappa}_0} < \phi < 3\frac{\sqrt{\mathcal{U}^2-1}}{\tilde{\kappa}_0}
\end{equation}
and extending periodically, we obtain a `cuspy' solution in which $\eta$ and $\eta_\phi$ are both continuous. These solutions are compared to their fully non-linear counterparts and the self-similar cuspy waves in figure \ref{NLTDcomp}. As expected, the agreement is good for larger values of $m$.

By allowing for a small discontinuity in $\eta_\phi$ (but keeping $\eta$ continuous) in the cubic solution (\ref{cubicsol}) at the periodic boundary, we may probe solutions with weak shocks. Strictly speaking, these solutions don't respect the Rankine--Hugoniot shock conditions, but the necessary correction (including one to the background state) is self-consistently neglected in the weakly non-linear approximation scheme so long as the shocks are weak. These shocks may be introduced by modifying the periodic domain to
\begin{equation}\label{shockdomain}
    -\frac{\lambda}{2}+\sqrt{\frac{3(\mathcal{U}^2-1)}{\tilde{\kappa}_0^2}-\frac{\lambda^2}{12}}< \phi <\frac{\lambda}{2}+\sqrt{\frac{3(\mathcal{U}^2-1)}{\tilde{\kappa}_0^2}-\frac{\lambda^2}{12}},
\end{equation}
where the wavelength of the solution is $\lambda$. A sketch of the surface density of these solutions over a wavelength is shown in figure \ref{wave_break} (bottom right). These shocking solutions both carry a larger wave action flux per wavelength than the smooth, cuspy solution, but also have a shorter wavelength at fixed $\mathcal{U}$. The wave action flux is related to the area enclosed in phase space by the solution (see figure \ref{wave_break} (left)). For waves carrying an action flux larger than that of the cuspy continuous wave, it may be seen that shocking solutions of the form sketched in figure \ref{wave_break} (with $\tilde{\mathcal{H}} = \mathcal{V}_m\left(\mathcal{U}\right)$) are inevitable.

In order that the Rankine--Hugoniot conditions be satisfied, shocks must be from supersonic to subsonic regions of phase space. The flow velocity normal to the shock is $\mathcal{U}(1+p)c_0$, which includes a contribution from the background orbital flow projected onto the shock's normal, as well as one from the radial velocity. The sound speed is $c_0/(1+p)^{(\gamma-1)/2}$, which places the sonic point at
\begin{equation}
    p_{\text{sonic}} = -1 + \mathcal{U}^{-2/(\gamma+1)},
\end{equation}
which is at the maximum of the potential $\mathcal{V}(p;\mathcal{U})$, and at the cusp of the continuous `cuspy' solutions. Supersonic and subsonic regions of phase-space are separated by the grey horizontal line in figure \ref{wave_break}. The conditions that the spiral angle proxy $\mathcal{U}$ as well as $\tilde{\mathcal{H}}$ be constant over a wavelength, combined with the necessity that any shocks be from supersonic to subsonic regions of phase-space, and the requirement that the mean density be $\Sigma_0$ (recall $\Sigma = \Sigma_0/(1+p)$) restrict phase-space trajectories of waves containing shocks to those of the kind sketched in figure \ref{wave_break} (centre-left). This picture does not change when considering the fully non-linear system (\ref{Spiraleqn}).

This is the possibility modelled by the cubic solution (\ref{cubicsol}) on the periodic domain (\ref{shockdomain}). The wave includes a smooth transonic flow, over the maximum of $\mathcal{V}(p;\mathcal{U})$, before a shock back to the subsonic region. From simulations presented in section \ref{LAS}, this seems indeed to be the typical behaviour of highly non-linear spiral waves. Of course, the cubic solution is negligent of the entropy generation in the shock and the slow modification to the background state as a consequence: strictly speaking it's not a valid solution; however, for weak shocks the entropy generation is cubic in the shock strength, and so may be self-consistently ignored for moderate times in the weakly non-linear regime. It can also be checked that the mass flux is continuous, and the momentum flux is continuous at leading order across the shock.

We therefore see that shock formation starts when
\begin{equation}
    \tilde{\mathcal{H}} = \mathcal{V}_m\left(\mathcal{U}\right),
\end{equation}
but solutions containing shocks still have $\tilde{\mathcal{H}} = \mathcal{V}_m\left(\mathcal{U}\right)$, though carry a larger wave action flux $\mathcal{B}$. It's insightful to compute within the weakly non-linear model the maximum wave action flux tenable by a smooth spiral with a given number of arms $m$. From now on we'll refer to this value as the `soft cap'. Taylor expanding and evaluating (\ref{WAint}) yields
\begin{equation}\label{Bmax}
    \mathcal{B}_{\text{max}} \sim \frac{6\mathcal{F}}{5(\gamma+1)^2}\left(\frac{\upi}{3}\right)^5\tilde{\kappa}_0^4m^{-4},
\end{equation}
where $\mathcal{F}$ is defined in equation (\ref{Fdef}), and the approximation is good for large $m$. Particularly striking in the formula (\ref{Bmax}) is the fourth power dependence on $\tilde{\kappa}_0$ as well as $1/m$. In particular, we see how important the dispersive factor $\tilde{\kappa}_0$ is in allowing for smooth waves; far outside a wave's corotation radius, $\tilde{\kappa}_0 \to 0$, and so shocking of the outer wake excited by a planet becomes inevitable. 

Furthermore the strong $m^{-4}$ dependence highlights the effective `non-linear cap' on the amplitude of spirals with many arms: to be able to transport the same wave action flux as a two-armed spiral, multiple very strong shocks are necessary, leading to rapid dissipation and leaving only a small wave remnant. We speculate that this effective cap on the amplitude of many-armed non-linear spirals contributes importantly to the observed prevalence of two-armed spirals across the diverse zoo of astrophysical systems, in addition to the `hard cap' introduced below.

\subsection{Ultra non-linear waves}\label{UNLW}

\begin{figure}\centering
  \includegraphics[width=0.99\linewidth]{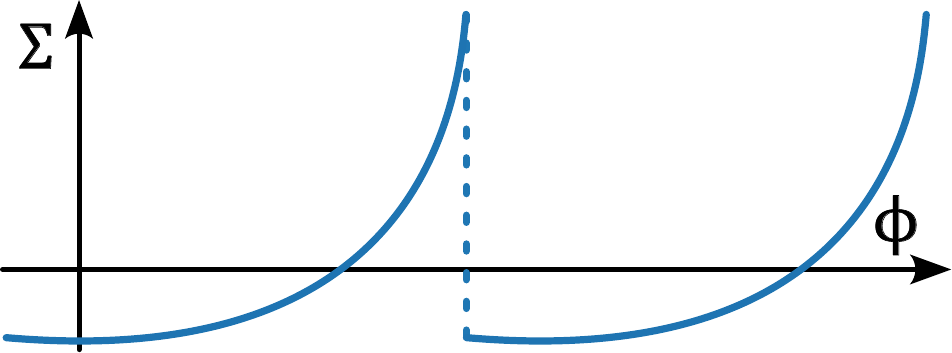}
  \includegraphics[width=0.99\linewidth]{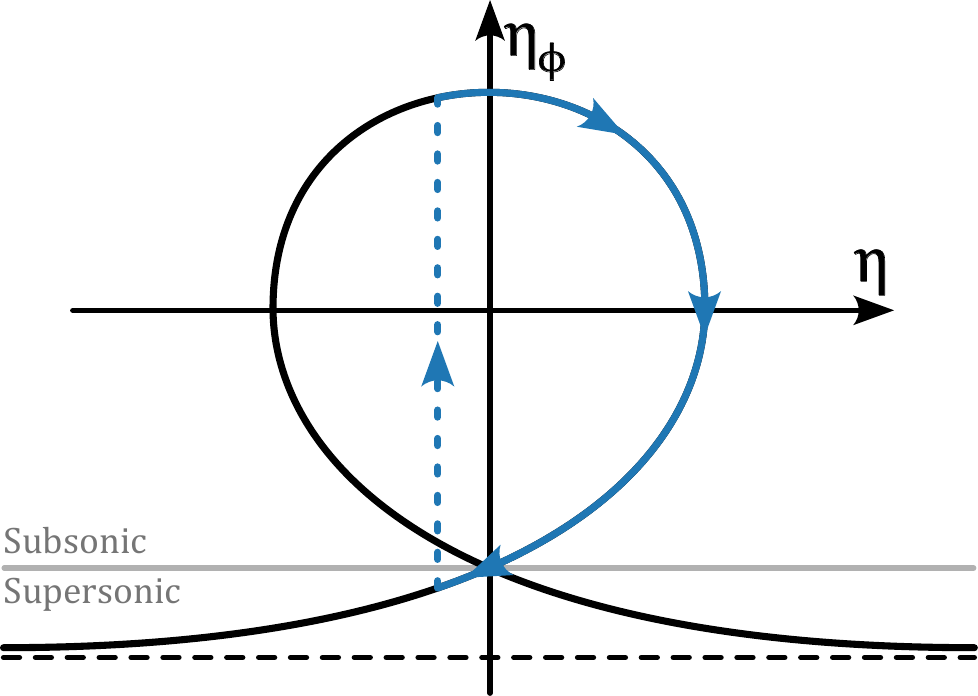}
  \caption{Wave profile and phase portrait of highly non-linear solutions to equation (\ref{Spiraleqn}). Evolution of the background disc as well as $\tilde{\mathcal{H}}$ and $\mathcal{U}$ due to the strong shocks is neglected, though reasonable qualitative agreement with simulations is retained (see section \ref{LAS} and in particular figure \ref{sim3bPS}).}
  \label{UNLPP}
\end{figure}
We now consider the very high amplitude case, involving strong shocks. We make the over-simplification of ignoring the secular evolution of the background state (and $\tilde{\mathcal{H}}$ and $\mathcal{U}$) due to the shocks, strictly necessary to restore consistency with the Rankine--Hugoniot conditions (we state the induced radial evolution of the wave action flux $\mathcal{B}$ due to shocks in section \ref{WAES}). We do however strictly enforce mass flux continuity, tangential velocity continuity (equivalent to continuity in $\eta$), and only permit (physical) shocks from supersonic to subsonic flow. The resulting solutions remain of the same character as those considered in section \ref{SFSW}, with $\tilde{\mathcal{H}} = \mathcal{V}_m(\mathcal{U})$ throughout the oscillation. They contain a smooth flow through the sonic point before a shock back to the subsonic section of the loop. The profile and phase space of such solutions are sketched in figure \ref{UNLPP}. In practice, momentum flux continuity and entropy generation are enforced by allowing for long radial variations in $\tilde{\mathcal{H}} = \mathcal{V}_m(\mathcal{U})$ as the wave propagates inwards (a force would be necessary to maintain the wave against dissipation otherwise), as well as slow modifications the background disc state.
\begin{table*}
    \centering
    \begin{threeparttable}
    \caption{Numerical `soft cap' and `hard cap' values for spiral angle-proxy $\mathcal{U}$ and radial angular momentum flux $\mathcal{B}$ for spirals with $m$ arms, $\gamma = 5/3$ and $\tilde{\kappa}_0 = 1$ (representing the case $\Omega_p \ll \Omega$ in a Keplerian disc). Smooth spirals are only possible below the soft cap, shocking spirals are possible between the soft and hard caps, and no spirals are possible beyond the hard cap.}
    \label{DRVALS}
    \setlength{\tabcolsep}{6pt} 
    \begin{tabular}{lcccccc}
        \toprule
        & \multicolumn{3}{c}{Soft cap} & \multicolumn{3}{c}{Hard cap} \\
        \cmidrule(lr){2-4}\cmidrule(lr){5-7}
        $m$ & $\mathcal{U}-1$ & $\mathcal{B}/\mathcal{F}$ & $\Delta\Sigma/\Sigma_0$ & $\mathcal{U}-1$ & $\mathcal{B}/\mathcal{F}$ & $\Delta\Sigma/\Sigma_0$ \\
        \toprule
        2   & $0.174$ & $0.0217$ & $0.128$ & $\infty$ & $\infty$ & $\infty$ \\
        3   & $0.0671$ & $3.21\times10^{-3}$ & $0.0499$ & $0.394$ & $0.114$ & $0.283$ \\
        4   & $0.0361$ & $9.27\times10^{-4}$ & $0.0270$ & $0.174$ & $0.0217$ & $0.128$ \\
        10  & $5.53\times10^{-3}$ & $2.16\times10^{-5}$ & $4.14\times10^{-3}$ & $0.0227$ & $3.65\times10^{-4}$ & $0.0170$ \\
        100 & $5.48\times10^{-5}$ & $2.13\times10^{-9}$ & $4.11\times10^{-5}$ & $2.19\times10^{-4}$ & $3.40\times10^{-8}$ & $1.65\times10^{-4}$ \\
        \bottomrule
    \end{tabular}
    \end{threeparttable}
\end{table*}

Waves of a higher amplitude, carrying a larger angular momentum flux, are possible for more loosely wound spirals, with larger $\mathcal{U}$, and correspondingly trace a larger loop in phase space (which approaches an ellipse as $\mathcal{U} \to \infty$). However, increasing $\mathcal{U}$ also increases the azimuthal extent of an oscillation.
\begin{figure}\centering
  \includegraphics[width=0.99\linewidth]{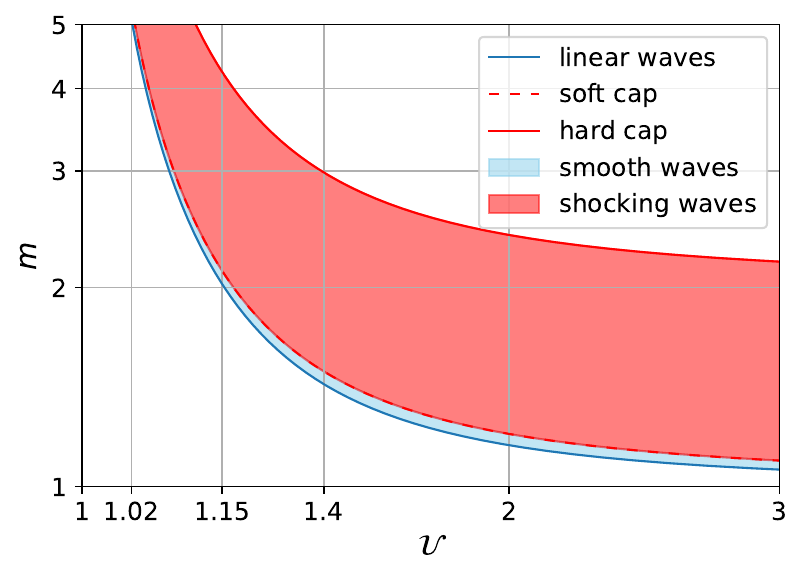}
  \caption{The non-linear dispersion relation for spiral waves. For fixed $m$, solutions with a higher phase speed $\mathcal{U}$ (which are more loosely wound) carry a larger radial wave action flux $\mathcal{B}$. Notice how only 2-armed solutions exist, with or without shocks, for $\mathcal{U}> 1.3944$ (the corresponding critical angular momentum flux is $\mathcal{B} = 0.1135\mathcal{F}$).}
  \label{NLDR}
\end{figure}

The shortest possible wavelength solution at a given value of $\mathcal{U}$ is the `half-loop' solution, in which the shock occurs just past the sonic point. This is an extreme case of the solution sketched on the phase portrait in figure \ref{UNLPP}. The wavelength of this solution places a calculable upper limit on the possible azimuthal wavenumber $m$ of a wave with a given phase speed $\mathcal{U}$, and by extension a hard upper limit on the radial angular momentum flux $\mathcal{B}$ which a spiral with $m$ arms may carry. Going forward, we'll refer to this limit as the `hard cap'. The non-linear dispersion relation for spiral waves is plotted in figure \ref{NLDR}. In the shaded region, amplitude and shock strength of the waves increases towards the right. $\mathcal{U}$ is related to the spiral angle made between the crests and the radial direction $\varphi$ via
\begin{equation}
    \tan \varphi = \frac{r(\Omega- \Omega_p)}{c_0 \mathcal{U}}.
\end{equation}

For fixed azimuthal wavenumber $m$, increasing $\mathcal{U}$ increases the amplitude of the solution. The solution develops a shock as it crosses the `soft cap' on wave amplitude, which strengthens until the hard cap half-loop solution is reached, beyond which no further solutions with $m$ spiral arms are possible. Table \ref{DRVALS} summarises numerical values associated with the soft and hard caps for various $m$ for waves far inside their corotation radii in Keplerian discs.

The weakly non-linear theory developed in section \ref{SFSW} remains valid for $m \gg 2$ even for solutions near the hard cap (in this limit the shock strength remains bounded by the size of the wave, which is assumed to be small). As a result, we can compute algebraically the asymptotic expression for the hard cap on tenable angular momentum flux by spirals with $m \gg 2$ arms. The result is
\begin{equation}\label{BHC}
    \mathcal{B}_{\text{hard cap}} \sim \frac{3\mathcal{F}}{5(\gamma+1)^2}\left(\frac{2\upi}{3}\right)^5\tilde{\kappa}_0^4m^{-4},
\end{equation}
around $16$ times the soft cap given in equation (\ref{Bmax}).

As $\mathcal{U} \to \infty$, the wave motion approximately resembles an epicycle away from a small region near the sonic point, and accordingly, for waves far inside their corotation radii in Keplerian discs (so that $\tilde{\kappa}_0 \approx 1$), the minimum angular extent of a single oscillation approaches $\pi$ (that is, the angular wavelength of half an epicycle). Two-armed spiral solutions are therefore possible for all $\mathcal{U} > 1.1547$ (as seen in figure \ref{NLDR}), and furthermore, based on the hard cap for $m=3$ listed in table \ref{DRVALS}, all solutions carrying radial angular momentum flux $\mathcal{B} > 0.114 \mathcal{F}$ must be two-armed, should the spiral angle $\varphi$ be constant with $\theta$. The restriction is more forgiving closer to corotation, where $\tilde{\kappa}_0 > 1$, as the wave is effectively more dispersive and correspondingly able to possess a higher amplitude. In the outer disc, which has $\tilde{\kappa}_0 < 1$, the caps become more restrictive, however one armed spirals also become possible here.

Simulation \texttt{2b}, detailed in table \ref{tab1} and discussed in section \ref{LAS}, probes this highly non-linear regime by including a planet with $M_p = 50M_\text{th}$ (the `thermal mass' here is defined by $M_\text{th} = h_p^3M_\star$). Particularly notable is the relative looseness of the inner spiral wake, which is comparable to that excited by the $1$ thermal mass planet in simulation \texttt{2c}, whose disc is twice as thick. This comparison is shown in figure \ref{sim2b3}. Despite the high wave amplitude, the measured spiral-angle proxy $\mathcal{U}\sim2-5$ is not so large as to completely invalidate the tight-winding assumptions necessary to derive the model (\ref{MasterEqn}), however this non-linear loosening certainly has important consequences for the interpretation of observations of companion-driven spirals in discs, for example in the protoplanetary discs MWC 758 \citep{benisty_asymmetric_2015} and HD100453 \citep{rosotti_spiral_2020}.

\subsection{Wave action and angular momentum evolution due to shocks}\label{WAES}

For very high amplitude shock waves, such as those arising in simulation \texttt{2b}, shown in figure \ref{sim2b3} (centre), the spiral wake quickly equilibrates into a two-armed pattern in which the spiral angle is constant with azimuth, but varies with radius. As discussed above, in this regime the entire azimuthal wave profile may be deduced from a single parameter: the radial wave action flux $\mathcal{B}$. In this section we exploit the equivalence between the wave-associated radial angular momentum flux and the wave action flux (which is established in appendix \ref{appx1}, and holds even when a small viscosity is introduced to resolve discontinuities across shocks) in order to write down the radial evolution of $\mathcal{B}$.

It may be shown that the radial angular momentum flux carried by a free shock wave with $m$ identical trailing spiral arms evolves according to \citep{2016ApJ...831..122R}
\begin{multline}\label{Bevoeqn}
    \frac{\p \mathcal{B}}{\p r} \equiv \frac{\p}{\p r} \left[r^2 \int\Sigma u v \dd \theta\right] \\= -\text{sgn}(\Omega_p - \Omega(r)) m r \Sigma_0 c_0^2 \psi_Q (d),
\end{multline}
with
\begin{equation}
    \psi_Q(d) = \frac{1}{\gamma - 1}\left[d^{-\gamma}\frac{(\gamma + 1)d - (\gamma - 1)}{(\gamma + 1) - (\gamma - 1)d} - 1\right],
\end{equation}
where the ratio between the post- and pre-shock surface density is
\begin{equation}
    d = \frac{\Sigma^+}{\Sigma_-} = \frac{1 + \eta_\phi^-}{1 + \eta_\phi^+} > 1.
\end{equation}
The right hand side of equation (\ref{Bevoeqn}) scales with the cube of the shock strength for weak shocks, and is positive for trailing inward-propagating waves, so $\mathcal{B}$ decreases inwards in this case. Note that our expression for $\mathcal{B}$ coincides with the definition of $F_\text{wave}$ used in studies of gap opening (e.g. \citet{cordwell_early_2024}).

Equation (\ref{Bevoeqn}) may be used to determine the global radial evolution of spiral shock waves, but only completes a full description when $\mathcal{U}$ is independent of $\theta$. Whilst this is a good model for the waves arising in simulation \texttt{2b} for $r \lesssim 0.7r_p$, shown in figure \ref{sim2b3} (centre), it's insufficient to model accurately the case $M_p \lesssim M_{\text{th}}$ (e.g. simulation \texttt{2c}, shown in figure \ref{sim2b3} (right)), in which $\mathcal{U}$ and the radial wave action flux density vary meaningfully with azimuth. We discuss briefly in section \ref{VSA} below how such variations may be calculated for smooth waves, though note that in general (without further analytical progress) it's necessary to solve equation (\ref{MasterEqn}) piecewise away from shocks, manually imposing the Rankine--Hugoniot jump conditions, and allowing for slow temporal variations in the background disc.

\subsection{Allowing for varying spiral angle}\label{VSA}

In practice, $\mathcal{U}$ and $\tilde{\mathcal{H}}$ vary slowly over the global extent of the disc due to a variety of factors. For example $\mathcal{U}$ may vary across a wave-packet made up of a number of different wavelengths, and $\tilde{\mathcal{H}}$ may be amplified or diminished by the profile of the background disc. In appendix \ref{appxb2} we derive the equations governing their global evolution in the case where the wave's profile remains smooth, equations (\ref{WWAJ}) and (\ref{Mod2}). These equations are useful for analysing the evolution and potential breaking of waves excited by low-mass planets, which we present high-resolution simulations of in section \ref{SAS}.

These are the Whitham modulation equations for the system, and form a hyperbolic pair of equations in $\mathcal{U}$ and $\tilde{\mathcal{H}}$ (having averaged over the small-scale wave dynamics). They consist of a non-linear wave action equation as well as a `conservation of waves' condition. Though slightly opaque, we make use of them in appendix \ref{appxB22} where we present a sketch proof that the inner spiral wakes excited by low-mass planets need not shock, that is, the evolution of $\mathcal{U}$ and $\tilde{\mathcal{H}}$ can be such that $\tilde{\mathcal{H}} < \mathcal{V}_m(\mathcal{U})$ everywhere, even in the absence of dissipation.

\section{Planet-driven spiral waves: numerical simulations and analysis}\label{PDSW}

\subsection{Numerical Procedure}

\begin{table*}
    \centering
    \begin{threeparttable}
    \caption{Parameter values and grid details for the numerical simulations presented in section \ref{PDSW}.}
    \label{tab1}
    \setlength{\tabcolsep}{5.0pt} 
    \begin{tabular}{lcccccccccccc}
        \toprule
        Run ID & $M_p/M_\text{th}$\tnote{a} & $h_p$ & $\gamma$ & $\Sigma_0(r)$ & $T_0(r)$ & $r_{\text{in}}$ & $r_{\text{out}}$ & $r_{\text{DZ,in}}$\tnote{b} & $r_{\text{DZ,out}}$ & $N_r \times N_\theta$ & CPH\tnote{c} & Duration\tnote{d} \\
        \toprule
        \texttt{test} & $0.01$ & $0.1$ & $1.4$ & $\propto r^{-1}$ & $\propto r^{-1}$ & $0.015r_p$ & $2.0r_p$ & $0.02r_p$ & $1.6r_p$ & $5400 \times 1000$ & $16$ & $10.0$ \\
        \midrule
        \texttt{1a} & $0.005$ & \multirow{3}{*}{$0.1$} & \multirow{3}{*}{$1.4$} & \multirow{3}{*}{$\propto r^{-1}$} & \multirow{3}{*}{$\propto r^{-1}$} & \multirow{3}{*}{$0.015r_p$} & \multirow{3}{*}{$2.0r_p$} & \multirow{3}{*}{$0.02r_p$} & \multirow{3}{*}{$1.6r_p$}& \multirow{3}{*}{$10800\times2000$} & \multirow{3}{*}{$32$} & \multirow{3}{*}{$10.0$}\\
        \texttt{1b} & $0.01$ & &  & & & & & & & & & \\
        \texttt{1c} & $0.04$ & &  & & & & & & & & & \\
        \midrule
        \texttt{2a} & $1.0$ & $0.05$ & \multirow{3}{*}{$1.001$} & \multirow{3}{*}{$\propto r^{-1/2}$} & \multirow{3}{*}{$\propto r^{-1/2}$} & \multirow{3}{*}{$0.12r_p$} & \multirow{3}{*}{$2.0r_p$} & $0.15r_p$ & \multirow{3}{*}{$1.6r_p$} & $10800\times2000$ & \multirow{3}{*}{$16$} & \multirow{3}{*}{$5.0$} \\
        \texttt{2b} & $50$ & $0.05$ & & & & & & $0.15r_p$ & & $10800\times2000$ & & \\
        \texttt{2c} & $1.0$ & $0.1$ & & & & & & $0.18r_p$ & & $2700\times1000$ & & \\
        \midrule
        \texttt{3} & $5.0$ & $0.05$ & $1.4$ & $\propto r^{-1}$ & $\propto r^{-1/2}$ & $0.035r_p$ & $2.0r_p$ & $0.045r_p$ & $1.6r_p$ & $21600\times2000$ & $16$ & $15.0$ \\
        \bottomrule
    \end{tabular}
    \begin{tablenotes}
        \item[a] $M_\text{th} = h_p^3M_\star$
        \item[b] Outer boundary of the inner damping zone.
        \item[c] Number of grid cells per disc scale-height at the planet's location.
        \item[d] Number of orbits elapsed at the planet's radial location during the simulation. For runs \texttt{test}, \texttt{1a}, \texttt{1b} and \texttt{1c}, 10 orbits at $r_p$ equates to 5400 orbits at the inner boundary.
    \end{tablenotes}
    \end{threeparttable}
\end{table*}

We performed global 2D hydrodynamic simulations of accretion discs hosting an embedded planet using Athena++ version 24.0 \citep{Stone2020}. Athena++ solves the equations for mass, momentum and energy conservation in conservative form using a Godunov scheme. We used the Roe approximate Riemann solver for flux calculations and a second-order Runge–Kutta (RK2) scheme for time integration. The equations evolved are
\begin{subequations}\label{athenaeoms}
\begin{align}
    &\p_t(\Sigma \vecu) + \nabla\cdot\left(\Sigma \vecu \vecu + P \mathbf{I}\right) = -\Sigma\nabla\Phi,\\
    &\p_t \Sigma + \nabla \cdot\left(\Sigma \vecu\right) = 0,\\
    &\p_t\left(\Sigma\mathcal{E}\right) + \nabla\cdot\left(\Sigma \vecu \left(\mathcal{E} + P\right)\right) = -\Sigma \vecu\cdot \nabla \Phi,
\end{align}
\end{subequations}
for energy density $\mathcal{E} = \frac{1}{2}\left|\vecu\right|^2 + \frac{1}{\gamma - 1}\frac{P}{\Sigma}$, and potential $\Phi = -\frac{G M_\star}{r} + \Phi_p$. In all simulations, we implement the prescription (including the indirect term) for the planet's potential
\begin{subequations}\label{phipresc}
\begin{align}
    & \Phi_p = -\frac{G M_p}{H_p}\frac{\ee^{\frac{1}{4}s^2}}{\sqrt{2\upi}}K_0\left(\tfrac{1}{4}s^2\right) + \frac{GM_p}{r_p^3}\vecr\cdot \vecr_p, \\
    & s^2 = \frac{\left|\vecr-\vecr_p\right|^2}{H_p^2} + \epsilon^2.
\end{align}
\end{subequations}
Here $H_p = H(r_p)$ is the scale height at the planet's orbital radius, and $\vecr_p$ denotes the planet's location, which executes fixed circular Keplerian orbits about the star. We found that best results were obtained by conducting the simulations in the frame corotating with the planet. We further smoothed the logarithmic singularity of the potential $\Phi_p$ by a small number $\epsilon$, which we chose to be the length of a grid cell in the azimuthal direction. We ramped up the planet's potential over 2 orbits in the case of runs \texttt{test}, \texttt{1a}, \texttt{1b} and \texttt{1c}, and over 5 orbits in the case of the higher mass planet in run \texttt{3}, before allowing sufficient time for the spiral wave pattern to establish in both cases. In runs \texttt{2a}, \texttt{2b} and \texttt{2c}, since small artefacts generated during the planet growth phase were of less concern to us, we chose to ramp up the planet's potential over the shorter time of 2 orbits. A shorter total run duration was also sufficient for the spiral wave pattern to become established in these cases due to the smaller radial domain (though the background discs were still slowly evolving at the end of these runs).

Equation (\ref{phipresc}) constitutes a mass-weighted vertical average of the Newtonian potential assuming an isothermal vertical structure with density $\rho \propto \exp\left(-z^2/2H_p^2\right)$. For simplicity, and following the direction of \citet{2025arXiv250904282C}, we use a constant scale height evaluated at the planet's radial location when evaluating the vertical average. Relative to a more carefully evaluated mass-weighted vertical average, this yields only small errors locally as well as over larger distances in the case of low-mass planets (as $H \approx H_p$ locally, and the Newtonian potential from which the averaged potential is derived becomes roughly independent of $z$ over larger distances). Whilst this prescription may only be quantitatively justified in the case of low-mass planets, for simplicity we also employ it for the more massive planets simulated in runs \texttt{2a}, \texttt{2b}, \texttt{2c} and \texttt{3}.

One might very reasonably alternatively have implemented the potential corresponding linearly to that which forces the 2D mode of the (adiabatic) disc (as suggested by \citep{brown_horseshoes_2024}). This constitutes a slightly different weighted average\footnote{leading to the modification $H_p \to \sqrt{\gamma}H_p$ in equation (\ref{phipresc})}, and we might in this case understand the 2D equations of motion (\ref{athenaeoms}) as representing the dynamics of a slice a fixed height above the disc's midplane instead of vertically integrated quantities.

We chose to simulate an energy equation (with no diabatic terms) over a locally isothermal equation of state since wave action flux is not conserved (in an arguably unphysical manner) in the latter system, as discussed in appendix \ref{WALISO}. In runs \texttt{2a}, \texttt{2b} and \texttt{2c} we chose $\gamma = 1.001$ to minimise shock heating and retain greater control over the aspect ratio $h$ (we had initially performed simulations with $\gamma = 1.4$, but found that shocks driven by a $50$ thermal mass planet rapidly doubled the temperature in the disc, which we had not allowed to cool).

We implemented wave damping zones near the inner and outer boundaries of the disc to prevent wave reflections. In the cases of runs \texttt{test}, \texttt{1a}, \texttt{1b} and \texttt{1c}, we damped only the radial velocity, whereas in runs \texttt{2a}, \texttt{2b}, \texttt{2c} and \texttt{3}, a sufficiently deep gap and strong zonal flow are generated in the damping zones due to the deposition of angular momentum that it is necessary to also damp the density towards its initial background value. This injection/removal of mass also allows for a moderate accretion flow to be driven by the wave without mass accumulating near the inner boundary. Further details are given in table \ref{tab1}.

In runs \texttt{test}, \texttt{1a}, \texttt{1b} and \texttt{1c}, it was important that the background disc's profile was chosen so that $\mathcal{F}\sim\text{const}$ as $r\to 0$, to avoid the effects of wave amplification. We therefore chose profiles $\Sigma_0(r) \propto r^{-1}$ and $T_0(r) \propto r^{-1}$. This relatively steep temperature profile results in relatively large velocity perturbations in the inner disc, requiring a small time step to satisfy the Courant–Friedrichs–Lewy (CFL) condition. We reduced some of the computational cost by allowing for a large maximum Courant number of $0.5$. We further employed the second order orbital advection scheme implemented in Athena++, based on the FARGO algorithm \citep{2000A&AS..141..165M}, which significantly reduces the computational cost. In runs \texttt{2a}, \texttt{2b}, \texttt{2c} and \texttt{3}, we chose a temperature profile perhaps more typical of a protoplanetary disc, $T_0(r) \propto r^{-1/2}$, which further reduced the computational cost by allowing for a larger time step.

All simulations were carried out on grids with logarithmic spacing in the radial direction, ensuring good resolution for the entirety of the inner spiral wakes. The azimuthal and radial resolutions were chosen so that the spiral wakes were approximately equally well resolved in both directions (note the domain extents in the $\theta$ and $x$ coordinates in figures \ref{1bdenspolar} (right) and \ref{sim2a} (bottom)). Analysis of the \texttt{test} run suggested that the higher resolution of 32 cells per disc scale-height was necessary in order to adequately resolve the wakes in runs \texttt{1a}, \texttt{1b} and \texttt{1c}. The analysis carried out in section \ref{SAS} also greatly benefits from the large radial domain simulated interior to the planet's orbit, with a very small inner disc edge at $r = 0.015 r_p$. A summary of many of the key details and parameters chosen in our simulations is given in table \ref{tab1}.

\subsection{Small-amplitude spirals: dispersion prevents shocking}\label{SAS}

\begin{figure*}\centering
  \includegraphics[width=0.49\linewidth]{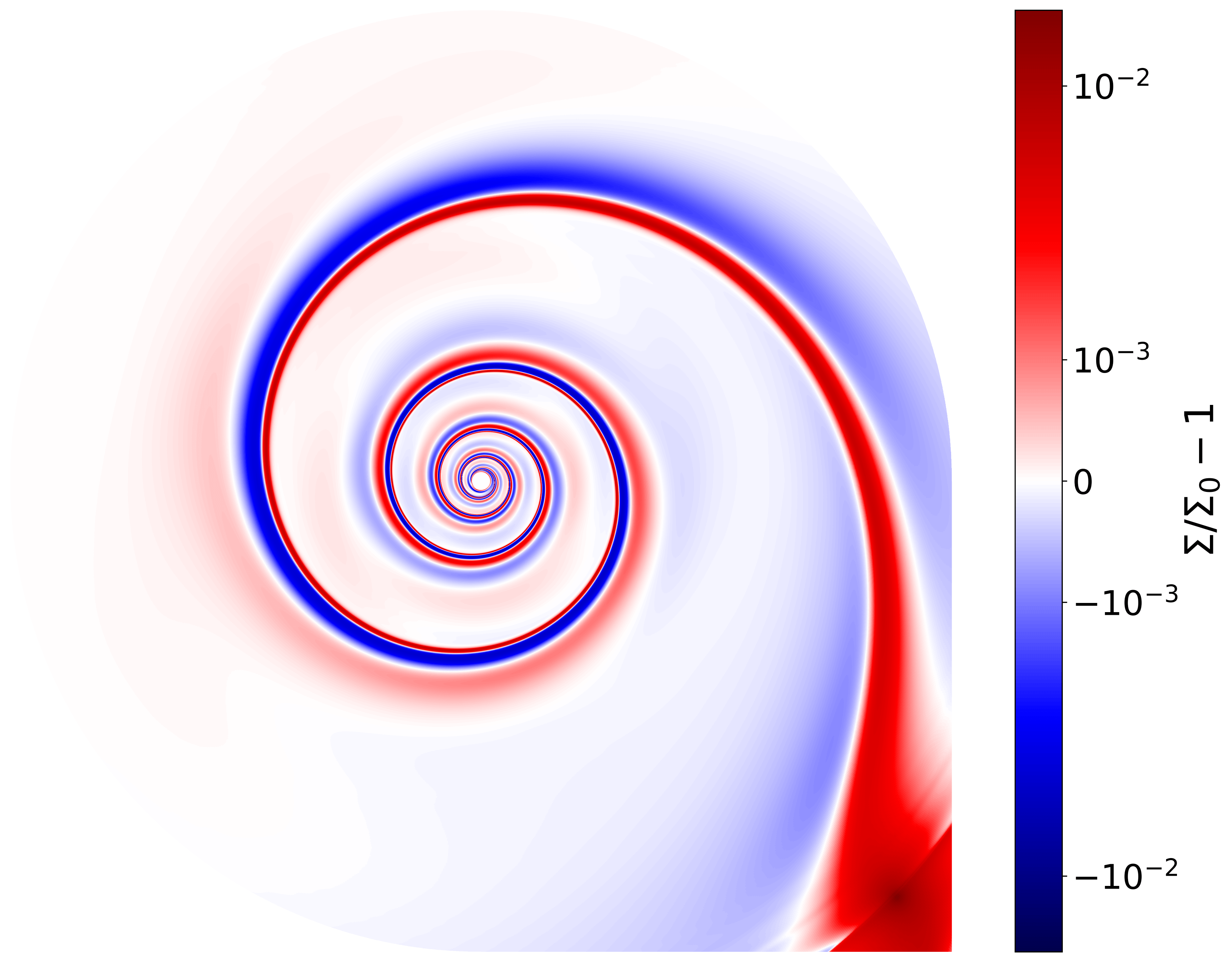}
  \includegraphics[width=0.49\linewidth]{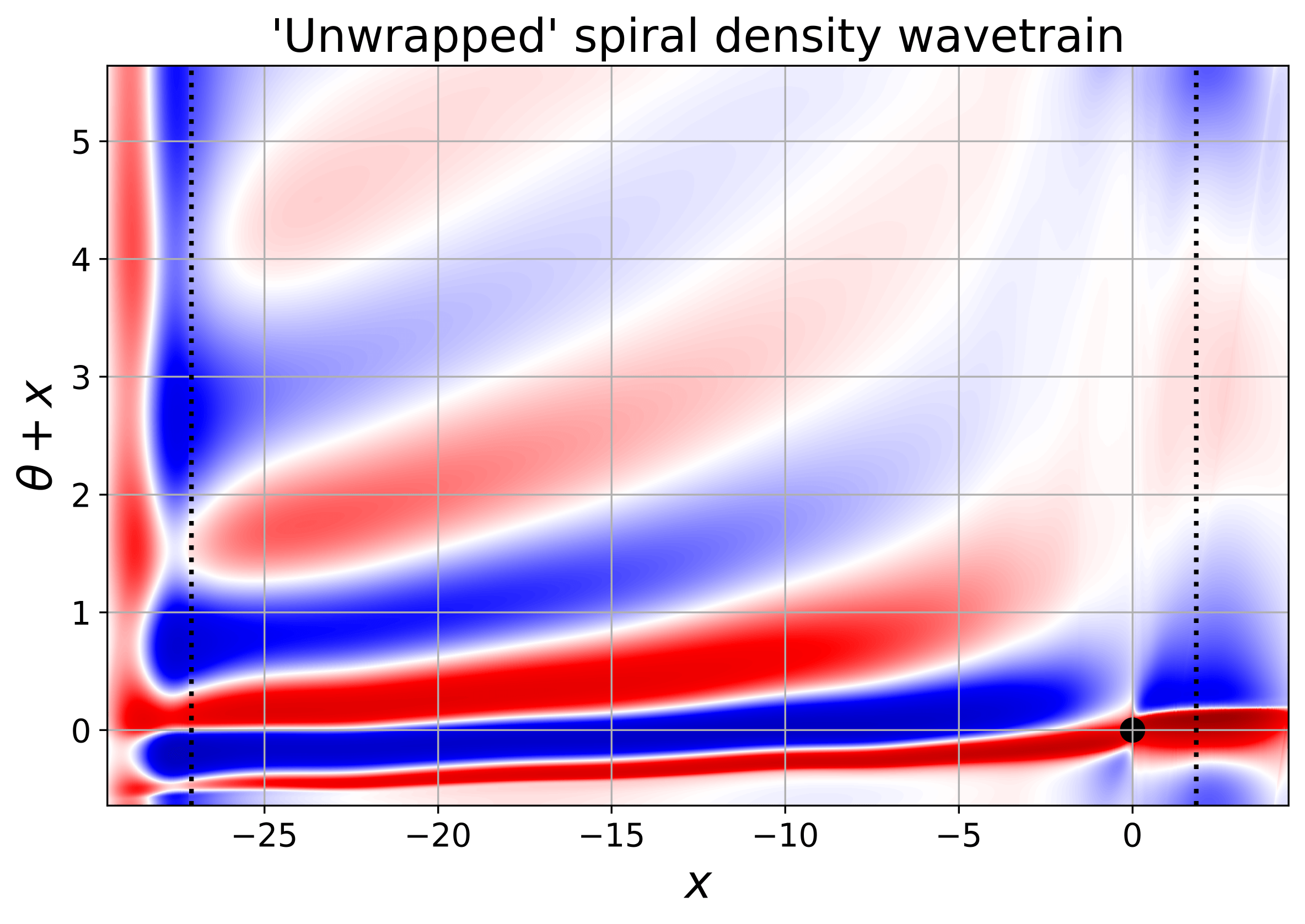}
  \caption{Plots of the fractional surface density perturbation from simulation \texttt{1b} after $t = 10.0$ orbits of the planet around the central star. Left: polar plot of the inner spiral wake excited by the planet ($M_p = 0.01M_\text{th}$), located in the bottom right corner. Right: the same spiral wave, now plotted in the coordinate system $(x,\theta+x)$, where $x$ is defined in equation (\ref{xdefn}). The planet is marked by a black dot at $x = 0$. The primary wave approximately follows the ray $\theta = - x$, so that the spiral wave appears `unwrapped' in this coordinate system. Dotted lines mark boundaries of damping zones.}
  \label{1bdenspolar}
\end{figure*}

\begin{figure}\centering
  \includegraphics[width=0.99\linewidth]{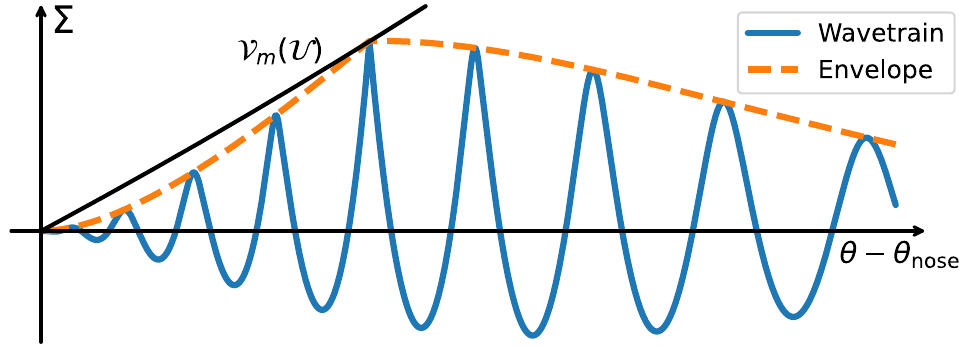}
  \vspace{-1em}
  \caption{A cartoon graphic of a non-linear modulated spiral wave train, representing for example an azimuthal cross-section of the simulation graphed in figure \ref{1bdenspolar} (though with many more arms); the nose of the wave train is on the left. The effective phase speed $\mathcal{U}$ increases to the right (so that crests approach the nose as the wave propagates), whilst conversely the ray trajectories are directed away from the nose, becoming increasingly tightly wound towards the tail. The cuspy waves arise when the modulating envelope $\tilde{\mathcal{H}}$ reaches $\mathcal{V}_m(\mathcal{U})$.}
  \label{wavetrain}
\end{figure}
\begin{figure*}\centering
  \includegraphics[width=0.49\linewidth]{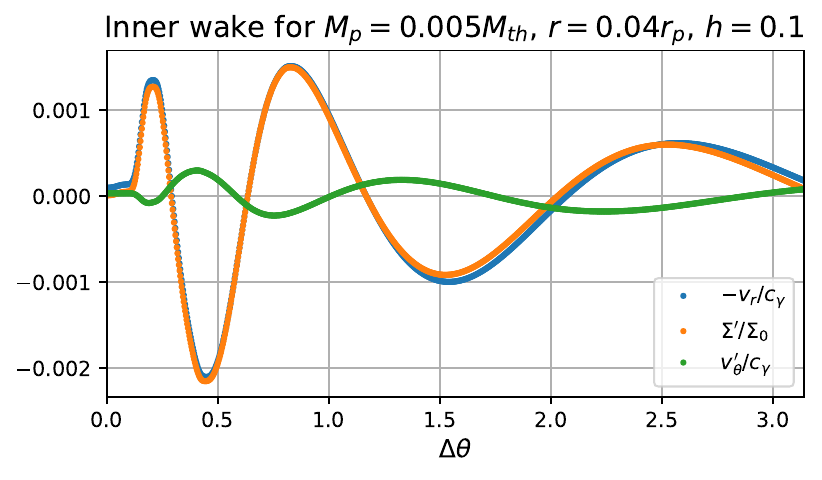}
  \includegraphics[width=0.49\linewidth]{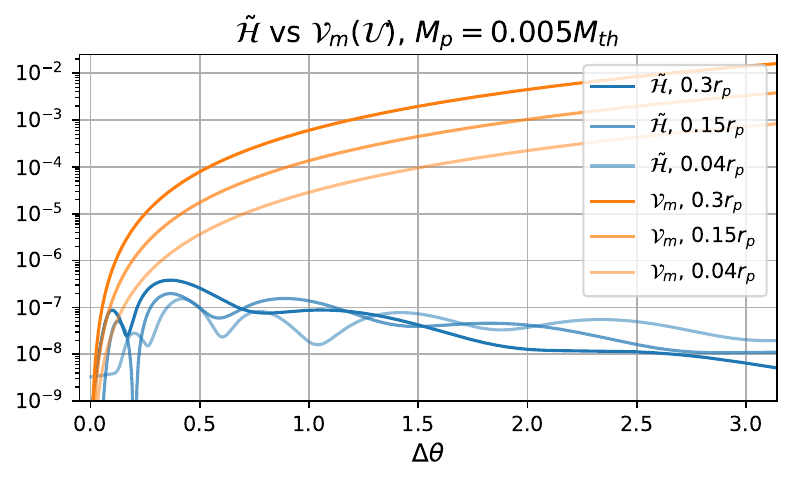}
    \includegraphics[width=0.49\linewidth]{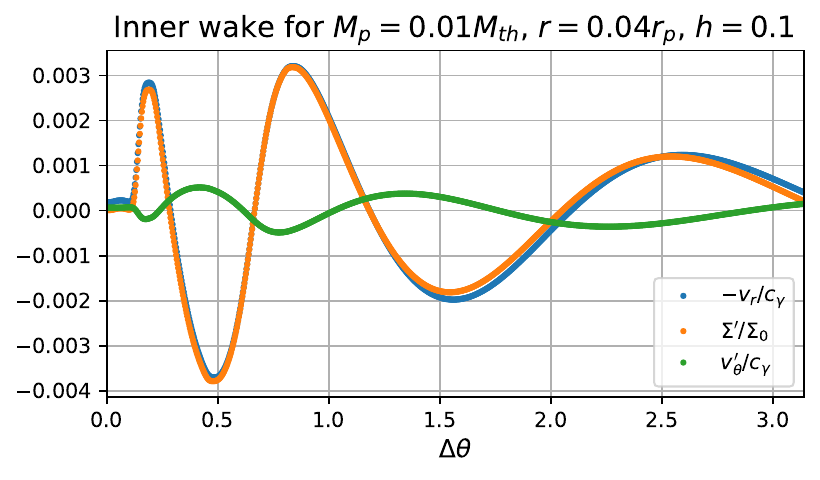}
  \includegraphics[width=0.49\linewidth]{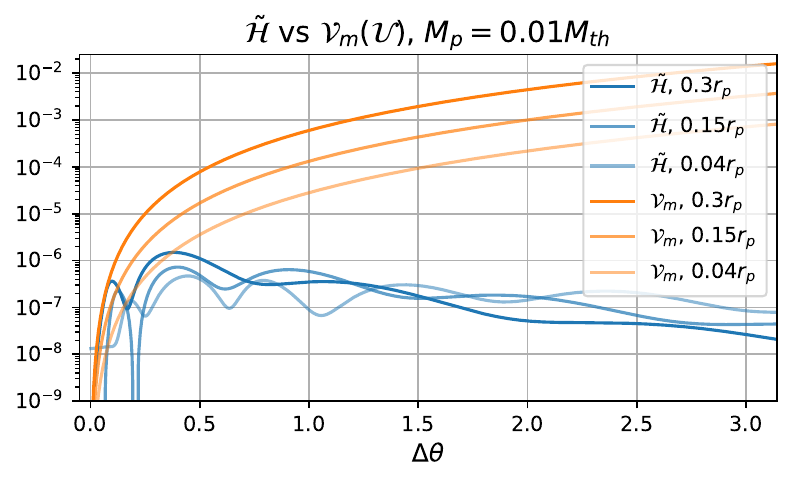}
    \includegraphics[width=0.49\linewidth]{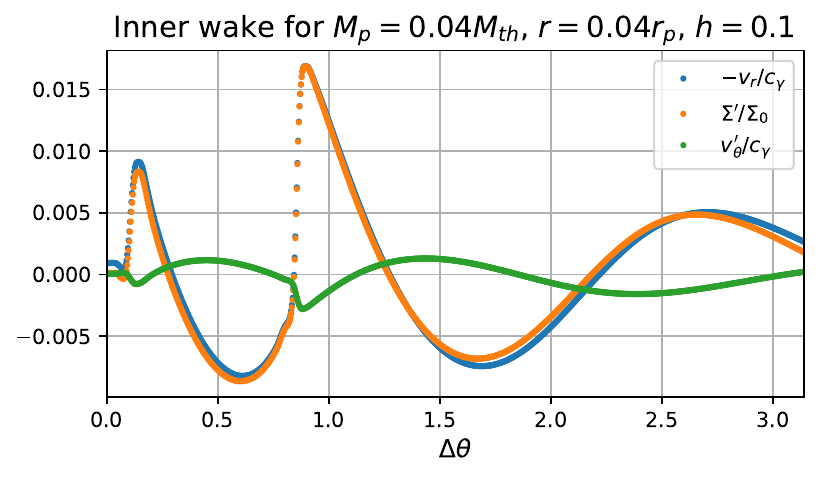}
  \includegraphics[width=0.49\linewidth]{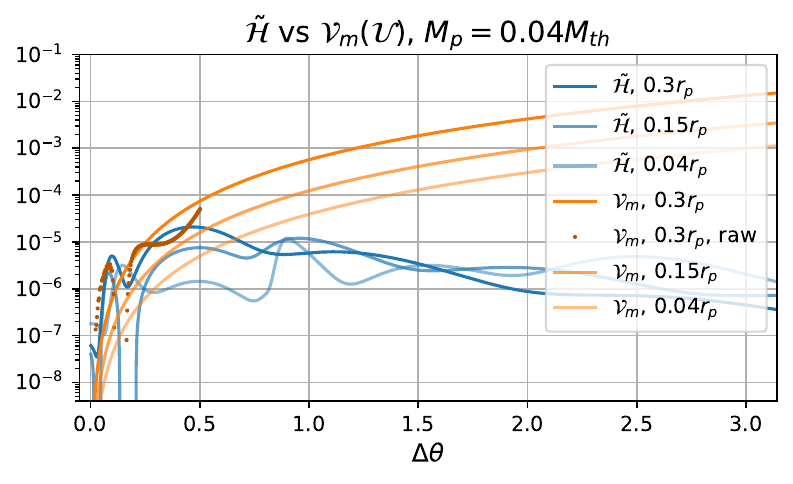}
  \vspace{-1em}
  \caption{Left: azimuthal cross sections of the fractional surface density perturbation and radial and azimuthal velocities from runs \texttt{1a}, \texttt{1b} and \texttt{1c} at $r = 0.04 r_p$ after $t = 10.0$ orbits of the planet around the central star. Two shocks are evident in the wake of the planet with mass $M_p = 0.04 M_\text{th}$ (bottom left), which may be identified via the associated wider gaps between neighbouring plotted points. Partial steepening is also present in the primary wave in run \texttt{1b} (centre left), consistent with the wave being very close to the threshold for shocking. The same steepening is absent from the primary wave in run \texttt{1a} (top left). Right: evolution of the Hamiltonian envelope $\tilde{\mathcal{H}}$ of the inner wake alongside the boundary for shock formation $\mathcal{V}_m(\mathcal{U})$ for the three simulations \texttt{1a}, \texttt{1b} and \texttt{1c}. $\mathcal{U}$ is estimated via a fitting procedure described in the text and figure \ref{Ufit}. The wave in run \texttt{1a} is below the threshold for shock formation at all radii, the wave in run \texttt{1b} appears marginal, and the threshold appears to have been reached (and $\mathcal{U}$ non-linearly modified) in run \texttt{1c}.}
  \label{WTG1}
\end{figure*}

We conducted three high-resolution inviscid simulations of low-mass planets embedded in 2D discs, namely runs \texttt{1a}, \texttt{1b} and \texttt{1c} in table \ref{tab1}, with planet masses $0.5\%$, $1\%$ and $4\%$ of a thermal mass respectively. The discs in these simulations have aspect ratio $h=0.1$, and surface density and sound speed profiles $\Sigma_0 \propto r^{-1}$ and $c_0 \propto r^{-1/2}$, so that $\mathcal{F} \sim \text{const}$ as $r \to 0$ (see equation (\ref{Fdef})), and correspondingly the inner spiral wakes are not amplified or diminished as they propagate inwards. In this case, the non-linear theory of \citet[equation (40)]{rafikov_nonlinear_2002} estimates that the inner wakes in the simulations \texttt{1a}, \texttt{1b} and \texttt{1c} will shock at radii $r \approx 0.34r_p$, $0.47 r_p$ and $0.67r_p$ respectively. The grid in these simulations extends from $r = 0.015r_p$ to $r=2r_p$. We tailored the simulations to best study the inner wake evolution, as the inner wake experiences stronger dispersion than the outer wake. Note that $\tilde{\kappa}_0$, which quantifies the relative importance of dispersive inertial effects, approaches $1$ far inside the wave's corotation radius, and $0$ far outside. We found additionally in section \ref{SFSW} using weakly non-linear theory that the maximum angular momentum flux tenable by a smooth spiral wave scales with $\tilde{\kappa}_0^4$.

The purpose of the simulations was to test numerically whether dispersion is able to prevent the inner spiral wakes from shocking, as suggested by the analysis in appendix \ref{appxB22}, and estimate the threshold planet mass below which the wake remains smooth indefinitely as it propagates inwards. As observed in sections \ref{S7} and \ref{ctss}, it's certainly possible that dispersion allows for smooth spiral solutions up to a maximum amplitude, at which they carry a radial angular momentum flux comparable to that of the inner spiral wave generated by a planet of mass $M_p = 0.24 M_\text{th}$. We expect the threshold mass to be far lower than this however; indeed, the \texttt{test} run gave preliminary indication that it likely lies near $0.01 M_\text{th}$ for $h = 0.1$.

The finite aspect ratio is also important here: if the disc were thinner, we'd expect the mass threshold to be lower. Indeed, in the shearing sheet it's expected that all waves eventually shock \citep{heinemann_weakly_2012}; the dispersion is not strong enough to overcome the transient growth of the wave amplitude due to the shear (see appendix \ref{appxc}).

Figure \ref{1bdenspolar} shows the fractional surface density perturbation from the run \texttt{1b}. On the left is a polar plot of the inner wake excited by the planet (which is situated in the bottom right of the image). The right hand plot shows the same data, but now plotted in terms of the coordinates which track the primary wave; the spiral therefore appears `unwrapped' in this coordinate system, and the dispersion of the wave into a multi-armed wave train is far clearer to see. This dispersion has been understood linearly by \citet{miranda_multiple_2019}, and can also be described linearly using the method of stationary phase shown in appendix \ref{appxB21}, which has the advantage of also predicting the behaviour of the wave amplitude. 

In this wave train regime, a cartoon sketch of which is depicted in figure \ref{wavetrain}, we imagine a sequence of waves trailing the primary, all modulated by a slowly varying envelope. In section \ref{SFSW} we demonstrated that when $\tilde{\mathcal{H}}$ reaches the function $\mathcal{V}_m(\mathcal{U})$ of the spiral-angle proxy $\mathcal{U}$, the wave begins to break and a shock forms. Of course, this picture is complicated by the azimuthal periodicity: very early on in the wave's evolution the tail of the wave train begins to overlap the nose. However, this effect seems surprisingly unimportant for the low-mass planet-driven wakes studied here. The tail's amplitude is sufficiently low in this case that it does not significantly disrupt the evolution of the nose, which is apparent from figures \ref{1bdenspolar} and \ref{WTG1}.
\begin{figure}\centering
  \includegraphics[width=1.05\linewidth]{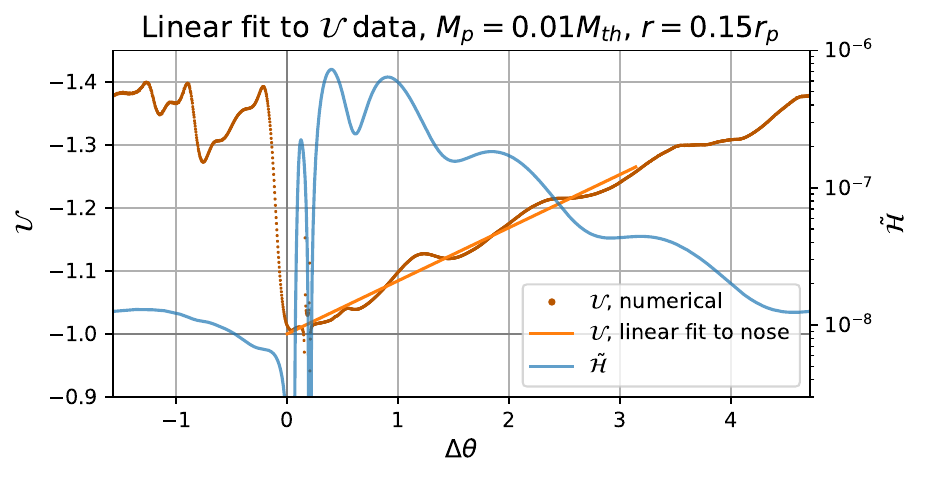}
  \caption{An example of the linear fitting procedure used to estimate the spiral-angle proxy $\mathcal{U}$ used in figure \ref{WTG1}. The wave is very close to the threshold for breaking in this example, so a faithful fit is important. The location of the nose is estimated from the profiles of $\tilde{\mathcal{H}}$ and $\mathcal{U}$, and a linear fit is taken through the first 1000 grid points after the nose.}
  \label{Ufit}
\end{figure}
\begin{figure}\centering
  \includegraphics[width=0.99\linewidth]{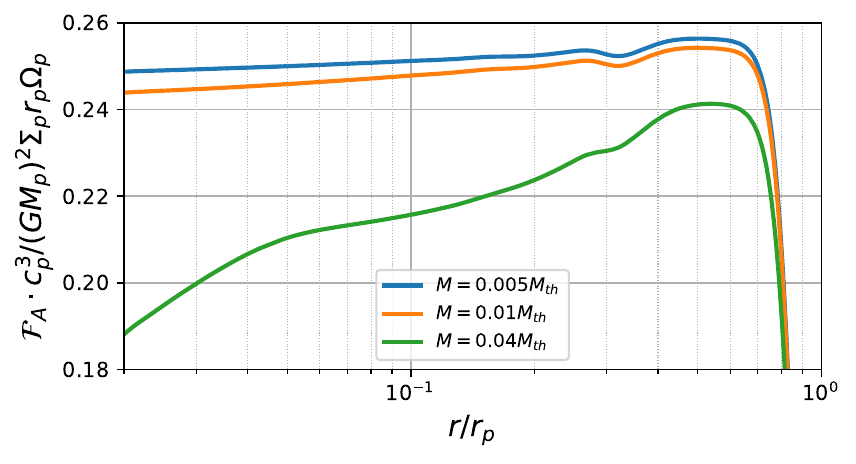}
  \includegraphics[width=0.99\linewidth]{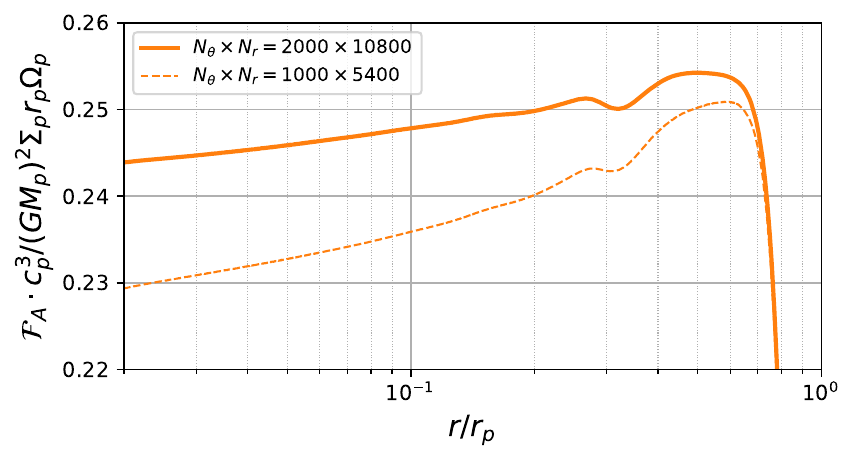}
  \vspace{-1em}
  \caption{Evolution of the angular momentum flux (AMF) carried by the inner spiral wakes in simulations \texttt{1a}, \texttt{1b} and \texttt{1c} (top), and \texttt{1b} and the \texttt{test} run (bottom). According to the exact identification with the wave action flux made in appendix \ref{appx1}, the AMF should be constant for smooth waves (neglecting any ongoing forcing by the planet). We see down-turns in the AMF profile in the simulation with $M_p = 0.04M_\text{th}$ after $r \approx 0.6 r_p$ and $r \approx 0.07r_p$, the initial formation locations of the primary and secondary shocks. Additionally, we see `torque wiggles' near $r = 0.3 r_p$ in all profiles, a feature of linear planet-disc interaction, appearing at the radial location predicted by \citet{2024MNRAS.529..425C}. We see a steady decline in the AMF profiles in the simulations with $M_p = 0.005M_\text{th}$ and $M_p = 0.01M_\text{th}$, consistent with numerical dissipation. The dissipation is stronger in the case $M_p = 0.01M_\text{th}$, which we attribute to the partial steepening of the primary wave, visible in figure \ref{WTG1} (centre left). Furthermore, the AMF in the \texttt{test} run (bottom graph) experiences roughly twice the level of dissipation of the run \texttt{1b} (which has double the resolution of the \texttt{test} run), suggesting a numerical rather than physical origin for the dissipation in run \texttt{1b}.}
  \label{AMFgraphs}
\end{figure}

It's worth pointing out that whilst it appears from figure \ref{1bdenspolar} (right) that the secondary, tertiary and quaternary spirals are gradually approaching the nose with decreasing $x$ (indeed the effective phase speed $\mathcal{U}$ is increasing in magnitude toward the tail, meaning that wave crests which follow $\dd x/\dd \theta = -\mathcal{U}$ are less tightly wound and gradually catch up with the nose), linear ray trajectories\footnote{note that the group velocity components are $\dd x/\dd t = \p \omega/\p k$ (for $k = c_0k_r/(\Omega-\Omega_p)$), $\dd \theta/\dd t = \p \omega/\p m$, therefore $\dd x/\dd \theta = \dd m/\dd k = -1/\mathcal{U}$} follow $\dd x/\dd \theta = -1/\mathcal{U}$. These rays are more tightly wound for larger $\left|\mathcal{U}\right|$, and so the wave packet is in fact dispersing in the opposite direction to the motion of the crests, namely away from the nose. 

Figure \ref{WTG1} (right) shows the evolution of the Hamiltonian $\tilde{\mathcal{H}}$ alongside the boundary for shock formation in the wave train regime $\mathcal{V}_m(\mathcal{U})$ for the three runs \texttt{1a}, \texttt{1b} and \texttt{1c}. Numerical data for $\tilde{\mathcal{H}}$ and $\mathcal{U}$ were extracted by computing $\eta$, $\eta_{X_0}$, $\eta_\theta$, $\eta_{X_0\theta}$ and $\eta_{\theta\theta}$ from the numerical data for $v_r$, $v_\theta'$, $\Sigma$ $\p_r v_r$ and $\p_r P$, making use of equations (\ref{sigmarel}), (\ref{xtoX0}), (\ref{urel}), (\ref{reomrel}) and (\ref{SAMcons}). These data were then used to estimate $\tilde{\mathcal{H}}$ and $\mathcal{U}$ at each individual grid cell\footnote{There was also a very small reflected wave present in the raw data which we removed by smoothing the data along the crests of the inward-propagating wave over one wavelength. We computed the expressions $\tilde{\mathcal{H}} = \frac{1}{2}\eta_\theta^2 + \frac{1}{2}\tilde{\kappa}_0^2 \eta^2 + \frac{1}{\gamma(\gamma-1)}\left[\frac{1+\gamma \eta_{X_0}}{(1 + \eta_{X_0})^\gamma} - 1\right]$, and $\mathcal{U} = -\frac{\eta_\theta^2 - \eta \eta_{\theta\theta}}{\eta_\theta\eta_{X_0} - \eta\eta_{\theta X_0}}$ using these smoothed data, resulting in a quantitatively similar but far higher quality result. The azimuthal wave profiles in figure \ref{WTG1} (left) were plotted using the raw data.}.

\begin{figure*}\centering
  \includegraphics[width=0.328\linewidth]{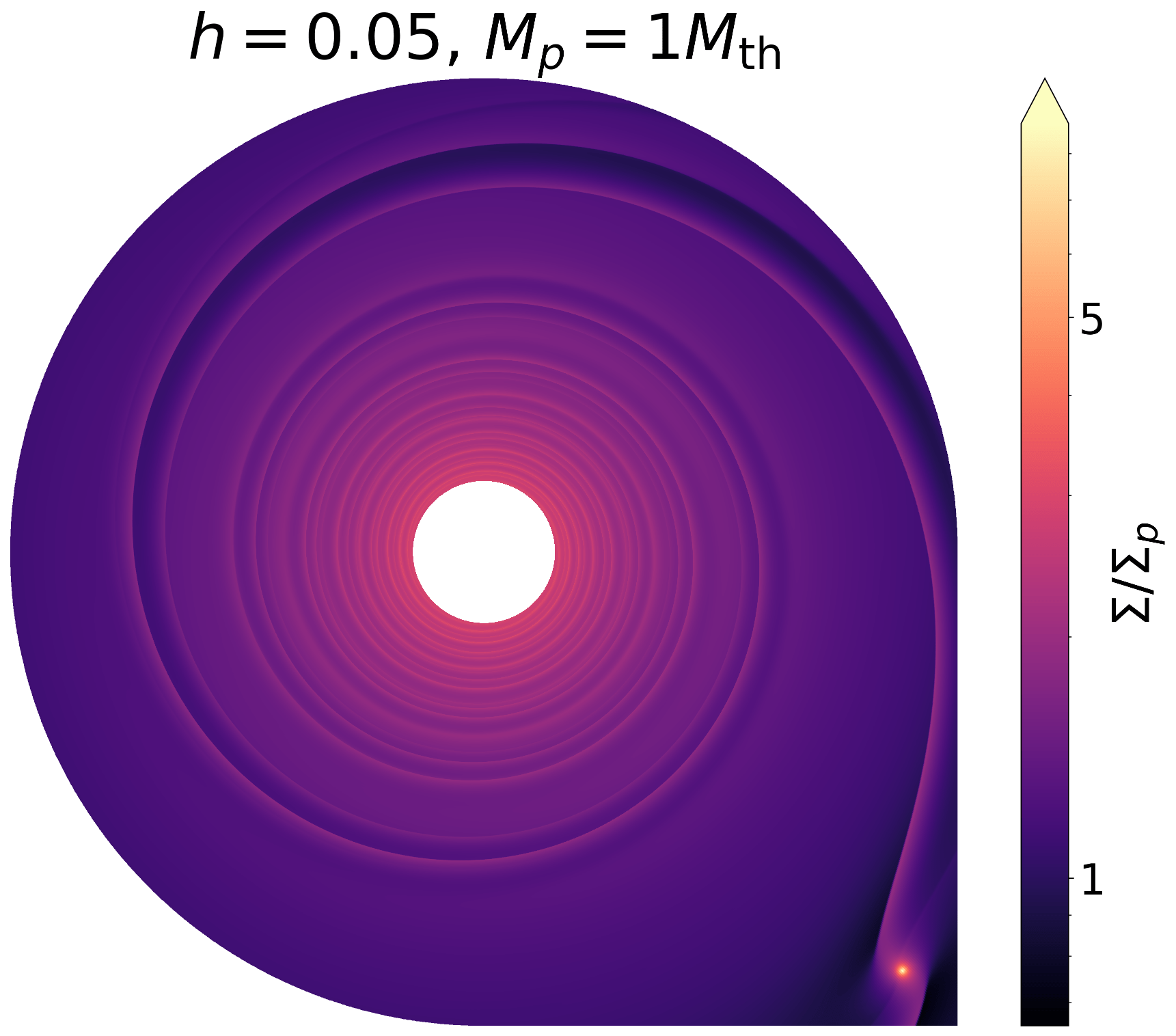}
  \includegraphics[width = 0.328\linewidth]{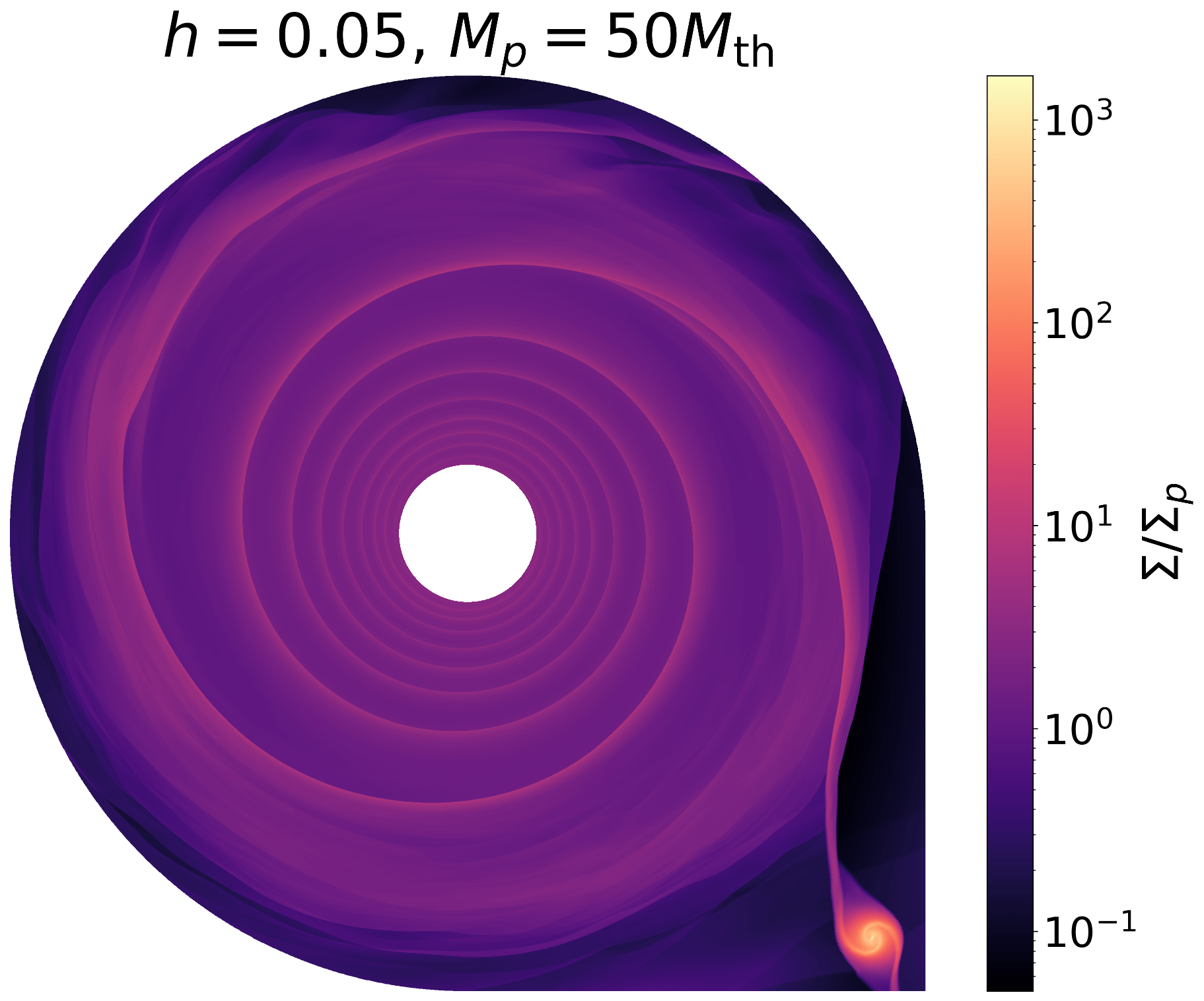}
  \includegraphics[width=0.328\linewidth]{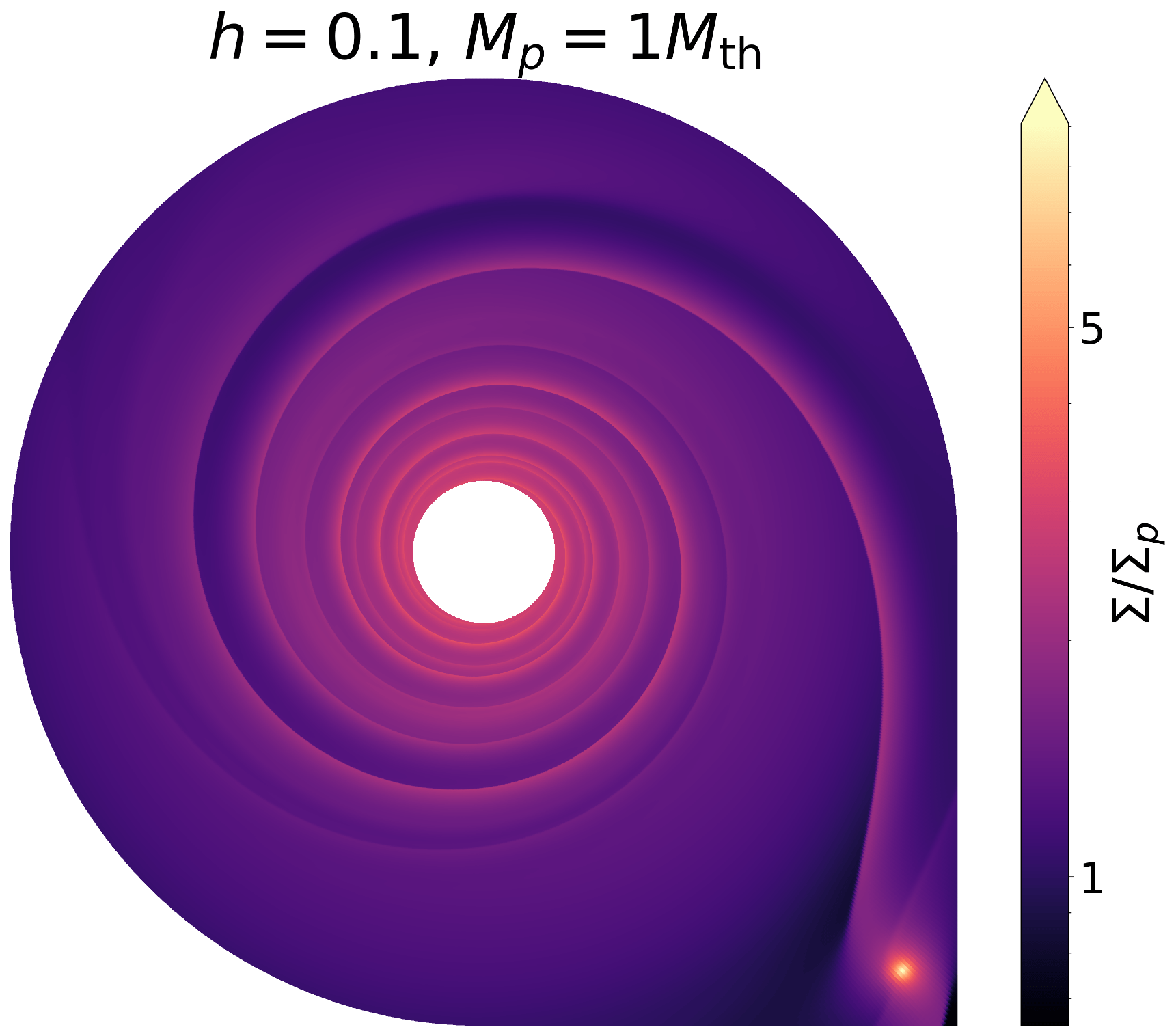}
  \includegraphics[width = 0.328\linewidth]{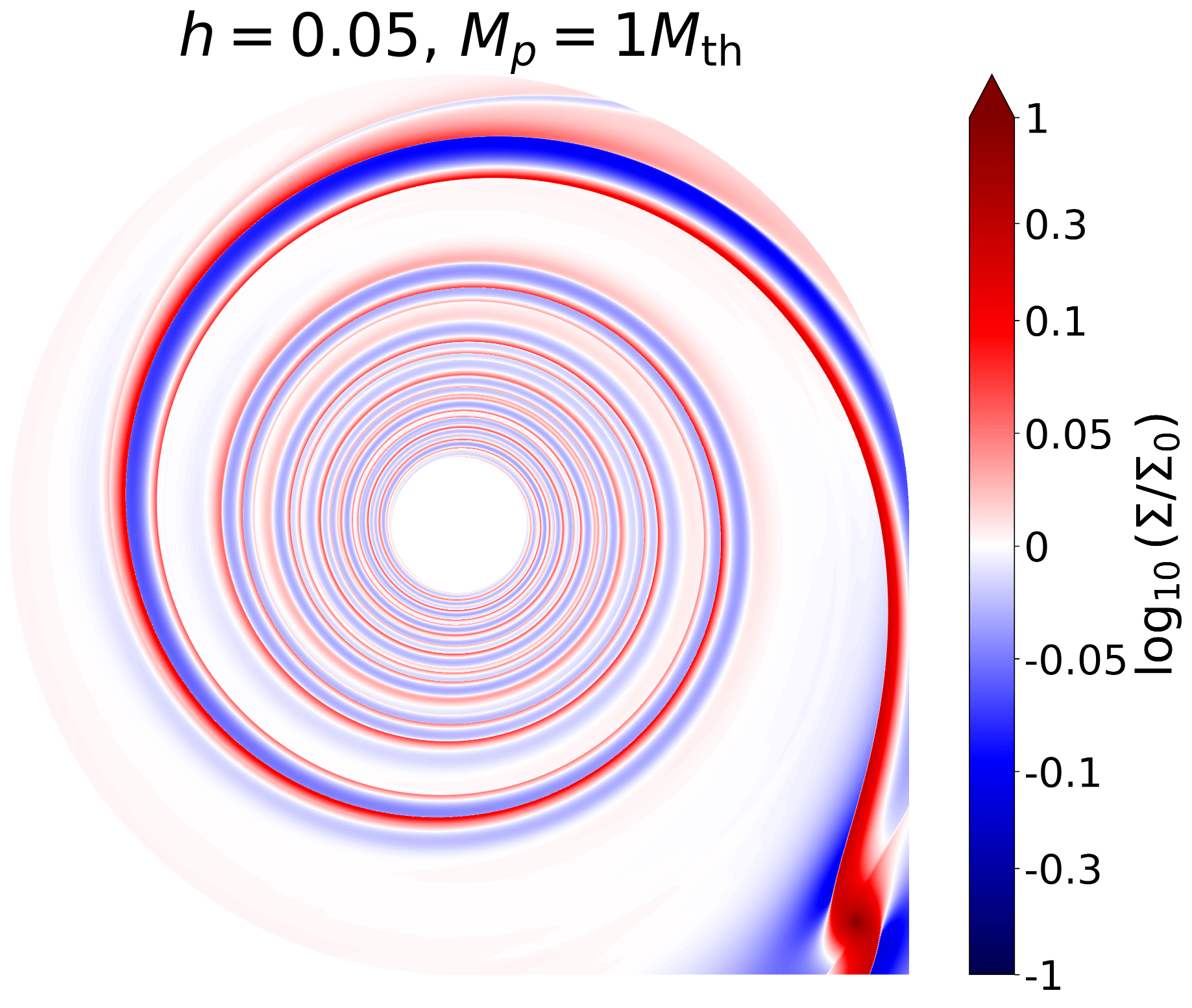}
  \includegraphics[width = 0.328\linewidth]{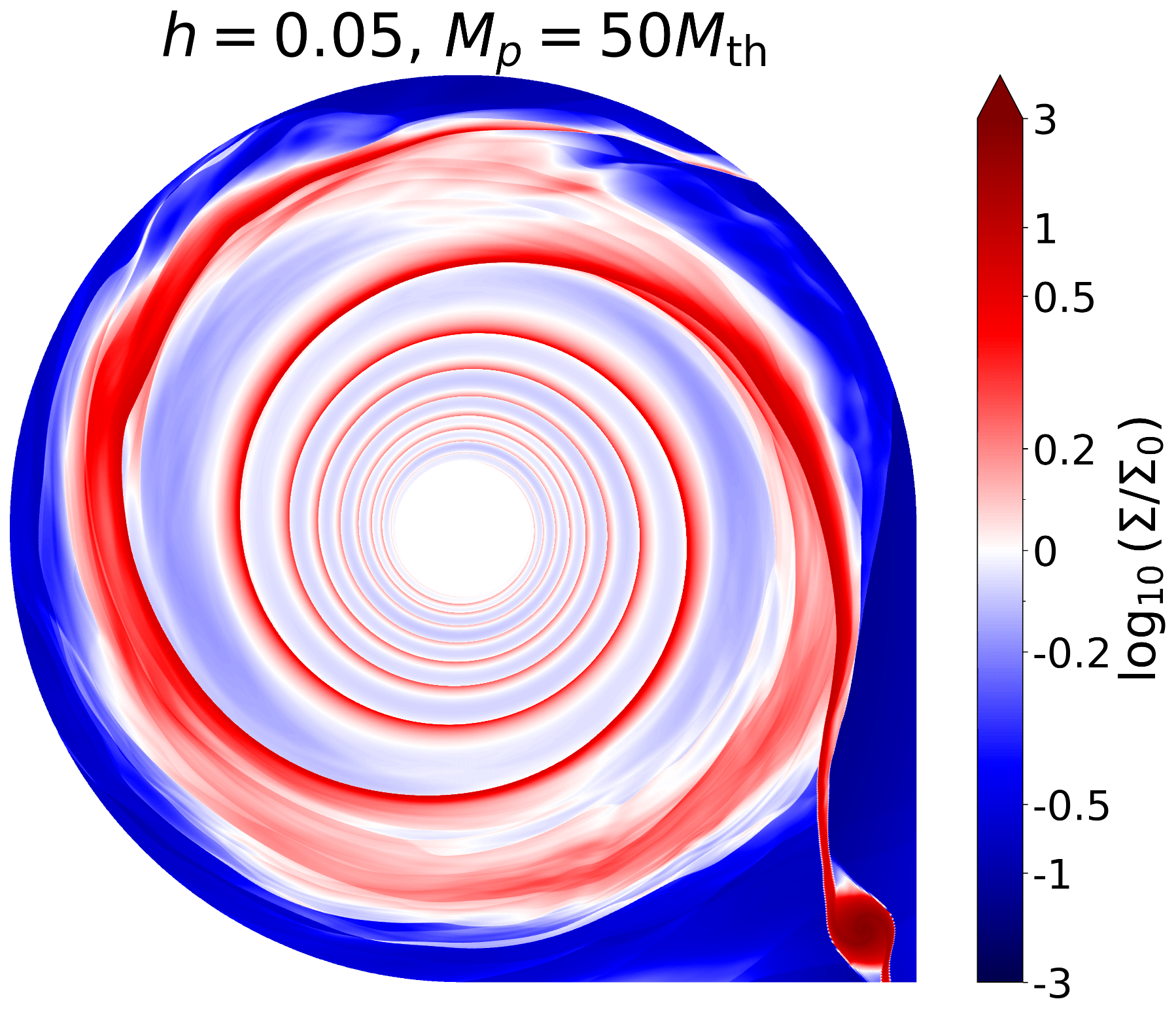}
  \includegraphics[width = 0.328\linewidth]{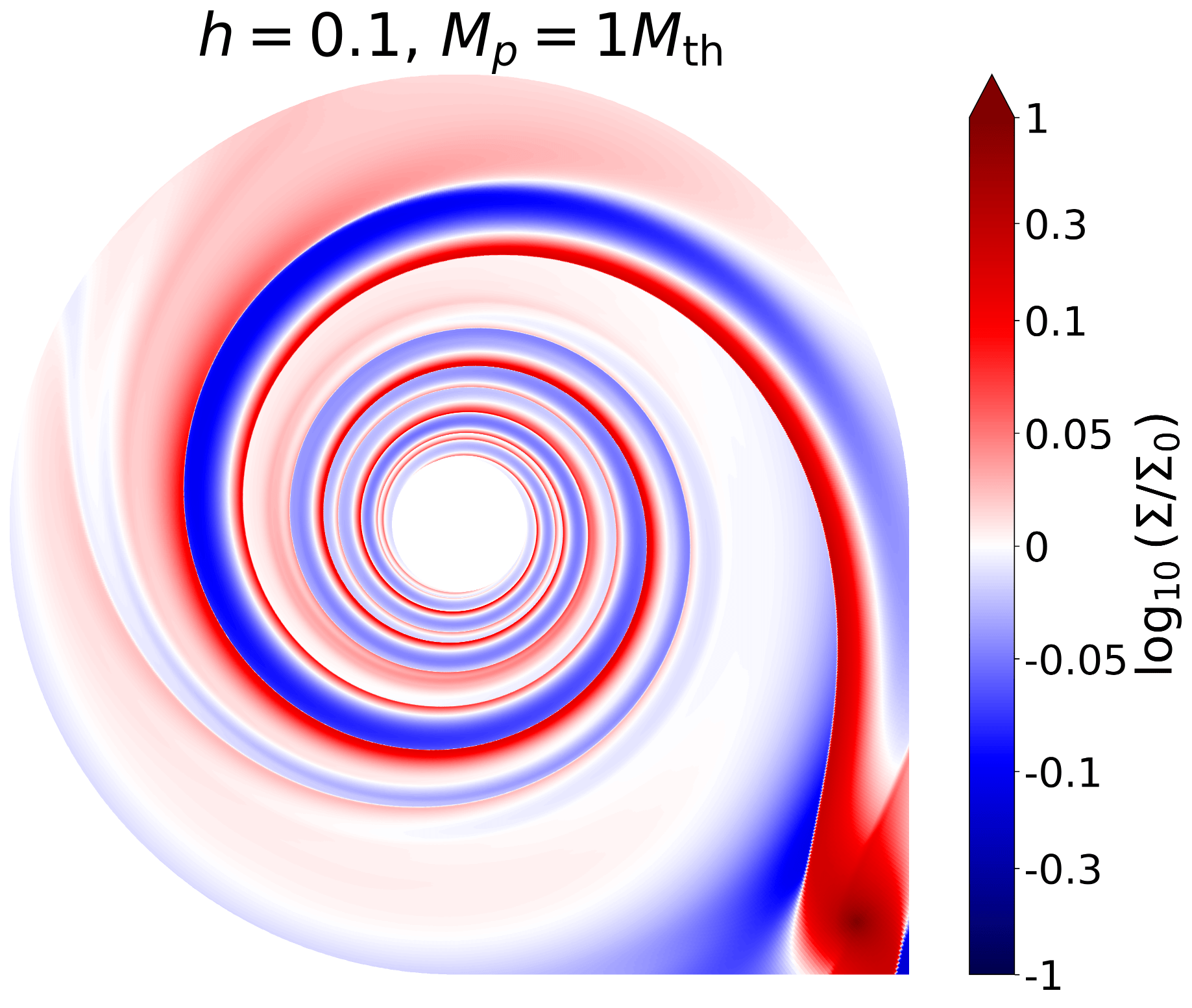}
  \includegraphics[width=0.329\linewidth]{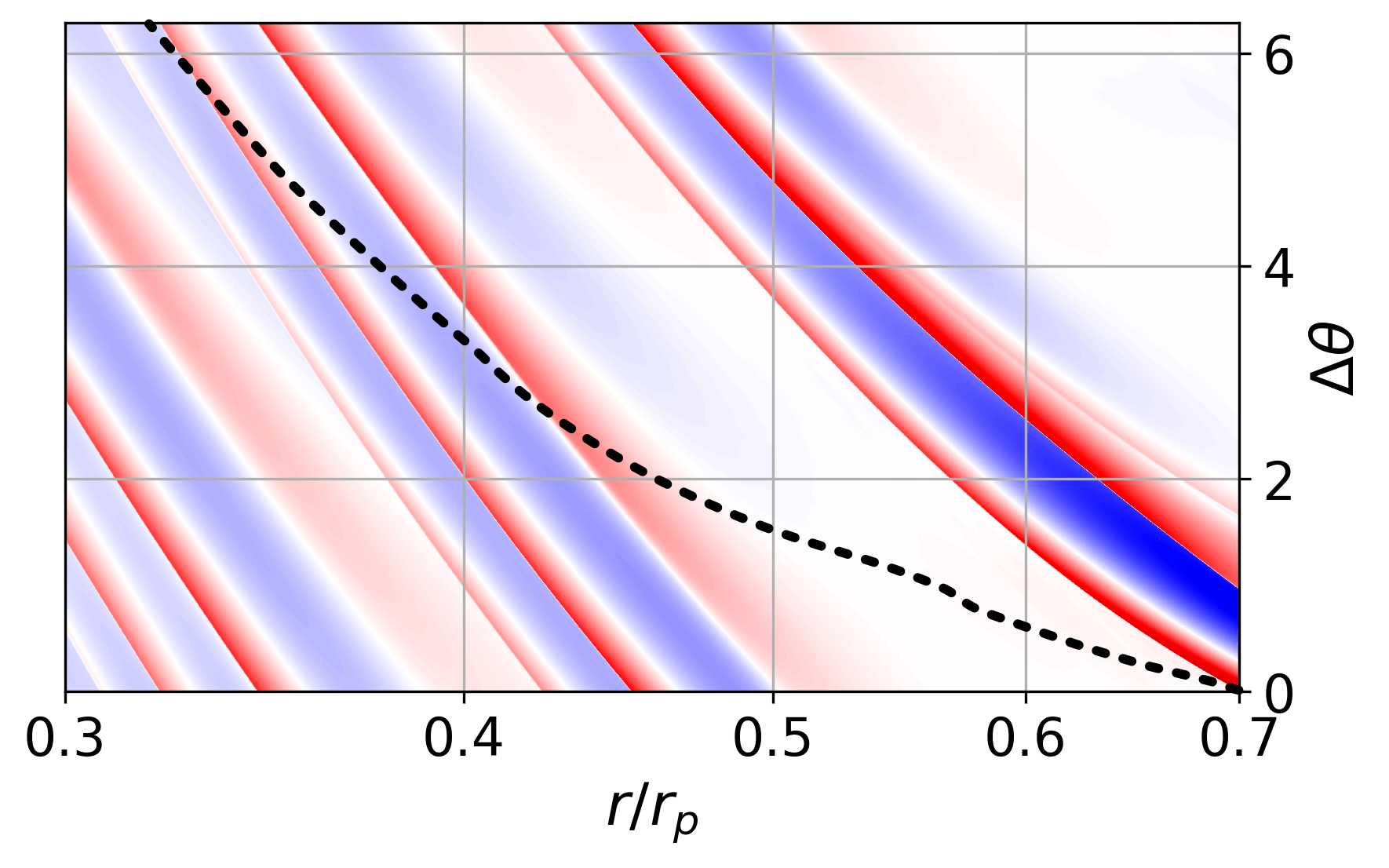}
  \includegraphics[width = 0.329\linewidth]{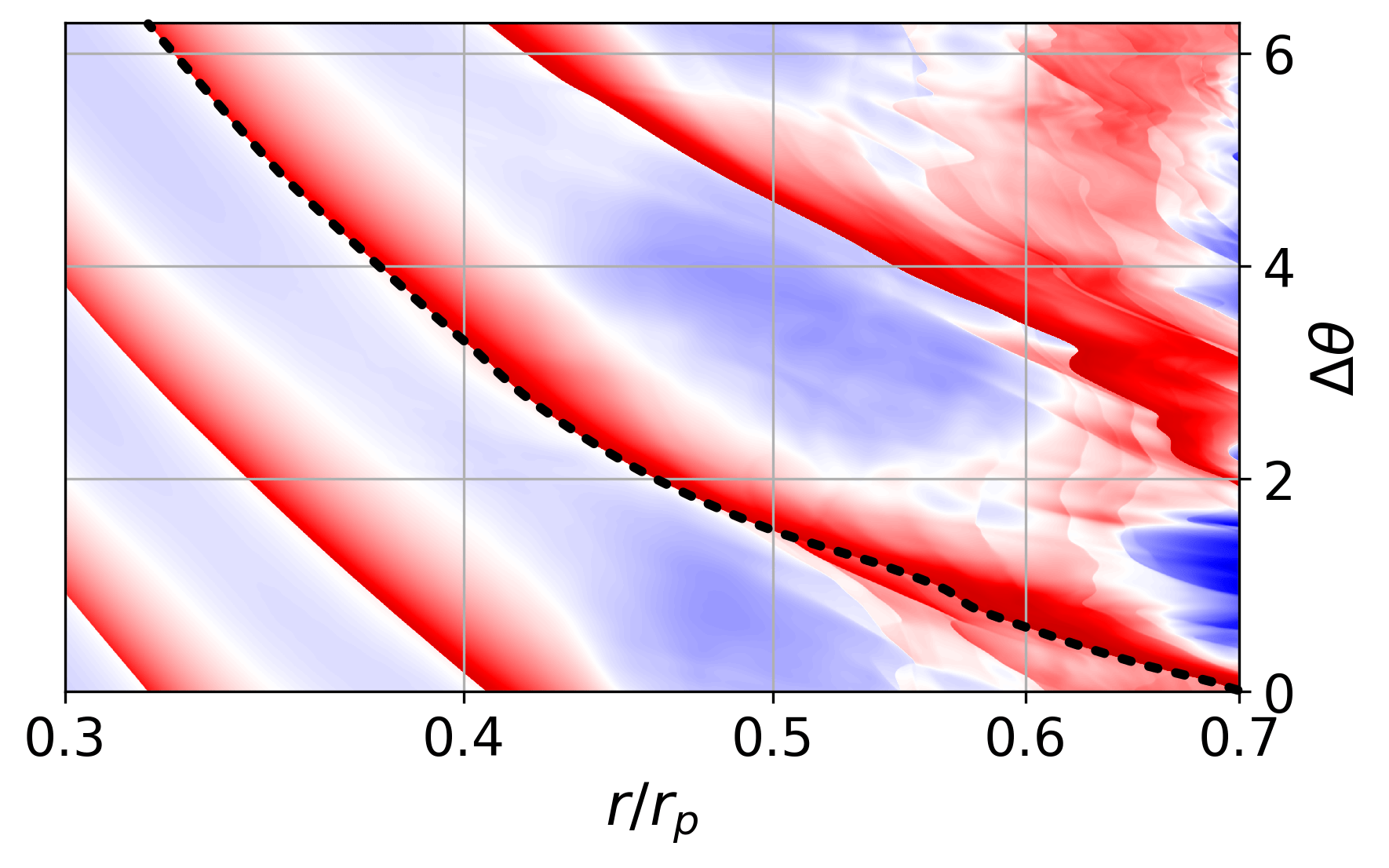}
  \includegraphics[width=0.329\linewidth]{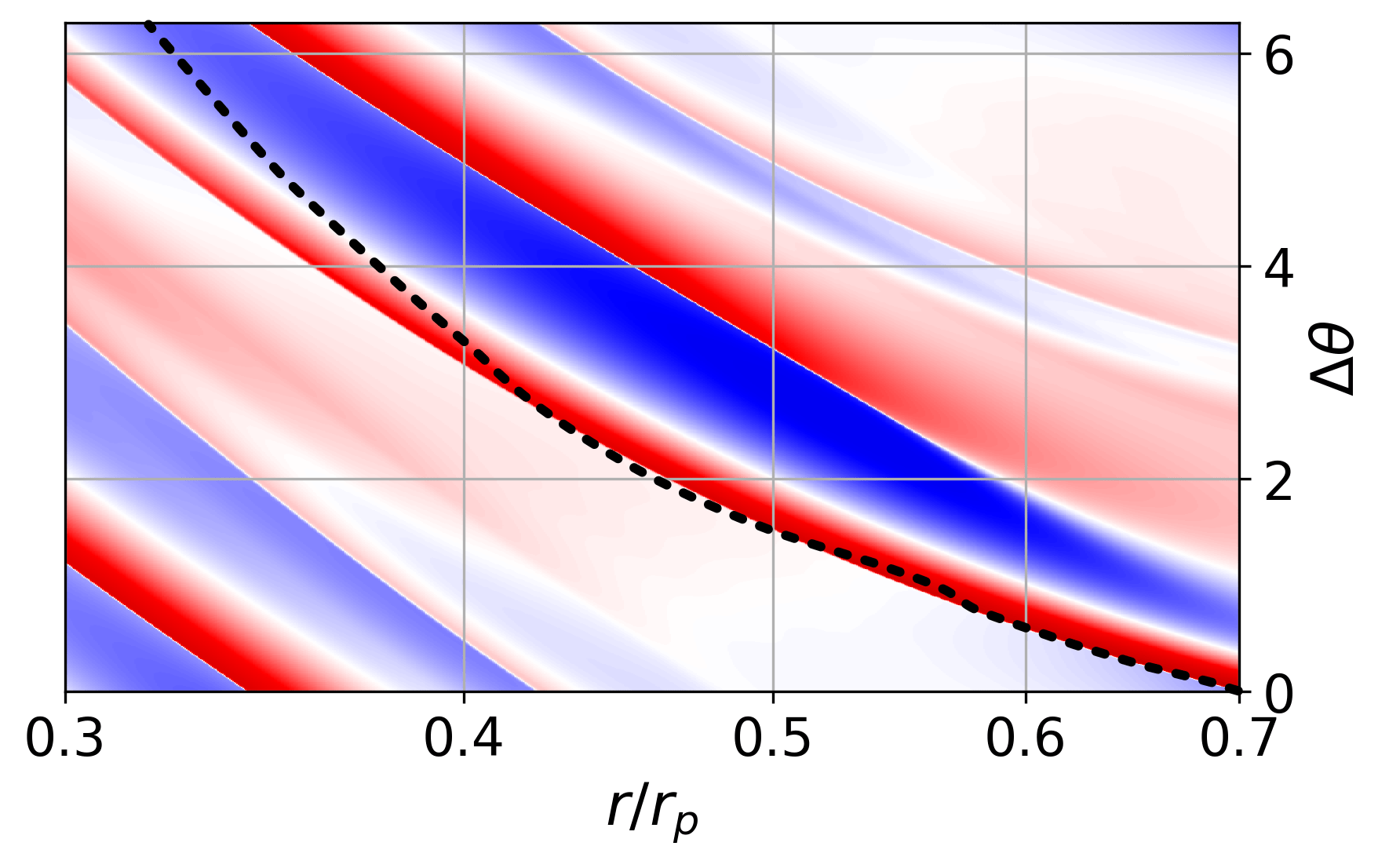}
  \vspace{-1em}
  \caption{Density plots from simulations \texttt{2a} (left), \texttt{2b} (middle) and \texttt{2c} (right) after $t = 5.0$ orbits of the planet around the central star respectively. The planet is located in the bottom right corner of the polar subplots. The very high amplitude wave driven by the $50$ thermal mass planet in run \texttt{2b} quickly disperses into a two armed pattern, which appears (initially) as loosely wound as the wake in run \texttt{2c}, despite the disc in \texttt{2c} being twice as thick. The black dotted line in the bottom subplots traces the primary shock wave in run \texttt{2b}, and is superposed on all three subplots to aid comparison. As the wave in run \texttt{2b} damps and decreases in amplitude towards the inner disc, it starts to become more tightly wound, and the spiral shape more comparable to the wake in run \texttt{2a}.}
  \label{sim2b3}
\end{figure*}

There was a further issue that due to the sensitive dependence of $\mathcal{V}_m$ on $(\mathcal{U}^2-1)$ near the nose of the wave train, just in front of the primary wave, a small numerical uncertainty in $\mathcal{U}$ translates to very large uncertainty in $\log\mathcal{V}_m$ near the nose. We dealt with this by fitting a trend-line\footnote{We note here that the expectation from linear ray theory is that $\mathcal{U}$ is linear in $\theta$ for radii $r \ll r_p$.} through the numerical data for $\mathcal{U}$ near the nose, passing through $\left|\mathcal{U}\right|=1$ at the nose. An example of this procedure is shown in figure \ref{Ufit}.

In run \texttt{1a} with $M_p = 0.005M_\text{th}$, we see from figure \ref{WTG1} (upper right), that $\tilde{\mathcal{H}}$ remains firmly below the shocking threshold for the duration of the inner wake evolution, decaying further away as the wave reaches inner radii. Furthermore, the angular momentum flux, shown in figure \ref{AMFgraphs}, appears only to experience a consistent gentle decay which we attribute to numerical dissipation. The smooth azimuthal wave profiles (see for example figure \ref{WTG1} (upper left)) also strongly suggest the absence of shocks in this simulation.

The run \texttt{1b} with $M_p = 0.01M_\text{th}$ appears marginal at all radii shown in figure \ref{WTG1}, with a single point on each profile touching the bounding envelope. Furthermore the azimuthal wave profile at $r=0.04r_p$ shows partial steepening of the primary wave, to which we attribute the increased level of numerical dissipation seen in the angular momentum flux profile in figure \ref{AMFgraphs}. It appears therefore that $M_p \approx 0.01M_\text{th}$ is very close to the threshold mass for shock formation (it's unclear from our data if it sits just above or below the threshold) in our inviscid disc.

Simulation \texttt{1c} with $M_p = 0.04M_\text{th}$ exhibits quite different behaviour. The primary spiral wave shocks at $r\approx0.6r_p$, in good agreement with the prediction of \citet{rafikov_nonlinear_2002}\footnote{The agreement is improved further when accounting for the difference in the prescriptions for the planetary potential. The Newtonian prescription $\Phi_p = -GM_p/\left|\vecr - \vecr_p\right|$ used by \citet{rafikov_nonlinear_2002} generates a wave with an amplitude roughly $40\%$ larger than the vertically averaged potential in equation (\ref{phipresc}). The close agreement in the primary spiral's shock location is likely a consequence of the primary wave being predominantly acoustic, consisting of wavenumbers $k_r \gtrsim 1/H$. In this regime the dispersion is weak. Dispersion is excluded from the theory of \citet{rafikov_nonlinear_2002} by the approximation made in his equation (A8), which effectively removes the Coriolis force.}. However, this non-dispersive theory is unaware of the secondary spiral, which develops a shock at $r\approx 0.07r_p$, corresponding to the second down-turn in the angular momentum flux graph in figure \ref{AMFgraphs}. We see both shocks in the azimuthal wave profile in figure \ref{WTG1} (bottom left). We also see that $\tilde{\mathcal{H}}$ exceeds the linearly fitted local shocking threshold $\mathcal{V}_m(\mathcal{U})$; however, a closer look at the raw data (see figure \ref{WTG1} (bottom right)) indicates that at points where the threshold appears to be exceeded, the phase speed $\mathcal{U}$ is modified so that $\tilde{\mathcal{H}} = \mathcal{V}_m(\mathcal{U})$, consistent with the family of solutions containing shocks discussed in section \ref{SFSW}.

\subsection{Large-amplitude spirals: the two-armed inner wake}\label{LAS}

\begin{figure}\centering
  \includegraphics[width = 0.99\linewidth]{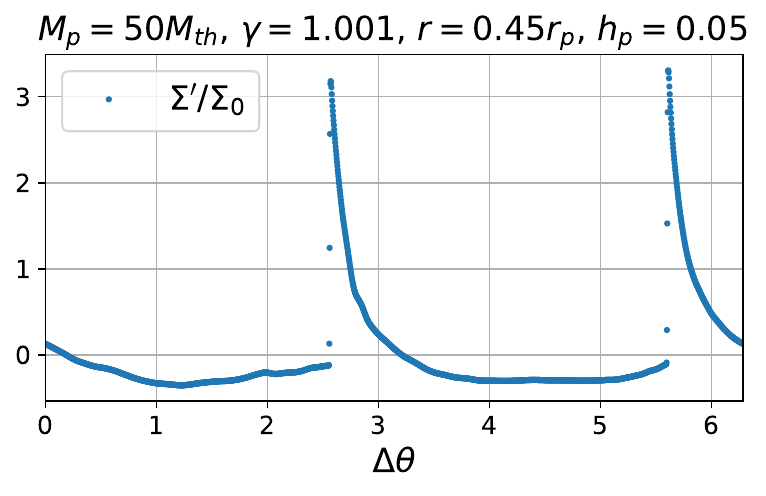}
  \includegraphics[width=0.99\linewidth]{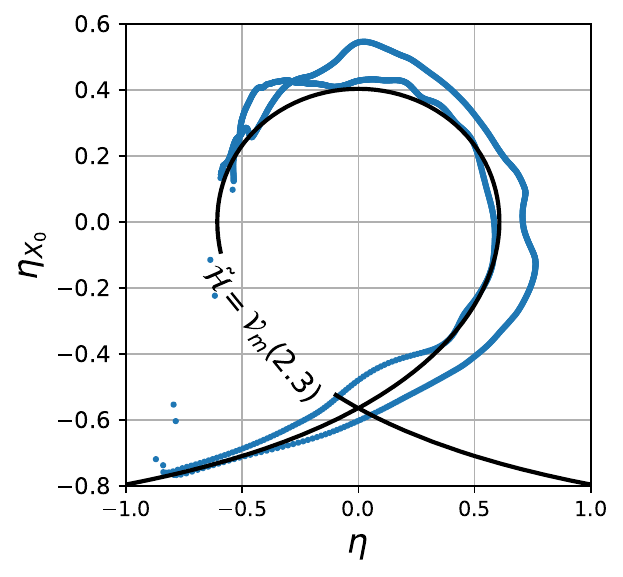}
  \vspace{-1em}
  \caption{Top: azimuthal cross-section of fractional surface density perturbation in the highly non-linear simulation \texttt{2b}. Bottom: the same data (from fixed radius $r = 0.45 r_p$ with $\tilde{\kappa}_0 = 1.43$) plotted in the phase space introduced in section \ref{SFPA}. Our model predicts that the data should lie on the loop $\tilde{\mathcal{H}} = \mathcal{V}_m(\mathcal{U})$; from the spiral's shape we've fitted $\mathcal{U} = 2.3$.}
  \label{sim3bPS}
\end{figure}

\begin{figure}\centering
  \includegraphics[width=0.99\linewidth]{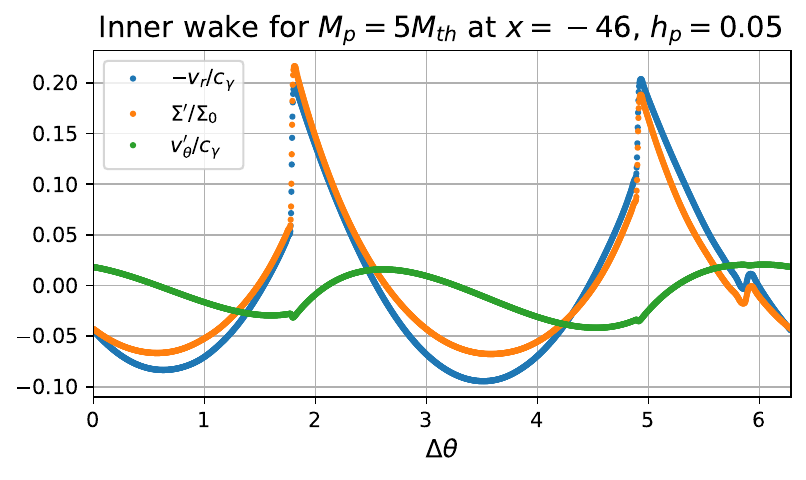}
  \includegraphics[width = 0.99\linewidth]{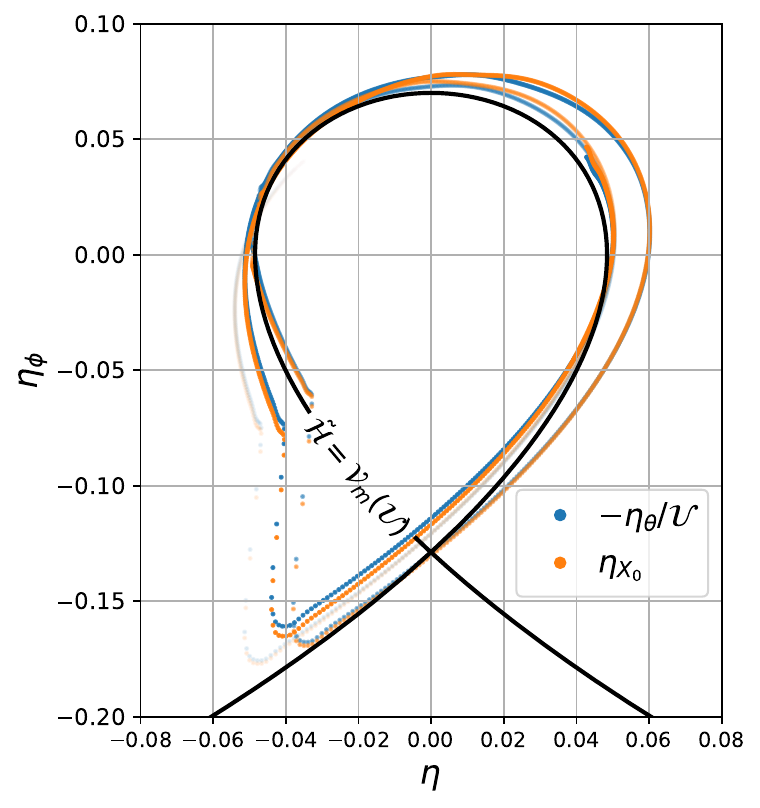}
  \vspace{-1em}
  \caption{Top: azimuthal cross-section of fractional surface density perturbation and radial and azimuthal velocities from run \texttt{3} at $x = -46$ (corresponding to $r = 0.067r_p$) (cf figure \ref{sim2a}, bottom). Note the resemblance to the wave profile predicted in section \ref{SFSW} and sketched in figure \ref{wave_break} (bottom right). Bottom: a radial slice of data (identifying $\eta_\theta = v_r/c_0$, $\eta_{X_0} = \Sigma_0/\Sigma - 1$) from the same simulation spanning the range $0.064r_p > r > 0.046 r_p$, plotted in the phase space outlined in section \ref{SFPA}. In both top and bottom plots, $\Sigma_0$ and $c_0$ are taken to be the azimuthally averaged surface density and sound speed rather than initial background values. The darker points represent data from larger radii, and the faintest point is taken from $r = 0.046r_p$. We've superposed the loop in phase space for phase speed $\mathcal{U} = 1.18$, and used the same value uniformly in the calculation of $\eta_\theta/\mathcal{U}$. The distance between data points for $\eta_{X_0}$ and $\eta_\theta/\mathcal{U}$ is a measure of how good the fit $\mathcal{U} = 1.18$ is.}
  \label{sim2aPS}
\end{figure}

\begin{figure*}\centering
  \includegraphics[width = 0.99\linewidth]{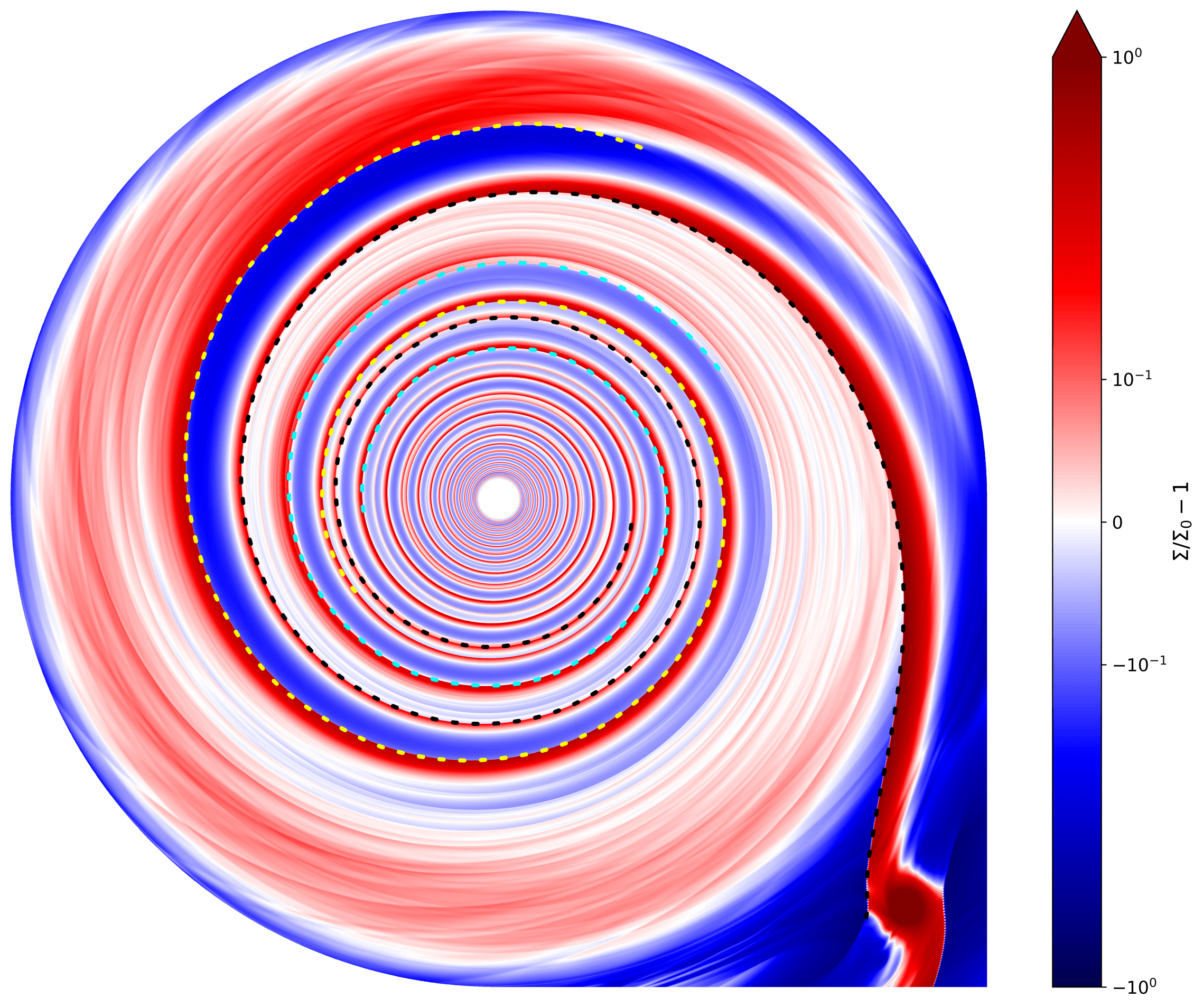}
  \includegraphics[width = 0.95\linewidth]{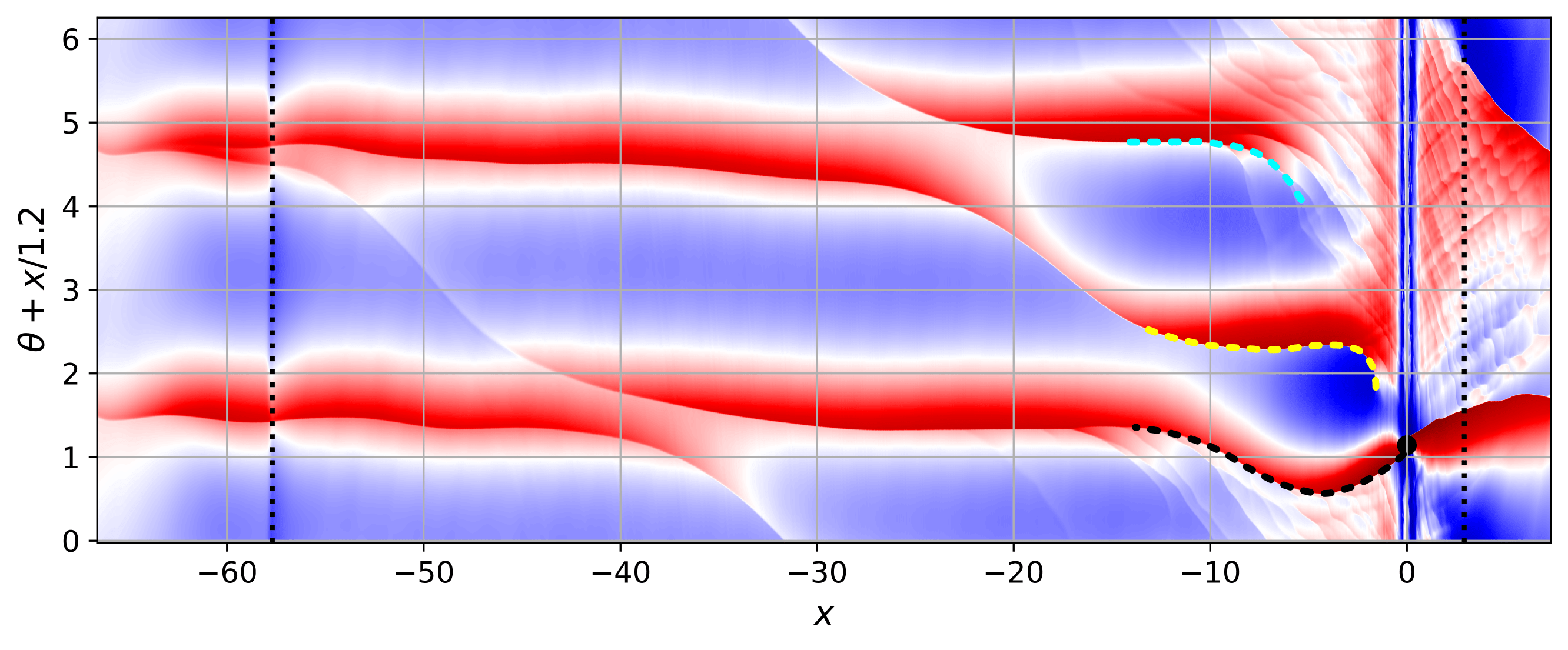}
  \vspace{-1em}
  \caption{Fractional surface density perturbation in the inner disc in simulation \texttt{3} after $t = 15.0$ orbits at $r_p$. Top: polar projection; the planet is situated in the bottom right corner, and excites an initially three-armed spiral wake. The three spiral shocks are highlighted by dotted lines; these arms gradually merge into two arms in the inner disc. Bottom: the same data (sharing the colour bar of the top plot) now plotted in the coordinate system $(x, \theta + x/\mathcal{U})$ for $\mathcal{U} = 1.2$. The planet is marked by the black dot at $x = 0$. $x$ was calculated via equation (\ref{xdefn}), taking $c_0(r)$ to be the azimuthal average of the sound speed (rather than its initial distribution), as shock heating has slightly warmed the disc. There is also a very small leading tightly wound spiral visible in the density distribution due to the presence of a small reflected wave. The vertical black dotted lines mark the boundaries of the wave damping zones.}
  \label{sim2a}
\end{figure*}

The non-linear theory of spirals developed in sections \ref{GDW} and \ref{SFPA} also makes quantitative predictions concerning the highly non-linear regime which we wish to test numerically, including the precise profile of tightly wound spirals with a given spiral angle, as well as the prediction that two-armed spirals are preferred non-linearly due to restrictive effective caps on the possible amplitude of spirals with more arms. We also wanted to investigate how much non-linearity can affect the looseness of spirals in practice.

In this section we present the results of four simulations, runs \texttt{2a}, \texttt{2b}, \texttt{2c} and \texttt{3}, which probe this non-linear regime. Figure \ref{sim2b3} shows density plots from the simulations \texttt{2a}, \texttt{2b} and \texttt{2c}. In run \texttt{2b} (which has $M_p = 50 M_\text{th}$, around $6$ Jupiter masses for $h_p = 0.05$), we see the wake excited by the planet quickly disperse into a two-armed spiral, comparable to that seen by \citet{2015ApJ...809L...5D} and in the and the discs MWC 758 and HD100453 \citep{benisty_asymmetric_2015,rosotti_spiral_2020}. The speed of the dispersion into two arms is striking, and we attribute this to the preference of very high amplitude waves for an angular wavelength of $\upi$, inferred via the non-linear dispersion relation discussed in section \ref{UNLW}. In addition, at this high planet mass, the hard cap becomes restrictive for waves with more than two arms: indeed we find numerically that the angular momentum flux $\mathcal{B} \approx 5 \mathcal{F}$ at $r = 0.45 r_p$ in run \texttt{2b} (corresponding to the cross-section plotted in figure \ref{sim3bPS}), which is well above the hard cap for $m \geqslant 3$. Remarkably, we see from this cross-section that the profile predicted by the non-linear model (\ref{Spiraleqn}) remains quantitatively good even at high amplitudes. 

Perhaps the most striking aspect of figure \ref{sim2b3} however is that the shape of the inner spiral wake in run \texttt{2b} is very similar (for $r \gtrsim 0.45r_p$) to that in run \texttt{2c}. This is surprising as we would expect spirals in thinner discs to be far more tightly wound \citep{ogilvie_wake_2002}. The crests of the very high amplitude waves excited by the $50$ thermal mass planet in simulation \texttt{2b} follow loose spirals as their large-amplitude shocks naturally travel radially at a couple times the sound speed. This may also be understood as a consequence of the non-linear dispersion relation shown in figure \ref{NLDR} and discussed in section \ref{UNLW}, and has been observed to some extent in a few previous studies (e.g. \citet[figure 9]{yuan_resonantly_1994} and \citet[figure 1]{2015ApJ...815L..21F}). This non-linear loosening effect may prove very important in understanding the surprisingly loose spirals in the discs MWC 758 and HD100453; we discuss this further in section \ref{WAP}.

Simulation \texttt{3} (shown in figure \ref{sim2a}) includes a planet of mass $M_p = 5M_\text{th}$ and covers a domain extending far inwards (to $r = 0.035r_p$) at a resolution of 16 cells per scale-height at the planet's location. The inner wake initially disperses into a three-armed spiral, before subsequently evolving into a two-armed spiral as the wave propagates inwards. We attribute the three-armed pattern close to the planet to the initial phase of the evolution taking place effectively within a shearing sheet centred on the planet (e.g. \citet{2017MNRAS.472.1432L}), in which we'd expect multiple arms to form via dispersion should the azimuthal extent of the sheet be very large (as is the case for very thin discs).

The evolution of the three-armed wake in figure \ref{sim2a} into a two-armed spiral pattern in the inner disc highlights further the natural preference for two-armed spirals. In this case, from figure \ref{sim2a} (bottom), it appears that the higher order azimuthal components of the solution quickly decay, leaving the gradually decaying two-armed remnant, and so we attribute this behaviour to the action of the soft cap on spiral wave amplitude discussed in sections \ref{SFSW} and \ref{UNLW}. It's worth noting that the transition from three to two arms happens far inside the 2:1 Lindblad resonance (located at $r = 0.63r_p$, which surprisingly is only at $x = -2$), whereas the transition seems to happen near $x = -12$ ($r = 0.285r_p$). That is, it seems that resonant enhancement of the $m=2$ component alone cannot explain the observed transition from two to three arms.

A cross-section of the wake in the inner disc is shown in figure \ref{sim2aPS}: there is a strong resemblance to the solution anticipated in section \ref{SFSW} and the profile sketched in figure \ref{wave_break} (bottom right). Here $\eta$, $\eta_{X_0}$, $\eta_\theta$ were extracted numerically using the same method as in section \ref{SAS}. Figure \ref{sim2aPS} shows that the numerical solutions sits remarkably close to the predicted solution curve $\tilde{\mathcal{H}} = \mathcal{V}_m(\mathcal{U})$ in phase space. Deviations are likely due to non-uniformity in $\mathcal{U}$ (due to the weak shocks and variations in the disc's profile), the finite aspect ratio, as well as the presence of a small reflected wave in the data. Nevertheless, we see the quantitative predictive power of the analytic description developed in section \ref{SFSW}.

\section{Discussion}\label{discussion}

\subsection{Connection to previous studies}

This paper was motivated by the question: `is it possible that dispersion is able to prevent the spiral wakes excited by planets from steepening into shocks?' We had the smooth yet non-linear waves studied by \citet{1985ApJ...291..356S}, \citet{yuan_resonantly_1994}, \citet{2009Icar..202..565L}, and in particular \citet{fromang_properties_2007} in mind when setting out on the project, and were curious whether the dispersive effects keeping the waves studied by \citet{fromang_properties_2007} from breaking acted similarly on planet-driven wakes.

\citet{heinemann_weakly_2012} derived a weakly non-linear theory for spiral waves in the shearing sheet. They found that dispersive effects were weak, and unable to prevent wave steepening into shocks, further showing asymptotic similarity to 1D non-linear gas dynamics, reminiscent of the theory derived by \citet{rafikov_nonlinear_2002} discussed below. This picture is consistent with the analysis in appendix \ref{appxc}, in which we find dispersion is unable to overcome the unbounded shear amplification in the linear regime in the shearing sheet. A global picture is necessary to capture the long length-scales and low angular wavenumbers $m$ meaningfully affected by dispersion; in inviscid, ultra-thin discs, all waves generated near their corotation radii are expected to steepen into shocks.

The derivation of the non-linear wave equation presented in section \ref{GDW} has much in common with derivation presented in \citet{1985ApJ...291..356S}. The key (non-cosmetic) differences include our extension to the globally valid wave equation (\ref{MasterEqn}) (the equations derived by \citet{1985ApJ...291..356S} and used by \citet{yuan_resonantly_1994} are valid over a short radial extent, and Taylor expanded in the vicinity of a Lindblad resonance) as well as our focus on the careful inclusion of pressure, capturing its effects on the disc's equilibrium structure as well as the non-linear wave equation (which we derived via entropy conservation). \citet{1985ApJ...291..356S} approached the problem from the perspective of particle dynamics (rather than hydrodynamics), imagining self-gravitation as the key non-linear effect.

Our equation (\ref{MasterEqn}) is equivalent to the wave equation derived by \citet{fromang_properties_2007} in the case $\Omega_p = 0$, $\gamma = 1$, $\Sigma_0 \propto r^{-1}$ and $T_0 \propto r^{-1}$ under the identification $\theta \leftrightarrow t$. For a thin disc therefore, the self-similar solutions found in section \ref{S7} also \emph{exactly} coincide with the axisymmetric travelling wave solutions derived by \citet{fromang_properties_2007}, which explains their resemblance. Physically, the connection is rooted in the analogy between time evolution in the axisymmetric case and the evolution following the fast orbital advection in the case of steady spiral waves.

This identification of the non-linear solutions to the spiral wave equation with exact Eulerian self-similar solutions (inspired by \citet{spruit_stationary_1987}) allowed for an explicit test of the thin disc and tight-winding assumptions which underpin the non-linear equation (\ref{MasterEqn}). The convincing fit even for moderate aspect ratio seen in figure \ref{NLTDcomp} suggests that the approach of section \ref{SFPA} may provide a suitable framework to generalise the results of \citet{spruit_stationary_1987} and \citet{hennebelle_spiral-driven_2016} (including accretion rates) to non self-similar disc profiles. In particular, spirals with very strong shocks which drive rapid accretion, such as those seen in our simulation \texttt{2b} (cf figure \ref{sim2b3}), are obscured from self-similar studies because wave amplification (due to radial variations in the background disc medium\footnote{which has $\mathcal{F}\propto \sqrt{r}$ (for solutions with constant mass flux) or $\mathcal{F}\propto 1$ (for solutions with constant angular momentum flux)}) is unable to balance the fast wave damping due to shocks in this case.

\citet{larson_non-linear_1990} addressed the problem of estimating the azimuthal profiles (and shock strengths) of spiral shock waves, and this served as inspiration for our analysis in section \ref{SFSW}. Our analysis extends that of \citet{larson_non-linear_1990} from weakly non-linear to fully non-linear waves, though still with formally weak shocks. 

Another important connection (which inspired the convenient radial coordinate transformation in section \ref{GDW}) is with \citet{rafikov_nonlinear_2002}, who pointed out the compelling analogy between the non-linear evolution of spiral wakes and wave steepening in one-dimensional non-linear gas dynamics. The main difference between our analysis and his is in the treatment of the Coriolis force in the radial equation of motion: replacement of the azimuthal velocity by a term proportional to the radial velocity in \citet[equation (A8)]{rafikov_nonlinear_2002} eliminates any dispersive behaviour. In the resulting system, all waves eventually steepen into shocks, and there is no dispersion into multiple arms. \citet{rafikov_nonlinear_2002} also made a prediction about the wakes from planets below a certain mass threshold not shocking, due to the wakes in discs with particularly steep surface density and temperature profiles only winding around $r=0$ finitely many times as they spiral inwards, meaning their evolution is essentially cut short before they have time to shock. This prediction notably doesn't apply to the disc profile simulated in section \ref{SAS} with $h = 0.1$, $T_0 \propto r^{-1}$, $\Sigma_0 \propto r^{-1}$, in which spirals encircle the origin infinitely many times, yet we nevertheless predict a threshold shocking mass of approximately $0.01 M_\text{th}$. Whilst we expect the inner spiral wake to shock for planets more massive than this threshold, for planets with mass in the range $0.01M_\text{th} \lesssim M_p \lesssim 0.05 M_\text{th}$, dispersion still has a significant effect on the shock location (which is pushed further away from the planet). For moderate planet masses $M_p \gtrsim 0.05 M_\text{th}$, the prediction made by \citet{rafikov_nonlinear_2002} for the shock location of the primary wave remains good, as the primary wave is predominantly acoustic, consisting of wavenumbers $k_r \gtrsim 1/H$ for which the dispersion is weak.

\subsection{Type I migration stalling}

Perhaps the most pertinent application of our results for low mass-planets is in the resonant migration stalling outside gaps or the inner edge of protoplanetary discs. \citet{2011ApJ...741..109T} showed that reflection of the inner wake at the disc's inner boundary can significantly enhance the inner Lindblad torques, should the interference with the reflected wake be constructive. This leads to a robust and stable mechanism for migration stalling, so long as the inner wake does not damp too quickly and a meaningful proportion of the wave is reflected at the inner boundary. \citet{2018MNRAS.473.5267M} followed up this study with numerical simulations involving Earths and super-Earths in viscous discs with aspect ratio $h = 0.05$, finding indeed that migration was consistently stalled, in qualitative agreement with \citet{2011ApJ...741..109T}. This effect is often artificially suppressed in numerical studies of planet migration due to the widespread adoption of strong wave damping zones (e.g. \citet{2006MNRAS.370..529D}).

We found in section \ref{SAS} that there is a threshold mass, $M_p \sim 0.01 M_\text{th}$ for $h=0.1$, below which the inner wake does not shock. In an inviscid disc, the inner wake therefore does not damp and the mechanism proposed by \citet{2011ApJ...741..109T} becomes particularly important, dependent now only on the damping rate due to other processes\footnote{which may include viscous and thermal effects as well as the parametric instability discussed below}, as well as the reflection efficiency. Our no-shock result highlights how important this mechanism may be for the stable migration stalling of low-mass planets and planetesimals, which consequently may be afforded a lifetime in the disc significantly longer than the migration timescale.

\subsection{Wider applications}\label{WAP}

Whilst our numerical investigation and many of our analytical considerations had the case of planet-driven spiral waves in mind, many of our findings also have applications in other systems which exhibit spiral waves. These include circumplanetary discs \citep{rivier_circum-planetary_2012,2016ApJ...832..193Z}, circumbinary discs \citep{1979MNRAS.186..799L,1994MNRAS.268...13S}, flybys \citep{2020A&A...639L...1M,2023MNRAS.521.3500S,2025MNRAS.tmp.1506P}, stars or black-holes embedded within AGN discs \citep{secunda_orbital_2019}, misaligned systems \citep{2025ApJ...980..259Z,2025MNRAS.542.1430R}, gaseous spirals near galactic centres \citep{2024MNRAS.528.5742S}, vortex-driven spirals \citep{2007MNRAS.381..809M,2010ApJ...725..146P}, as well as spirals generated via instabilities, such as the GI and MRI \citep{heinemann_excitation_2009,bethune_spiral_2021}. The non-linear spiral wave equation (\ref{MasterEqn}) and its solutions detailed in section \ref{SFPA} may provide valuable insight into preferred wave profiles, shock strengths, spiral angles\footnote{Notably, high amplitude spirals with large shocks are significantly more loosely wound than their linear counterparts.}, as well as the non-shocking of spirals with moderate amplitude (with $\Delta\Sigma / \Sigma_0$ up to $13\%$, as seen in figure \ref{SSSpirals}) across the diverse range of systems mentioned above.

A more specific example concerns the spiral waves excited by vortices (e.g. \citet{2010ApJ...725..146P}). Whilst the threshold planet mass (found in section \ref{SAS}) below which the inner wake does not shock is low, and certainly lies below the mass at which gap opening becomes an important effect, there may be a more restrictive threshold for shock formation in the waves excited by vortices (which in general have a shallower initial profile compared to waves excited by the singular potential of an embedded planet, see for example \citet[figure 3]{2025ApJ...979..244M}). As in our study, this would need investigating in a global disc model, and could have important consequences in diagnosing the origin of gaps and rings, should vortex-induced waves be less prone to steepen into shocks.

Whilst the coherent three-armed planet-induced spiral seen in figure \ref{sim2a} highlights the possibility of more intricate companion-driven spiral patterns in protoplanetary discs, the very highest amplitude (and therefore most discernable in nature) driven inward-propagating waves remain nevertheless two-armed. As discussed in section \ref{UNLW}, the preferred angular wavelength of ultra-high amplitude spiral waves approaches $\upi$, and spirals with three or more arms can only exist, even with shocks, up to a maximum amplitude. There is no such hard cap on the amplitude of two-armed spiral waves. As well as this hard cap on the amplitude of non-linear spirals, there's a much lower soft cap for each $m$, above which spiral waves with $m$ arms must have shocks in their profile, and experience dissipation (conversely, below the soft cap, smooth non-linear waves are possible, such as the exact solutions found in self-similar discs in section \ref{S7}). Both the soft and hard caps on angular momentum flux scale with $m^{-4}$ for $m \gg 1$, with the consequence that two-armed spirals are strongly preferred among waves of high amplitude.

Furthermore, these highly non-linear spirals are far more loosely wound than their linear counterparts. This effect appears in the non-linear dispersion relation discussed in section \ref{UNLW}, and may be understood to be a consequence of higher amplitude shock waves having large pre-shock radial velocities, meaning that their shocks (along with their wave crests and troughs) travel radially at a speed noticeably faster than the background sound speed. This is observed in simulation \texttt{2b}, which is detailed in section \ref{LAS} and shown in figure \ref{sim2b3} (centre). Despite the disc in run \texttt{2b} having half the thickness of the disc in run \texttt{2c}, with the consequence that spirals ought to wind twice as many times around the origin in run \texttt{2b}, we see the inner wakes assume roughly the same shape (initially) in both simulations. This is due to the $50$ thermal mass planet in run \texttt{2b} exciting a spiral of far higher amplitude than the $1$ thermal mass planet in run \texttt{2c}. This effect is also implicit in the self-similar solutions found by \citet{spruit_stationary_1987}, though not directly commented on, and is visible in \citet[figure 9]{yuan_resonantly_1994} and \citet[figure 1]{2015ApJ...815L..21F}). It has also been previously understood to be caused by higher amplitude N-waves expanding more rapidly azimuthally in the context of the weakly non-linear theory developed by \citet{rafikov_nonlinear_2002} (e.g. \citet{2015ApJ...813...88Z,2021MNRAS.508.2329C}).

The non-linear loosening of spirals has important consequences for the interpretation of spiral waves observed in protoplanetary discs, for example in the discs MWC 758 \citep{benisty_asymmetric_2015}, HD100453 \citep{rosotti_spiral_2020} and around the T-Tauri star Haro 6-13 \citep{2025ApJ...988..106H}. In particular, the surprisingly loose spirals in the scattered light emission from the disc MWC 758 led \citep{benisty_asymmetric_2015} to conclude that the disc is unusually warm and thick, with $h \sim 0.2$ (the largest value permissible in their model). This value arose from the best fit of the observed spirals to linear wave characteristics; however, as we've discussed, high levels of non-linearity can lead to spirals which are far more open (by factors of $2$ to $3$) than their linear counterparts. This effect is almost certainly relevant for the disc HD100453 and the Haro 6-13 system too.

\citet{rosotti_spiral_2020} point out that the sound speed is higher in the warmer upper layers of the disc which are probed by scattered light imaging, which may lead to spirals appearing more loosely wound; however, it's unclear whether such a prominent coherent wave mode should be expected to exist purely in the disc's upper layers. Indeed, a temperature profile which increases in the vertical direction away from the midplane causes acoustic waves in the upper layers to refract downwards (akin to sound waves in the desert after sunset). This leads to greater confinement of the f-mode near to the disc's midplane, as found by \citet{2015ApJ...814...72L}. Furthermore, \citet[figure 6]{bae_observational_2021} see no sign of a separate loosely wound acoustic wave in the upper layers of the disc in their simulations, despite having implemented a vertical temperature profile which increases with height.

\subsection{Parametric instability}

\citet{fromang_properties_2007} studied the parametric instability of axisymmetric non-linear density waves to lower-frequency inertial waves. They found a growth rate (in the low-amplitude regime) proportional to both the amplitude of the density wave and its radial wavenumber. Given the exact mathematical analogy between our non-linear spiral waves (in the inner disc) and the axisymmetric waves considered by \citet{fromang_properties_2007}, it's natural to expect an analogous parametric instability to act importantly on spiral waves, especially considering that the primary spiral arms typically have very high radial wavenumbers. 

\citet{bae_self-destructing_2016} studied this analagous instablility, referring to it as the spiral wave instability (SWI). They presented a suite of three-dimensional simulations of forced two-armed non-linear spiral waves, focusing on their inner wakes. They found the instability was robust to a variety of physical models, and gave an illuminating but heuristic theoretical analysis of the three-mode coupling which gives rise to the instability. The transformation introduced in section \ref{GDW} opens the possibility of making this theoretical analysis precise. 

Furthermore, the non-dissipation of low-amplitude spiral waves by other means signals the potential relevance of the SWI for the wave damping of such spirals. Despite its low amplitude, the highly acoustic nature of the primary spiral wave generated by a low-mass planet may mean the SWI nevertheless has a non-negligible growth rate. This regime has yet to be studied.

\subsection{Caveats}

Perhaps the most important caveat to our study is our neglect of the disc's vertical structure, alongside all dynamics in the vertical direction. In practice, the upper layers in protoplanetary discs (a few pressure scale-heights above the mid-plane) are around a factor of two warmer than the mid-plane \citep{dartois_structure_2003,law_mapping_2024}. Whilst we might expect density waves to remain coherent in such a medium (as acoustic waves in the upper layers would tend to refract downwards towards the midplane), the full extent of their dynamics is certainly not capturable with the 2D equations which we adopted. The disc's thermal structure may also be complicated by shock heating in the case of very high amplitude waves.

Our ignorance of the vertical dimension also manifests in the neglect of inherently 3D wave modes, including the inertia-gravity waves responsible for the spiral wave instability \citep{fromang_properties_2007,bae_self-destructing_2016}, as well as an unjustified treatment of the planetary potential for super-thermal planets. In this case ($M_p \gtrsim M_\text{th}$) the vertical structure of the disc is significantly modified by the presence of the planet, so that the vertical projection or average resulting in the prescription (\ref{phipresc}) is no longer valid. This is, however, only a problem in the immediate vicinity of the planet, as the prescription (\ref{phipresc}) approaches the Newtonian point-mass potential a few scale-heights from the planet.

There's a wealth of further important physics omitted from our study, which includes (but is not limited to) stellar irradiation, cooling, radiative transfer and turbulent viscosity, which affects the dissipation of spiral waves as they propagate radially (see for example \citet{2020ApJ...904..121M}), as well as magnetic fields and self-gravity, which can have noticeable effects on wave profiles, propagation and dispersion should the plasma beta or Toomre $Q$ parameters be $\lesssim10$.

Within the confines of our model, there are further caveats to mention. Firstly, the result that the inner spiral wake excited by low-mass planets $M_p \lesssim 0.01 M_\text{th}$ does not shock also depends on the specific disc model, including its aspect ratio\footnote{Indeed, in the limit $h \to 0$, the initial wave evolution occurs within an infinite shearing sheet, in which we found that dispersion is unable to overcome the strong shear amplification of waves (see for example appendix \ref{appxc} and \citet{heinemann_weakly_2012}). The inner wakes of planets in thinner discs are more susceptible to shocking.} $h$ and the profile of $\mathcal{F} = r\Sigma_0 c_0^3/(\Omega-\Omega_p)$ (as well as the adiabatic index $\gamma$). Most notably, if $\mathcal{F} \to 0$ as $r \to 0$, the level of non-linearity of a wave is amplified without limit as it propagates towards $r = 0$. In this case, unless the disc has a finite inner radius to truncate this amplification, shock formation is guaranteed. Dispersion is nevertheless still important for shock prevention if the wake reflects on the inner boundary: the amplification will be reversed for the reflected wave, and dispersion will continue to oppose wave steepening. This may be a relevant consideration for low mass planets close to an inner boundary or gap (for example planets stalled by resonant amplification of the inner Lindblad torque \citep{2011ApJ...741..109T}), in which case $\mathcal{F}$ may not vary by much between $r_\text{in}$ and $r_p$. 

Our analysis in section \ref{LAS} and appendix \ref{appxb2} of the spiral wave trains excited by low-mass planets is also ignorant of the non-linear superposition of the nose of the spiral wave train with its tail, which we simply neglected. More generally, the expected outcome of two interacting non-linear spiral waves with different spiral angles or pattern speeds is unclear, and would benefit from future attention.

We've also only analysed analytically the free wave solutions to equation (\ref{MasterEqn}). It's worth pointing out that a generalisation of equation (\ref{MasterEqn}) to include forcing must be careful in its treatment of corotation (which, akin to linear treatments, appears as a singularity).

Finally, we derived the key governing equation (\ref{MasterEqn}) assuming a smooth flow solution; however, we gave much subsequent attention to shock waves. In practice, a self-consistent solution containing shocks requires solving (\ref{MasterEqn}) piecewise away from shocks, with appropriate jump conditions imposed across shocks (so that the Rankine--Hugoniot relations are respected). The associated radial evolution of the wave action flux\footnote{which decreases secularly as the wave propagates away from its source due to dissipation in the shocks} is given in section \ref{WAES}. This provides sufficient information for a global shock wave solution in the case that the spiral angle is independent of $\theta$ (a property typical of very high amplitude forced waves). Shocks also introduce a (slow) time-evolution of $\mathcal{F}$ as well as the sound speed $c_0$ due to processes such as heating and vortensity production leading to gap-opening. These must be separately calculated, and may be manually (self-consistently) included so long as the associated timescale is indeed slow compared to the orbital timescale.

\subsection{Possible extensions}

A number of extensions may prove insightful.

Our results highlight the necessity of further numerical investigation into resonant type I migration stalling (which may provide a robust mechanism for the migration halting of low-mass planets over long timescales). Such work should extend the study by \citet{2018MNRAS.473.5267M} to very low-mass planets $M_p \lesssim 0.01 M_\text{th}$ in low-viscosity discs, carefully mitigating the effects of numerical dissipation.

Self-gravity may be included in the local equation (\ref{GNLDWEOM}) using a similar method to \citet{1985ApJ...291..356S}, and it may be possible to generalise (\ref{MasterEqn}) to include self-gravity in a global wave equation. Parts of our study may also be extendable 3D; for example considering the self-similar solutions of section \ref{S7} as functions of $\psi$ and a vertical coordinate, now solving coupled partial differential equations.

Further analysis of outward-propagating spiral waves, as well as one armed spirals in non-Keplerian discs would also complement our study, which has focused on inward-propagating spirals in (near-) Keplerian discs; and an understanding of the outcome of the non-linear superposition of two spiral waves should also be illuminating.

\section{Summary and astrophysical outlook}\label{summary}

We distilled the 2D equations of hydrodynamics into a simple yet accurate global non-linear equation for spiral waves in a thin, non-self-gravitating disc: equation (\ref{MasterEqn}). The generality is substantial, with applications to many astrophysical systems hosting spiral waves.

We found (extending work by \citet{spruit_stationary_1987} and \citet{hennebelle_spiral-driven_2016}) that exact self-similar steady 'equilibrium' smooth waves exist: spiral waves can be non-linear without necessarily shocking. This is possible up to maximum radial wave action flux, $\mathcal{B}_\text{max} = 0.022c_0^3\Sigma_0 r/\Omega$. In appendix \ref{appx1}, we showed the \emph{non-linear} equivalence of this radial wave action flux with the angular momentum flux of density waves; this result is important for understanding the amplitude evolution of non-linear waves and how they secularly shape their host discs, and has not yet been explicitly demonstrated in the astrophysical literature. There is no such \emph{non-linear} equivalence in the locally isothermal disc model, and we demonstrate the mechanism for angular momentum exchange between waves and their host discs in this model in appendix \ref{WALISO}.

In the same way that smooth non-linear waves are possible, we sketched a proof in appendix \ref{appxB22} of the result that dispersion may also prevent the inner spiral wake excited by a low-mass planet from shocking. We followed this up with very high-resolution simulations, presented in section \ref{SAS}, which indicate that the wake remains smooth for $M_p \lesssim 0.01 M_\text{th}$. The details of this result depend on the specific disc model and its aspect ratio; we expect a lower threshold for thinner discs. The possibility that the inner spiral wake excited by a low-mass planet does not rapidly dissipate, and may reflect with moderate efficiency at the inner disc edge, suggests that the resonant enhancement of the inner Lindblad torque suggested by \citet{2011ApJ...741..109T} is an important effect, and may robustly (and stably) halt type I migration.

We used equation (\ref{MasterEqn}) to find highly non-linear spiral wave solutions containing shocks, which show striking agreement to simulations with massive planets above a thermal mass, detailed in section \ref{LAS}. The radial evolution of their wave action flux due to dissipation in the shocks (calculated by \citet{2016ApJ...831..122R}), is given in terms of variables appearing in our non-linear model in section \ref{WAES}. Studying this regime revealed that high amplitude waves can be far more loosely wound than their linear counterparts, and further brought to light theoretical soft and hard caps on the amplitude of non-linear inward-propagating spirals with a given number of arms $m$. Spiral waves must contain shocks if their amplitude falls above the soft cap, and are simply impossible if their amplitude is greater than the hard cap, which is finite for $m \geqslant 3$, infinite for $m = 2$, and proportional to $m^{-4}$ for $m \gg 2$. Highly non-linear waves therefore, towards which we are observationally biased, strongly favour a two-armed, loosely wound spiral pattern. We speculate that this mechanism contributes importantly to the prevalence of two-armed spirals observed across astrophysical systems, despite many of these systems possessing no inherent twofold symmetry.

\section*{Acknowledgements}

This research was supported by the Science and Technology Facilities Council (STFC) through an STFC PhD studentship (grant number 2750631). JJB would like to thank Amelia Cordwell, Juliana Kwan, Henrik Latter and Roman Rafikov for helpful discussions. Simulations were conducted using the Swirles high performance computing facility at DAMTP, University of Cambridge.

\section*{Data availability}

The data underlying this article will be shared on reasonable request to the corresponding author.

\bibliographystyle{mnras}
\bibliography{bibmaster}

\appendix

\section{On wave action and angular momentum}\label{appx}

\renewcommand{\theequation}{A\arabic{equation}}
\setcounter{equation}{0}

Whilst the total angular momentum of the disc (and its contents) about the central star is conserved, this constraint does not exclude the possibility that, nor prescribe when, waves may exchange angular momentum with the mean circular orbital flow. Indeed, non-linear waves are known in general to `deposit' angular momentum into their parent discs, leading to important morphological evolution of the disc including gap opening \citep{goldreich_disk-satellite_1980,1984ApJ...285..818P,rafikov_planet_2002,cordwell_early_2024}. A non-linear measure of wave activity, the `wave action', alongside an appropriate conservation law is therefore necessary to disentangle and describe the waves and their evolution. 

In this appendix, motivated in part by the striking numerical agreement found in section \ref{ctss} between the expressions (\ref{AMFint}) and (\ref{WAint}), we aim to verify (non-linearly) the correspondence between the radial wave action flux and the angular momentum flux of waves in circular accretion discs in the case where no dissipation is present. We will further express these quantities using the notation of the paper in the section \ref{appx2}, which will aid in the verification of the global non-linear theory developed in section \ref{GDW}. Whilst the connection is arguably well known, and well documented for linear waves (e.g. \citet{1979ApJ...233..857G,rafikov_nonlinear_2002}), treatments which include the non-linear case appear to be sparsely discussed in the astrophysical literature. The connection is subtle\footnote{and at its heart due to the rotational symmetry of the mean orbital flow}: indeed the azimuthal fluxes of wave action and angular momentum do not coincide even for smooth waves. 

One of the difficulties faced by non-linear analyses is in distinguishing exactly the waves present in a given flow from an underlying mean flow. For this reason we'll introduce and use the Generalised Lagrangian Mean (GLM) framework of \citet{andrews_exact_1978,andrews_wave-action_1978}, as in our context it allows for this wave-mean flow separation to be carried out precisely. An excellent introduction to GLM theory is also given in \citet[chapter 10]{buhler_waves_2014}.

\subsection{The angular momentum-wave action flux equivalence}\label{appx1}

It's worth briefly introducing the GLM formalism and notation required to describe our non-linear waves and their activity, which we'll interpret in the context of this paper. The key idea of GLM theory is to define a Lagrangian-mean operator which averages over fluid elements (or an ensemble of possibilities available to a given fluid element), rather than an Eulerian coordinate-based average. One important advantage of this approach is the resulting simplicity of the material derivative, which may be exploited in the derivation of an appropriate wave action equation.

We introduce the displacement field $\bmath{\xi}$ via
\begin{equation}
    \vecr = \vecr_0 + \bmath{\xi}(r_0,\theta_0,t),
\end{equation}
where $\vecr_0$ denotes the mean position (which lies on a mean trajectory) of a fluid element whose actual position is $\vecr$. We further assume all fields depend differentiably on a phase parameter $\alpha$, which we imagine as labelling the different points along the wave's profile (representing the various possibilities within some phase space available to a given fluid element). In practice we'll take $\alpha$ to be $r_0$ or $\theta_0$. Following \citet{andrews_exact_1978}, we introduce the phase average $\overline{\left(\cdots\right)}$ over $\alpha$. Now, as $\overline{\left(\cdots\right)}$ is an average which tracks fluid elements, the material derivative of any field $\varphi$ has the simple Lagrangian formulation\footnote{For a proof of this, see \citet{andrews_exact_1978}.}
\begin{equation}
    \left(\frac{\DD \varphi}{\DD t}\right)^{\bmath{\xi}} = \bar{\DD}^L \varphi^{\bmath{\xi}},
\end{equation}
and $\bmath{\xi}$ inherits the important properties
\begin{equation}
    \bar{\bmath{\xi}} = \bmath{0}, \quad \bar{\DD}^L\bmath{\xi} = \vecu^l \equiv \vecu^{\bmath{\xi}} - \bar{\vecu}^L,
\end{equation}
where $\bar{\DD}^L = \p_t + \bar{\vecu}^L\cdot \nabla$ is the Lagrangian-mean material derivative, $\vecu^{\bmath{\xi}}$ is the fluid velocity at $\vecr = \vecr_0 + \bmath{\xi}$, and $\bar{\vecu}^L$ and $\bar{\vecu}^l$ are the mean and perturbed velocity fields.

In general, we adopt the notation
\begin{equation}
    \bar{\varphi}^L \equiv \overline{\varphi^{\bmath{\xi}}} \equiv \overline{\varphi\left(\vecr_0+\bmath{\xi},t\right)}, \quad \varphi^l \equiv \varphi^{\bmath{\xi}} - \bar{\varphi}^L
\end{equation}
for any field $\varphi$, and additionally we denote derivatives $\p/\p x_i$ and $\p/\p\alpha$ by $(\cdots)_{,i} = $ and $(\cdots)_{,\alpha}$. Here, $\p/\p x_i$ represents a derivative taken relative to the Lagrangian-mean coordinate system (e.g. $(r_0,\theta_0)$). Since mean values are independent of $\alpha$, it follows that $\overline{\varphi^{l}} = 0$.

We now have the machinery necessary to demonstrate the equivalence of the radial angular momentum flux and wave action flux in 2D accretion discs at our disposal. Before we turn it on, we should briefly discuss in a mathematical sense what we're trying to prove. Specifically, we'll show that it's possible to find an $\alpha$ such that the pressure-associated radial angular momentum flux density expressed in the Lagrangian coordinate system $(r_0,\theta_0)$ (and averaged over nearby fluid elements) is equal to the pressure-related wave action flux density\footnote{Since the Lagrangian coordinates (often) trace flow streamlines, the advective flux of angular momentum naturally vanishes in this perspective. The equivalence to the more familiar Eulerian advective flux of angular momentum through a radial boundary follows from the identification of this flux with the pressure-communicated angular momentum flux across a nearby streamline. This is most easily done if the flow is smooth and steady, and streamlines are closed.}:
\begin{equation}
    \hat{z}_i\varepsilon_{ijk}\overline{r_jP^{\bmath{\xi}}K_{kr_0}} = \overline{P^{\bmath{\xi}}\xi_{j,\alpha}K_{j r_0}}.
\end{equation}
The slightly obscure form of the left hand side of the above equation, representing the angular momentum flux, is a consequence of the coordinate change from Eulerian to Lagrangian coordinates (this introduces the tensor $\bmath{K}$, defined below, related to the inverse of the Jacobian of the map $\vecr_0 \to \vecr$). It's also important to note that the $\alpha$--derivative appearing in the expression for the wave action flux is partial, not covariant. In the case of no net mass flux, $\bar{u}^L_r = 0$, as is the case in the non-dissipative system assumed here, it then follows that total (azimuthally integrated) radial fluxes of angular momentum and wave action are equal also.

We begin with the angular momentum equation in Eulerian coordinates (e.g. equation (\ref{AME})):
\begin{equation}
    \frac{\DD}{\DD t}\left(\bmath{r} \times \vecu\right) + \frac{1}{\Sigma}\nabla \cdot \left(\bmath{r} \times P \bmath{I}\right) = \bmath{0}.
\end{equation}
We take the Lagrangian mean (that is, average over fluid elements), and express the result in the mean flow Lagrangian coordinate system\footnote{\citet[section 10.2.6]{buhler_waves_2014} provides an excellent account of how to carry out this transformation.},
\begin{equation}
    \bar{\DD}^L\varepsilon_{ijk}\left({r_0}_j\bar{u}_k^L + \overline{\xi_ju_k^l}\right)+ \frac{1}{\bar{\Sigma}^{L}}\left(\varepsilon_{ijk}\overline{r_jP^{\bmath{\xi}}K_{km}}\right)_{,m} = 0,
\end{equation}
where $\bmath{K}$ is the matrix of cofactors of the Jacobian of the map $\vecr_0 \to \vecr_0 + \bmath{\xi}$. It holds that
\begin{equation}\label{Kdef}
    \left(\delta_{jk} + \xi_{j,k}\right)K_{ji} = J\delta_{ik},
\end{equation}
where $J$ is the Jacobian determinant. The mean angular momentum flux density is therefore
\begin{equation}
    \mathcal{F}_i = \hat{z}_j \varepsilon_{jkm}\overline{r_k P^{\bmath{\xi}}K_{mi}},
\end{equation}
which, writing $r_k = {r_0}_k + \xi_k$ gives
\begin{equation}\label{AMFistep}
    \mathcal{F}_i = \hat{z}_j \varepsilon_{jkm}\overline{\xi_kP^{\bmath{\xi}}K_{mi}} + \hat{z}_j \varepsilon_{jkm}{r_0}_k\overline{P^{\bmath{\xi}}K_{mi}}.
\end{equation}
We now use equation (\ref{Kdef}) to write $K_{mi} = - \xi_{n,m}K_{ni} + J\delta_{mi}$, which when substituted into the second term in (\ref{AMFistep}) yields
\begin{multline}\label{AMFistep2}
    \mathcal{F}_i = \hat{z}_j \varepsilon_{jkm}\left[\overline{\xi_kP^{\bmath{\xi}}K_{mi}} - {r_0}_k\overline{\xi_{n,m}P^{\bmath{\xi}}K_{ni}}\right] \\+ \varepsilon_{ijk}\hat{z}_j {r_0}_k\overline{P^{\bmath{\xi}}J},
\end{multline}
which may be recognised as\footnote{It's useful to observe that the \emph{partial} derivative with respect to $-\theta_0$ of the vector $\bmath{\xi}$ written in coordinate-free notation is $\bmath{\xi}_{,-\theta_0} \equiv -\frac{\p\xi_{r_0}}{\p \theta_0}\vece_{r_0} -\frac{\p\xi_{\theta_0}}{\p \theta_0}\vece_{\theta_0} = - \hat{\bmath{z}}\cdot\left(\vecr_0 \times \nabla\right)\bmath{\xi} + \hat{\bmath{z}}\times\bmath{\xi}$.}
\begin{equation}
    \mathcal{F}_i = \overline{P^{\bmath{\xi}}\xi_{j,-\theta_0}K_{ji}} + \bar{\Sigma}^{L}\varepsilon_{ijk}\hat{z}_j {r_0}_k\overline{P^{\bmath{\xi}}/\Sigma^{\bmath{\xi}}}.
\end{equation}
Now, the first term we recognise as the wave action flux density when $\alpha = -\theta_0$ is identified\footnote{\citet{andrews_wave-action_1978} point out that the wave action equation also becomes the angular pseudo-momentum equation when the identification $\alpha = -\theta_0$ is made, though this differs from the mean angular momentum equation considered here.}, and the second is a purely azimuthal contribution to the angular momentum flux density. This then is precisely what we wanted to show: that the radial wave action flux is equal to the radial angular momentum flux, averaged over fluid elements! The same is not true for the azimuthal fluxes. This occurs as a consequence of the more general principle that the off-diagonal terms of the (pressure-related) pseudo-momentum flux tensor and the mean-momentum flux tensor agree (see for example \citet[section 10.3]{buhler_waves_2014}).

The flux of angular momentum through a radial boundary in the Eulerian $(r,\theta)$ system is given by an azimuthal integral of $r\Sigma u v$: the pressure-communicated flux is entirely azimuthal. In the case where the same flux penetrates a nearby streamline (now communicated by pressure forces), on account of the above analysis we may directly equate
\begin{equation}
    \int r^2 \Sigma u v \dd \theta = -\int r_0 P^{\bmath{\xi}} \bmath{\xi}_{,\theta_0}\cdot \bmath{K}\cdot \vece_{r_0} \dd\theta_0.
\end{equation}
Consequently, the radial angular momentum flux of smooth, steady waves in a conservative system\footnotemark, which expressed in the standard Eulerian notation is $\int r^2 \Sigma u v \dd \theta$, is exactly (non-linearly) conserved! This then is the reason for the numerical agreement we saw between the expressions (\ref{AMFint}) and (\ref{WAint}) found in section \ref{ctss}. It remains however to evaluate the latter integral in terms of quantities known to us through the model (\ref{GNLDWEOM}). We do this in the next section.

\subsection{The non-linear wave action equation for spiral waves}\label{appx2}

\citet{andrews_wave-action_1978} showed that the wave action of non-linear waves in a conservative system\footnotemark[\value{footnote}]\footnotetext{This notably excludes the `locally isothermal' disc model, which does not conserve entropy materially. See section \ref{WALISO} for a more detailed discussion.} obeys the conservation law
\begin{equation}\label{AMWA}
    \bar{\Sigma}^{L}\bar{\DD}^L\mathcal{A} + \nabla\cdot\bmath{\mathcal{B}} = 0,
\end{equation}
where $\bar{\Sigma}^L = \Sigma_0(r_0)$ is the mean surface density in our disc, and expressions for the wave action density $\mathcal{A}$ and flux density $\bmath{\mathcal{B}}$ are given in equations (\ref{WAdef}) and (\ref{WAFdef}). In this and the subsequent appendix \ref{appxb1}, we aim to verify that the model (\ref{MasterEqn}) with Lagrangian (\ref{Lagrangian}) is sufficient to reproduce (globally) each term this equation in the case of smooth non-linear spiral waves in a thin disc. In this way, equation (\ref{MasterEqn}) predicts accurately the wave action, phase and local shape of spiral waves globally, and therefore provides the leading asymptotic description of these waves for thin discs.

Equation (\ref{AMWA}) applies generally to non-linear waves solving the 2D Euler equations; our strategy will be to simplify each term with the knowledge that the waves obey equation (\ref{GNLDWEOM}) locally. The Lagrangian-mean state in our non-dissipative system is closely related to the axisymmetric reference disc described in section \ref{GDW} (which we allowed waves to perturb). It has surface density $\bar{\Sigma}^{L} = \Sigma_0(r_0)$ and mean-flow velocity\footnote{This may be verified for $\left|\bmath{\xi}\right| \ll r$ by directly averaging equation (\ref{lagpertvel}), making use of equation (\ref{azivel}) and the relation $u = \bar{\DD}^L \xi_r$ (see the footnote on page \pageref{MatDer}).} $\bar{\vecu}^L = r_0(\Omega(r_0)-\Omega_p)\vece_{\theta_0}$.

We'd like first to show that the expression for the wave action density may be simplified to $\mathcal{A} = \overline{u\xi_{r,\alpha}}$, where $u$ is the (Eulerian) radial velocity.

We take the expression \citep[equation (2.7a)]{andrews_wave-action_1978}
\begin{equation}\label{WAdef}
    \mathcal{A} = \overline{{\bmath{\xi}}_{,\alpha}\cdot({\bf{u}}^l + {\bmath{\Omega}}_p \times {\bmath{\xi}})},
\end{equation}
and begin by examining $\vecu^l = \vecu^{\bmath{\xi}} - \bar{\vecu}^L$:
\begin{equation}
    \vecu^l = u \vece_r + \left[r\left(\Omega(r)-\Omega_p\right) + v\right]\vece_\theta - r_0\left(\Omega(r_0) - \Omega_p\right)\vece_{\theta_0},
\end{equation}
which, observing that $\vece_\theta\cdot\vece_{r_0} \approx - \xi_\theta/r$, may be approximated at leading order as
\begin{multline}\label{lagpertvel}
    \vecu^l \approx \left[u - (\Omega-\Omega_p)\xi_\theta\right] \vece_{r_0} \\+ \left[r\left(\Omega(r)-\Omega_p\right) - r_0\left(\Omega(r_0) - \Omega_p\right) + v\right]\vece_{\theta_0}.
\end{multline}
Now, specific angular momentum conservation, expressed in the text in equation (\ref{SAMcons}), may be written as
\begin{equation}\label{samcons}
    r(v + r\Omega(r)) = r_0^2\Omega(r_0),
\end{equation}
so that, correct at our desired order, 
\begin{align}
    \vecu^l \cdot \vece_{\theta_0}&= r_0\Omega_0(r_0)\left(\frac{r_0}{r}-1\right) - \xi_r\Omega_p \\ &\approx - \left(\Omega + \Omega_p\right)\xi_r.\label{azivel}
\end{align}
Substituting into equation (\ref{WAdef}) gives
\begin{multline}
    \mathcal{A} = \overline{\xi_{r,\alpha}(u - \Omega \xi_\theta) + \xi_{\theta,\alpha}(-\Omega\xi_r)} \\= \overline{\xi_{r,\alpha}u} - \overline{\Omega\left(\xi_r\xi_\theta\right)_{,\alpha}} = \overline{\xi_{r,\alpha}u},
\end{multline}
as desired. We similarly manipulate $\bmath{\mathcal{B}}$ into a tractable form in terms of variables appearing in the reduced model of section \ref{GDW}. We again begin with the expression from \citet[equation (2.7b)]{andrews_wave-action_1978}:
\begin{equation}\label{WAFdef}
    \mathcal{B}_j = \overline{P^{\bmath{\xi}}\xi_{i,\alpha}K_{ij}}.
\end{equation}
The cofactor tensor $\bmath{K}$ expressed in the $(r_0,\theta_0)$ coordinate system reads
\begin{equation}
    \bmath{K} = \begin{pmatrix} 1 & -\frac{\p \xi_\theta}{\p r_0}\\ 0 & 1 + \frac{\p \xi_r}{\p r_0}\end{pmatrix} + \cdots.
\end{equation}
It follows that
\begin{equation}\label{Bgen}
    \bmath{\mathcal{B}} = \overline{P^{\bmath{\xi}}\xi_{r,\alpha}}\vece_{r_0} + \overline{P^{\bmath{\xi}}\left(-\xi_{\theta,r_0}\xi_{r,\alpha} + (1+\xi_{r,r_0})\xi_{\theta,\alpha}\right)}\vece_{\theta_0}.
\end{equation}
Choosing again $\alpha = -\theta_0 \approx -\theta$, the total radial wave action flux through a given radius $r_0$ is therefore
\begin{multline}\label{Bexpr}
    \mathcal{B} = -r_0\int_{0}^{2\upi}P\xi_{r,\theta}\dd \theta \\= -\frac{r_0 \Sigma_0 c_0^3}{\Omega_0 - \Omega_p}\int_{0}^{2\upi}\frac{\eta_\theta}{\gamma}\left(\frac{1}{(1+\eta_{X_0})^\gamma} - 1\right)\dd \theta,
\end{multline}
which now concludes (in combination with appendix \ref{appx1}) the analytical verification of the equivalence of the expressions (\ref{AMFint}) and (\ref{WAint}).

If we further consider spiral waves described locally as a function of the phase $\phi = X_0 - \mathcal{U} \theta$ (for slowly varying phase speed $\mathcal{U}$, and $\theta \approx \theta_0$), the terms $-\xi_{\theta,r_0}\xi_{r,\alpha} + \xi_{r,r_0}\xi_{\theta,\alpha}$ sum to zero, and the expression for $\bmath{\mathcal{B}}$ simplifies to
\begin{equation}
    \bmath{\mathcal{B}} = \overline{P^{\bmath{\xi}}\xi_{r,-\theta_0}}\vece_{r_0} + \overline{P^{\bmath{\xi}}\xi_{\theta,-\theta_0}}\vece_{\theta_0}.
\end{equation}

Remarkably, it turns out that for tightly wound density waves, the azimuthal component of $\bmath{\mathcal{B}}$ vanishes at leading order. Whilst slightly fiddly to prove, this fact is important as it demonstrates that the azimuthal pressure gradient which we'd neglected in the derivation of the wave equation (\ref{GNLDWEOM}) also has no contribution to the wave action equation, and so in a sense is uniformly unimportant. Firstly, note that (recall equation (\ref{azivel}))
\begin{equation}\label{xiazi}
    \frac{\DD \bmath{{\xi}}}{\DD t} \cdot \vece_{\theta_0} = (\Omega - \Omega_p)\left(\xi_{\theta,\theta} + \xi_r\right) = -(\Omega + \Omega_p)\xi_r
\end{equation}
\begin{equation}
    \implies (\Omega - \Omega_p)\xi_{\theta,\theta} = -2\Omega\xi_r,
\end{equation}
a useful equation in its own right, allowing for the calculation of the azimuthal displacement from a solution to equation (\ref{MasterEqn}). Again considering waves which are functions of the phase $\phi = X_0 - \mathcal{U} \theta$, and approximating $\p_{-\theta_0} \approx -\p_\theta\big|_{r_0}$, it follows that
\begin{equation}\label{aziPint}
    \overline{P^{\bmath{\xi}}\xi_{\theta,-\theta_0}} \propto I = \int_0^{\lambda}\frac{\eta}{\gamma}\left(\frac{1}{(1+\eta_{\phi})^\gamma}-1\right)\dd \phi,
\end{equation}
for wavelength $\lambda$. We use the leading order spiral wave equation (\ref{Spiraleqn}) to write $\eta$ in terms of $\eta_{\phi}$ and $\eta_{\phi\phi}$. It's then possible to write the integrand in the above equation as an exact derivative:
\begin{multline}\label{Iint}
    I = \frac{1}{\tilde{\kappa}_0^2}\int_0^{\lambda}\p_\phi\Bigg[\frac{\mathcal{U}^2}{\gamma}\left(\frac{1}{(\gamma-1)(1+\eta_\phi)^{\gamma-1}} + \eta_\phi\right) \\- \frac{1}{2\gamma^2}\left(\frac{1}{(1+\eta_{\phi})^\gamma}-1\right)^2\Bigg]\dd \phi = 0.
\end{multline}
The integral therefore vanishes in the case of smooth solutions. We can think of this as a result of the pressure and displacement being out of phase in a non-linear sense. That is to say, the azimuthal pressure gradient which we found to be subdominant for the local wave behaviour also does not contribute to the global wave action equation: its small local contributions don't accumulate into a noticeable effect on a global scale!

The (GLM) wave action equation for a steady tightly wound density wave becomes 
\begin{equation}\label{GLMWAsimp}
    \frac{\p}{\p \theta}\left(\Sigma_0(\Omega-\Omega_p)r_0\overline{\xi_{r,\theta}u}\right) + \frac{\p}{\p r_0}\left(r_0\overline{P^{\bmath{\xi}}\xi_{r,\theta}}\right)=0.
\end{equation}
In appendix \ref{appxb} below, we'll verify that equation (\ref{GLMWAsimp}) naturally arises from the reduced model (\ref{MasterEqn}). That is, to faithfully model global spiral waves in discs, we only need knowledge of the leading order wave equation as well as the background profile of the disc. This information is also captured in the Lagrangian density in equation (\ref{Lagrangian}), which is uniformly the leading approximation to the Lagrangian density of 2D hydrodynamics in the case of tightly wound smooth spiral waves (away from shocks).

\subsubsection{Wave action conservation in locally isothermal discs}\label{WALISO}

An instructive and illuminating, though tangential, use of the formalism just introduced is found in considering the wave action equation in the `locally isothermal' approximation. The following analysis highlights the spurious origins of the unusual physical behaviour exhibited by waves in the approximation, whose angular momentum flux is not conserved. In this thermodynamic prescription, the sound speed is a fixed function of (Eulerian) radius, $c_s(r)$, and the density and pressure are related via $P = c_s^2(r)\Sigma$. \citet{2016ApJ...832..166L} and \citet{2020ApJ...892...65M} found that for linear waves it is the ratio of the angular momentum flux and squared sound speed which is conserved.

In the absence of external and dissipative forces, the wave action equation reads \citep{andrews_wave-action_1978}:
\begin{equation}\label{WAiso}
    \Sigma_0\bar{\DD}^L\mathcal{A} + \nabla\cdot\bmath{\mathcal{B}} = - \Sigma_0\overline{(P^{\bmath{\xi}})_{,\alpha}/\Sigma^{\bmath{\xi}}},
\end{equation}
and the term on the right hand side of the equation vanishes when $P = P(r_0,\Sigma)$. However, instead we have $P(r^{\bmath{\xi}},\Sigma) = c_s^2(r)\Sigma$, and we may not simply directly integrate the term to give zero.

We may gain some further insight into the behaviour of the forcing on the right hand side by considering the linear approximation. Namely,
\begin{align}
    &\overline{(P^{\bmath{\xi}})_{,\alpha}/\Sigma^{\bmath{\xi}}} = \overline{(c_s^2(r)\Sigma^{\bmath{\xi}})_{,\alpha}/\Sigma^{\bmath{\xi}}} \nonumber \\
    = \;&\overline{(c_s^2(r))_{,\alpha}} + \overline{c_s^2(r)\left(\ln\left(\Sigma^{\bmath{\xi}}/\Sigma_0\right)\right)_{,\alpha}} = \overline{c_s^2(r)\left(\ln\left(\Sigma^{\bmath{\xi}}/\Sigma_0\right)\right)_{,\alpha}} \nonumber \\
    \approx \;&\overline{(c_s^2(r_0) + (c_s^2)'\cdot \xi_r)(-\xi_{r,r})_{,\alpha}} \approx -(c_s^2)'\overline{\xi_r\xi_{r,r\alpha}} \nonumber \\
    = \; &(c_s^2)'\overline{\xi_{r,r}\xi_{r,\alpha}}.
\end{align}
That is, the wave exerts a torque on the background disc proportional to its radial action flux density and the gradient of the sound speed, since (using equation (\ref{Bgen})) we see
\begin{align}
    \mathcal{B}_r = & \;\overline{P^{\bmath{\xi}}\xi_{r,\alpha}} = \overline{c_s^2(r)\Sigma^{\bmath{\xi}}\xi_{r,\alpha}} \nonumber\\
    \approx & \;\Sigma_0\overline{(c_s^2(r_0) + (c_s^2)'\cdot \xi_r)(1 - \xi_{r,r})\xi_{r,\alpha}} \nonumber \\
    \approx & -\Sigma_0 c_s^2 \overline{\xi_{r,r}\xi_{r,\alpha}},
\end{align}
as $\overline{\xi_r\xi_{r,\alpha}}=0$. Azimuthally integrating equation (\ref{WAiso}) and using $\mathcal{B} = r_0\int\mathcal{B}_r\dd \theta$, we deduce that linearly
\begin{equation}
    \frac{1}{r_0}\p_{r_0}(\mathcal{B}) = \frac{\p_{r_0}(c_s^2)}{c_s^2}\frac{\mathcal{B}}{r_0}
\end{equation}
\begin{equation}\label{BLIcons}
    \implies\p_{r_0}(\mathcal{B}/c_s^2) = 0.
\end{equation}
That is, we recover the result found by \citet{2016ApJ...832..166L} (and demonstrated explicitly by \citet{2020ApJ...892...65M}) when the total radial wave action flux $\mathcal{B}$ is identified with the angular momentum flux. This analysis highlights in particular the direct coupling between the wave and the mean orbital flow facilitated by the locally isothermal prescription. In imposing that the pressure and surface density be related by a function of Eulerian radius, rather than intrinsic Lagrangian coordinate, the wave ends up doing net work on the mean flow over the course of an oscillation as it constantly adjusts to the prescription.

This is markedly different to the behaviour of an adiabatic prescription with $\gamma=1$, in which the temperature of a given fluid element is fixed,
\begin{equation}
    \frac{\DD}{\DD t}\left(\frac{P}{\Sigma}\right) = 0 \implies P^{\bmath{\xi}} = \frac{P_0(r_0)}{\Sigma_0(r_0)}\Sigma^{\bmath{\xi}} = c_s^2(r_0)\Sigma^{\bmath{\xi}},
\end{equation}
with the consequence that smooth waves will conserve their radial wave action (and angular momentum) flux non-linearly.

Finally, it's worth mentioning on account of the above calculation that we only expect the conservation law (\ref{BLIcons}) to hold linearly, in contrast to the exact non-linear conservation of the wave action and angular momentum flux of smooth waves in adiabatic discs.

\section{Whitham modulation equations}\label{appxb}

\renewcommand{\theequation}{B\arabic{equation}}
\setcounter{equation}{0}

In this appendix we verify that the wave action equation arising directly from the Lagrangian formulation (\ref{Lagrangian}) is equal to that arising from the full 2D system, namely equation (\ref{GLMWAsimp}). To derive it, we employ the toolkit developed by \citet[chapter 14]{whitham_linear_1974}. We'll then combine this wave action equation with conservation of waves (equation \ref{CW}) to derive the canonical modulation equations governing the distribution of the wave packet's amplitude and its local pitch angle. It should be noted that the modulation equations are equally applicable to the axisymmetric travelling waves considered by \citet{fromang_properties_2007}, under the re-identification of $\theta$ with time and $X_0$ with radius.

\subsection{Whitham wave action equation}\label{appxb1}

An elegant justification of the variational method used below is given in \citet[chapter 14.4]{whitham_linear_1974}. We first suppose the displacement $\eta(\psi,\Theta, \mathcal{X}; \varepsilon)$ to be a function of the phase $\psi = \varepsilon^{-1}\Psi(\Theta,\mathcal{X})$ as well as slow coordinates $\Theta = \varepsilon \theta$, $\mathcal{X} = \varepsilon X_0$, where $\varepsilon$ is a small parameter. We'll assume that the background state varies on long length-scales, with $\mathcal{X}$, and that $\eta$ is $2\upi$-periodic in $\psi$.

We define the azimuthal and radial wavenumbers via
\begin{equation}
    \quad m(\Theta,\mathcal{X}) = \Psi_\Theta = \psi_\theta, \quad k(\Theta,\mathcal{X}) = \Psi_{\mathcal{X}} = \psi_{X_0}.
\end{equation}
In this way, the phase $\psi$ locally resembles `$m\theta + k X_0$', and we also obtain the identity, which we'll call `conservation of waves',
\begin{equation}\label{CW}
    \frac{\p k}{\p \Theta} = \frac{\p m}{\p \mathcal{X}}.
\end{equation}
The Lagrangian density appearing in equation (\ref{Lagrangian}) now depends on the displacement and its derivatives as $\mathcal{L}\left(\eta_\theta \equiv m\eta_\psi + \varepsilon\eta_\Theta,\eta_{X_0} \equiv k\eta_\psi + \varepsilon\eta_\mathcal{X}, \eta, \mathcal{X}\right)$, which at 0th order in $\varepsilon$ reads
\begin{multline}
    \mathcal{L} =\mathcal{F}\bigg\{\frac{1}{2}m^2\eta_\psi^2 -\frac{1}{2}\tilde{\kappa}_0^2 \eta^2 \\- \frac{1}{\gamma(\gamma-1)}\left[\frac{1}{(1+k\eta_\psi)^{\gamma-1}}-1\right] - \frac{k\eta_\psi}{\gamma}\bigg\}.
\end{multline}
Whitham showed that, remarkably, the non-linear wave action equation only involves terms from this 0th order (in $\varepsilon$) approximation to be the system.\footnote{This holds even when the Lagrangian density is allowed to depend explicitly on $\mathcal{X}$. Note that equation (14.42) in \citet{whitham_linear_1974} remains true in this case.} The Hamiltonian $\mathcal{H}(\Theta,\mathcal{X}) = \mu \eta_\psi - \mathcal{L}$ (which only varies over long length-scales) in this 0th order system is
\begin{multline}
    \mathcal{H}(\Theta,\mathcal{X}) = \mathcal{F}\bigg\{\frac{1}{2}m^2\eta_\psi^2 + \frac{1}{2}\tilde{\kappa}_0^2 \eta^2 \\+ \frac{1}{\gamma(\gamma-1)}\left[\frac{1+\gamma k \eta_\psi}{(1 + k \eta_\psi)^\gamma} - 1\right]\bigg\},
\end{multline}
and the canonical momentum $\mu$ is given by
\begin{equation}
    \mu = \frac{\p \mathcal{L}}{\p \eta_\psi} = \mathcal{F}\left\{m^2\eta_\psi + \frac{k}{\gamma}\left[\frac{1}{(1 + k\eta_\psi)^\gamma} - 1\right]\right\}.
\end{equation}
We define the averaged Lagrangian $\bar{\mathcal{L}}$ to be the mean value of $\mathcal{L}$ (at 0th order in $\varepsilon$) over one period, neglecting any slow variations in $k$, $n$, $\mathcal{H}$ or the background state. It follows that
\begin{equation}\label{avgLag}
    \bar{\mathcal{L}} = \frac{1}{2\upi}\int_0^{2\upi} \mathcal{L} \dd \psi = \frac{1}{2\upi}\int_0^{2\upi} \mu \eta_\psi - \mathcal{H} \dd \psi = \frac{k\mathcal{J}}{2\pi} - \mathcal{H},
\end{equation}
where the action $\mathcal{J}$ (whose utility will become apparent later) is
\begin{equation}
    \mathcal{J} = \oint \frac{\mu}{k} \dd \eta = \frac{\mathcal{F}(\mathcal{X})}{\tilde{\kappa}_0(\mathcal{X})}\tilde{\mathcal{J}}\left(\mathcal{U},\tilde{\mathcal{H}}\right).
\end{equation}
The contour integral is taken clockwise in $\eta \mu$-space. It may be seen that the `reduced action' $\tilde{\mathcal{J}}$ depends only on the phase speed $\mathcal{U}= -m/k$ and reduced Hamiltonian $\tilde{\mathcal{H}} = \mathcal{H}/\mathcal{F}$, by writing directly
\begin{multline}\label{Jexpr}
    \mathcal{J} = -\oint\frac{\eta}{k}\dd \mu = -\oint\frac{\eta}{k}\frac{\dd \mu}{\dd (k \eta_\psi)}\dd (k\eta_\psi) =\\ 2\frac{\mathcal{F}}{\tilde{\kappa}_0}\int_{p_\text{min}}^{p_\text{max}} \hspace{-2mm}\sqrt{2\big(\tilde{\mathcal{H}} - \mathcal{V}(p;\mathcal{U})\big)}\left[\mathcal{U}^2 - \frac{1}{(1 + p)^{\gamma+1}}\right]\dd p.
\end{multline}
Here $p = \eta_\phi = k\eta_\psi$, and
\begin{equation}
    \mathcal{V}(p;\mathcal{U}) = \frac{1}{2}\mathcal{U}^2p^2 + \frac{1}{\gamma(\gamma - 1)}\left(\frac{1 + \gamma p}{\left(1 + p\right)^\gamma}-1\right).
\end{equation}
$p_\text{min}$ and $p_\text{max}$ solve $\mathcal{V}(p_{\text{min/max}};\mathcal{U}) = \tilde{\mathcal{H}}$, and so are also themselves functions of $\tilde{\mathcal{H}}$ and $\mathcal{U}$. Note that our definitions of $\tilde{\mathcal{H}}$ and $\mathcal{U}$ in this appendix coincide with the definitions introduced in section \ref{SFPA}.

Wave action conservation may be expressed as
\begin{equation}\label{WhithamWA}
    \frac{\p}{\p \Theta}\left(\bar{\mathcal{L}}_m\right) + \frac{\p}{\p \mathcal{X}}\left(\bar{\mathcal{L}}_k\right) = 0,
\end{equation}
where the partial derivatives with respect to $m$ and $k$ are taken at constant $\Theta$ and $\mathcal{X}$. More explicitly,
\begin{multline}\label{WhithamWAintform}
    \frac{\p}{\p \Theta}\left[\frac{\mathcal{F}}{2\upi}\int_0^{2\upi}m \eta_\psi^2\dd\psi\right] +\\ \frac{\p}{\p \mathcal{X}}\left[\frac{\mathcal{F}}{2\upi}\int_0^{2\upi}\frac{\eta_\psi}{\gamma}\left(\frac{1}{(1+k\eta_\psi)^\gamma}-1\right)\dd\psi\right] = 0.
\end{multline}
Substituting the relations $(\Omega-\Omega_p)\xi_{r,\theta} = u = c_0 \eta_\theta = c_0 m \eta_\psi$, $P^{\bmath{\xi}} = \frac{\Sigma_0 c_0^2}{\gamma}\frac{1}{(1+k\eta_\psi)^\gamma}$, $\p_{r_0} = \frac{\Omega-\Omega_p}{c_0}\p_{X_0}$ and $\Theta = \varepsilon\theta$, $\mathcal{X} = \varepsilon X_0$, and re-expressing the average over $\theta_0$ as one over $\psi$, we see that equation (\ref{GLMWAsimp}) exactly becomes equation (\ref{WhithamWAintform}). That is to say, the simple Lagrangian (\ref{Lagrangian}) faithfully captures the global evolution of 2D spiral waves.

\subsection{Modulation equations and possibility of no shock formation}\label{appxb2}

We now show how conservation of waves and the wave action equation may be used to determine the evolution of the slowly varying quantities $\mathcal{H}$ and $\mathcal{U}$ in the case of smooth waves. Equation (\ref{avgLag}) allows us to express the wave action equation (\ref{WhithamWA}) in terms of the action $\mathcal{J}$ and phase speed $\mathcal{U}$ as
\begin{equation}\label{WWAJ}
    \frac{\p}{\p \Theta}\left(\mathcal{J}_\mathcal{U}\right) + \frac{\p}{\p \mathcal{X}}\left(\mathcal{U} \mathcal{J}_\mathcal{U} - \mathcal{J}\right) = 0,
\end{equation}
where the partial derivative with respect to $\mathcal{U}$, $\mathcal{J}_\mathcal{U}$, is taken at constant $\mathcal{X}$ and $\mathcal{H}$. This form of the wave action equation is particularly useful as it no longer depends on any of the small-scale wave dynamics, only the slowly varying functions $\mathcal{U}$ and $\mathcal{H}$, as well as $\mathcal{X}$. An explicit expression for the function $\mathcal{J} = (\mathcal{F}/\tilde{\kappa}_0)\tilde{\mathcal{J}}(\mathcal{U},\tilde{\mathcal{H}})$ is given in equation (\ref{Jexpr}).

Combining equation (\ref{WWAJ}) with equation (\ref{CW}) yields nearly enough information to be able to solve for the spatial distributions of $\mathcal{H}$ and $\mathcal{U}$, which would determine a global wave train solution (up to a reference phase value). A final relation between the variables is necessary to close the system. Linearly, this would be the usual dispersion relation. The desired non-linear extension is a relation recognisable from classical dynamics:
\begin{equation}
    k = 2\upi \frac{\p \mathcal{H}}{\p\mathcal{J}}\bigg|_{\mathcal{U},\mathcal{X}},
\end{equation}
which may be directly verified in our case by observing
\begin{multline}
    \lambda = \int \frac{\dd \psi}{k} = \oint \frac{\dd \eta}{k\eta_\psi} = - \frac{1}{\tilde{\kappa}_0}\oint\frac{1}{p}\frac{(\p_p\mathcal{V}) \dd p}{\sqrt{2\big(\tilde{\mathcal{H}} - \mathcal{\mathcal{V}}(p;\mathcal{U})\big)}} \\= \frac{\p}{\p \tilde{\mathcal{H}}}\left\{\frac{2}{\tilde{\kappa}_0}\int_{p_\text{min}}^{p_\text{max}} \sqrt{2\big(\tilde{\mathcal{H}} - \mathcal{V}\big)}\left[\mathcal{U}^2 - \frac{1}{(1 + p)^{\gamma+1}}\right]\dd p\right\} \\= \mathcal{J}_\mathcal{H},
\end{multline}
where we've also used that the integrand vanishes at the lower and upper integration limits $p_\text{min/max}$. Equation (\ref{CW}) may then be rewritten as:
\begin{equation}\label{Mod2}
    \frac{\p \mathcal{J}_\mathcal{H}}{\p \Theta} + \mathcal{U}\frac{\p \mathcal{J}_\mathcal{H}}{\p \mathcal{X}} - \mathcal{J}_{\mathcal{H}}\frac{\p \mathcal{U}}{\p \mathcal{X}} = 0.
\end{equation}
Equations (\ref{WWAJ}) and (\ref{Mod2}) now form a solvable hyperbolic system for the slowly varying functions $\mathcal{U}(\Theta,\mathcal{X})$ and $\mathcal{H}(\Theta,\mathcal{X})$ which modulate the system. We refer to them collectively as the modulation equations. Together they describe the dispersion of the non-linear wave-packet, determining the evolution of the modulating envelope as well as the pitch angle of the spiral wave's crests. In the proceeding subsections, we aim to use equations (\ref{WWAJ}) and (\ref{Mod2}) to find sufficient conditions for the phase speed and Hamiltonian of a spiral wave train to satisfy $\tilde{\mathcal{H}} < \mathcal{V}_m(\mathcal{U})$ throughout the disc, with the consequence that the wave never shocks.

\subsubsection{Non-linear spiral waves may never shock: linear motivation}\label{appxB21}

We now set $\mathcal{F} \equiv \text{const}$ and $\tilde{\kappa}_0 \equiv 1$ to exclude effects of amplification, and simplify the system. The assumed value $\tilde{\kappa}_0 = 1$ applies to waves well within their corotation radii in Keplerian discs. This choice distils the competition between non-linearity and dispersion from the tendency of the background disc to amplify or diminish the wave.

We further, for simplicity, ignore the azimuthal periodicity, and hereafter imagine $\theta \in \mathbb{R}$. Figures \ref{1bdenspolar} and \ref{WTG1} suggest for low-amplitude spiral waves, the self-overlap of the tail of the wave train with its nose due to the azimuthal periodicity is not particularly consequential (at least for moderate $X_0$).

We gain some intuition for the problem by considering the linearisation of equation (\ref{GNLDWEOM}),
\begin{equation}
    \frac{\p^2\eta}{\p \theta^2} + \eta - \frac{\p^2 \eta}{\p X_0^2} = 0,
\end{equation}
with corresponding dispersion relation
\begin{equation}
    m^2 = 1 + k^2.
\end{equation}
Restricting our analysis to trailing inward-propagating waves ($m(k) = +\sqrt{1+k^2}$), we obtain the general linear solution:
\begin{equation}
    \eta = \int_0^\infty F(k)\cos(k X_0 + m(k) \theta) \dd k,
\end{equation}
which, by the method of stationary phase, has the asymptotic form
\begin{equation}\label{etaSP}
    \eta \sim F(k)\sqrt{\frac{2\upi m'(k)}{-X_0\left|m''(k)\right|}}\cos\left(kX_0 + m(k)\theta + \frac{\upi}{4}\right)
\end{equation}
on rays with $m'(k)\theta = - X_0$ as $X_0 \to -\infty$. This equation captures linearly the dispersion of the initial excited wave into multiple spiral arms, corresponding to the various peaks of the cosine function (for a more detailed treatment see \citet{miranda_multiple_2019}). The nose of the wave train propagates radially at the sound speed, giving $\theta_\text{nose} = -X_0 + \text{const}$, and the various spiral arms occupy $\theta > \theta_\text{nose}$.

\begin{figure}\centering
  \includegraphics[width=0.79\linewidth]{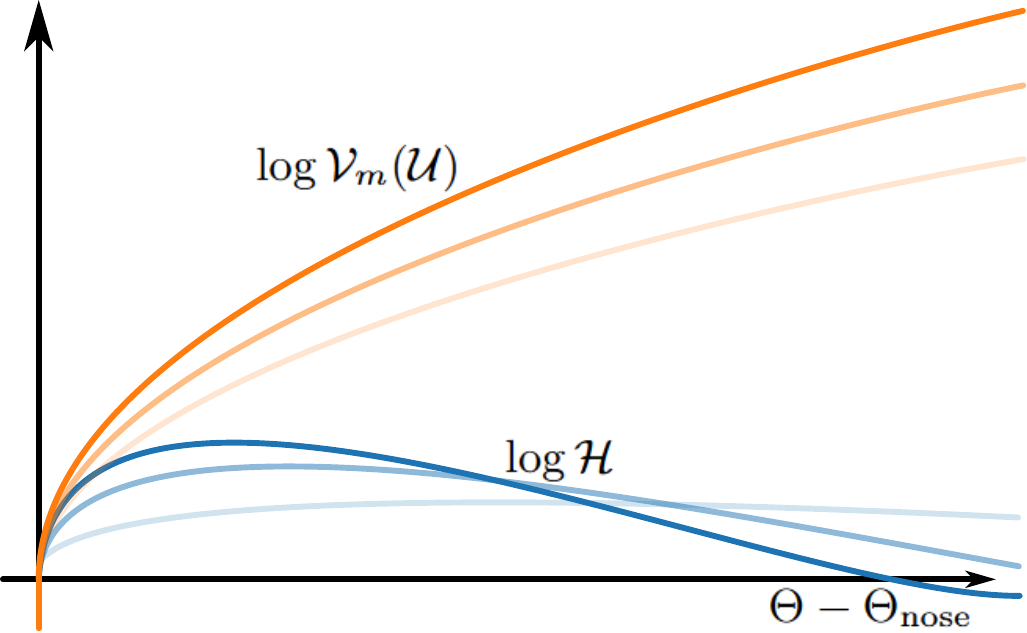}
  \vspace{-1em}
  \caption{Sketch of the expected evolution of $\mathcal{H}$ and $\mathcal{V}_m(\mathcal{U})$ near the nose of a wave train. Lighter colours indicate later points in the wave's evolution (corresponding to radii further from the source). Note the tendency of $\mathcal{H}$ to shrink further below the shocking threshold $\mathcal{V}_m(\mathcal{U})$.}
  \label{HVsketch}
\end{figure}

In terms of the spiral angle or `phase speed' $\mathcal{U}<0$, equation (\ref{etaSP}) may be re-expressed as
\begin{equation}\label{Lindisp}
    \eta \sim \frac{-\mathcal{U}F\left[1/\sqrt{\mathcal{U}^2-1}\right]}{(\mathcal{U}^2-1)^{3/4}}\sqrt{\frac{2\upi}{-X_0}}\cos\left(\frac{X_0 - \mathcal{U}\theta}{\sqrt{\mathcal{U}^2 - 1}} + \frac{\upi}{4}\right).
\end{equation}
Identifying $(2\tilde{\mathcal{H}})^{1/2}$ with the amplitude of $\eta$ in the above expression, we see that $\tilde{\mathcal{H}} \sim \mathcal{X}^{-1}$ on rays with $\Theta =  \mathcal{U}\mathcal{X}$, however $\mathcal{U}(k)$ (and correspondingly $\mathcal{V}_m(\mathcal{U})$) remains constant on these rays. This typical behaviour is sketched in figure \ref{HVsketch}.

The condition $\tilde{\mathcal{H}} < \mathcal{V}_m(\mathcal{U})$ is therefore naturally satisfied at large negative $X_0$; however, we're faced with a singularity as $X_0 \to 0$. We must remember however that the asymptotic relation in equation (\ref{Lindisp}) is only valid for large $X_0$: if the wave may propagate far enough radially (several wavelengths) and reach the regime (\ref{Lindisp}) without having shocked then we can reasonably expect the wave to remain below the threshold for breaking indefinitely (provided $\tilde{\mathcal{H}} < \mathcal{V}_m(\mathcal{U})$ for all $\theta$ once this regime is reached).

One might be concerned about whether it is natural for the condition $\tilde{\mathcal{H}} < \mathcal{V}_m(\mathcal{U})$ to be satisfied near the nose of the wave train, where $\left|\mathcal{U}\right| \to 1$ and $\mathcal{V}_m(\mathcal{U}) \to 0$. It's worth pointing out however that for smooth initial disturbances, $F(k)$ is exponentially small for large $k$, and correspondingly $\mathcal{H}$ is typically exponentially small very near to the nose of the wave train.

The result we prove in the following subsection generalises this analysis to the non-linear setting, and reads informally: suppose an inward-propagating non-linear spiral wave has dispersed into a wave train in which $\tilde{\mathcal{H}} < \mathcal{V}_m(\mathcal{U})$ for all $\theta$ (at some fixed radius $r_0$). It follows that the wave will never break as $X_0 \to -\infty$.

\subsubsection{Sketch-proof that inward-propagating non-linear spiral waves may never shock}\label{appxB22}

In this section we find sufficient conditions for $\tilde{\mathcal{H}}$ to remain below $\mathcal{V}_m(\mathcal{U})$ as $X_0 \to - \infty$ (with the consequence that the wave remains smooth throughout the inner disc) in a disc model with $\mathcal{F} \equiv \text{const}$ and $\tilde{\kappa}_0 \equiv 1$. Since we've assumed $\mathcal{F} \equiv \text{const}$, for notational convenience we no longer distinguish $\mathcal{H}$ and $\tilde{\mathcal{H}}$ in this section.

\begin{figure}\centering
  \includegraphics[width=0.79\linewidth]{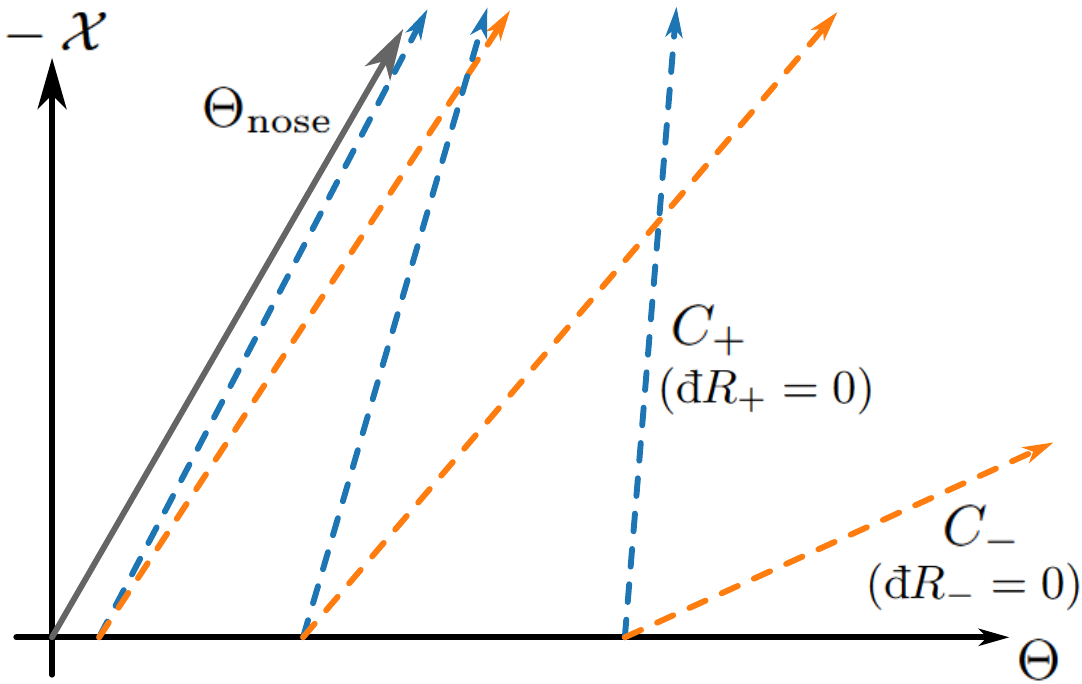}
  \vspace{-1em}
  \caption{$C_+$ and $C_-$ characteristics in the $\Theta\mathcal{X}$ plane. Characteristics at the nose of the wave train are parallel to the trajectory of the nose itself, and both $C_+$ and $C_-$ characteristics fill the region to the right of the nose. $\left|\mathcal{U}\right|$ increases towards the tail of the wave train, consistent with the condition $\p_\Theta R_+<0$ for small $\mathcal{H}$.}
  \label{Wchars}
\end{figure}

We first undertake some useful manipulations (their use will become apparent subsequently). The hyperbolic problem consisting of equations (\ref{WWAJ}) and (\ref{Mod2}) may be recast in the characteristic form:
\begin{equation}\label{WCHAR}
    \frac{\dbar R_\pm}{\dd \mathcal{X}}\bigg|_{C_\pm} = 0, \quad \frac{\dd}{\dd \mathcal{X}}\bigg|_{C_\pm} \equiv \p_\mathcal{X} + \Theta'_\pm\p_\Theta,
\end{equation}
where
\begin{subequations}
\begin{equation}
    \frac{1}{\Theta'_\pm} = \mathcal{U}-\frac{\mathcal{J}_\mathcal{H}}{\mathcal{J}_{\mathcal{H\mathcal{U}}}\mp\sqrt{\mathcal{J}_{\mathcal{U}\mathcal{U}}\mathcal{J}_{\mathcal{H}\mathcal{H}}}},
\end{equation}
\begin{equation}\label{diffR}
    \dbar R_\pm = \sqrt{-\mathcal{J}_{\mathcal{H}\mathcal{H}}} \dd \mathcal{H} \pm \sqrt{-\mathcal{J}_{\mathcal{U}\mathcal{U}}} \dd \mathcal{U}.
\end{equation}
\end{subequations}
That is, $R_\pm$ is conserved on $C_\pm$ characteristics, which are curves satisfying $\frac{\dd \Theta_\pm}{\dd \mathcal{X}} = \Theta'_\pm$. In general the differential $\dbar R_\pm$ is not \emph{proper}, indeed when expressed as above there may be no such function $R_\pm$ satisfying (\ref{diffR}), nevertheless, equation (\ref{WCHAR}) still holds. A sketch of these characteristics for $\mathcal{U} < 0$ is shown in figure \ref{Wchars}.

Combining both the $+$ and $-$ forms of equation (\ref{WCHAR}), it may be shown that
\begin{multline}\label{HevoC-}
    \frac{\dd \mathcal{H}}{\dd \mathcal{X}}\Big|_{C_-} = \sqrt{\frac{\mathcal{J}_{\mathcal{U}\mathcal{U}}}{\mathcal{J}_{\mathcal{H}\mathcal{H}}}}\frac{\dd \mathcal{U}}{\dd \mathcal{X}}\Big|_{C_-} = -\frac{\Theta'_+-\Theta'_-}{2\sqrt{-\mathcal{J}_{\mathcal{H}\mathcal{H}}}}\p_\Theta R_+ \\ = -\left[\frac{\mathcal{J}_\mathcal{H}\sqrt{-\mathcal{J}_{\mathcal{U}\mathcal{U}}}}{\left(\mathcal{U}\mathcal{J}_{\mathcal{H}\mathcal{U}} - \mathcal{J}_{\mathcal{H}}\right)^2 - \mathcal{U}^2\mathcal{J}_{\mathcal{H}\mathcal{H}}\mathcal{J}_{\mathcal{U}\mathcal{U}}}\right]\p_\Theta R_+ \\ = - \left(\mathcal{U}^2-1\right)^{3/4}\sqrt{\frac{\mathcal{H}}{2\upi}}\Big[1 + \mathcal{O}(\mathcal{H})\Big]\p_\Theta R_+.
\end{multline}
The final (linear) estimate in the above equation is included to give context, and may be arrived at by noting that $\mathcal{J} = 2\upi \mathcal{H} \sqrt{\mathcal{U}^2-1} + \mathcal{O}(\mathcal{H}^2)$ for small $\mathcal{H}$, that is, low-amplitude solutions. The denominator in the second line of (\ref{HevoC-}), which is certainly positive for small $\mathcal{H}$, vanishes when one of the $C_\pm$ characteristics becomes a purely azimuthal ray; we take for granted that this doesn't happen. Furthermore, the numerator is positive, as the wavelength $\mathcal{J}_\mathcal{H}>0$. (It follows from the properties of $\mathcal{V}(p;\mathcal{U})$ that both $\mathcal{J}_{\mathcal{U}\mathcal{U}}$ and $\mathcal{J}_{\mathcal{H}\mathcal{H}}$ are negative\footnote{this is a typical result which we won't verify here, e.g. \citet[section 4]{whitham_non-linear_1965}, \citet[chapter 15.3]{whitham_linear_1974}} for all $\mathcal{H} < \mathcal{V}_m(\mathcal{U})$, which further ensures the hyperbolic nature of the system). All this is to say, the bracketed term in the second line of (\ref{HevoC-}) is well-defined and positive for all $\mathcal{H} < \mathcal{V}_m(\mathcal{U})$.

We're now ready to state the conditions for our sketch-proof. The following are sufficient:

Suppose that at some initial $\mathcal{X} = \mathcal{X}_i$, we have $\mathcal{H} < \mathcal{V}_m(\mathcal{U})$ for all $\Theta$, and further $\p_\Theta R_+ < 0$ for all $\Theta$. (Here for simplicity we ignore the periodicity in $\Theta$, and imagine $\Theta \in \mathbb{R}$). It may then be shown that $\mathcal{H} < \mathcal{V}_m(\mathcal{U})$ for all $\Theta$ and $\mathcal{X}< \mathcal{X}_i$.

The sketch-proof now proceeds. Assuming that $C_+$ characteristics both fill the $\mathcal{X}\Theta$ plane and don't cross, it follows that $\p_\Theta R_+ < 0$ for all $\Theta$ and $\mathcal{X} < \mathcal{X}_i$.

Now considering each $C_-$ characteristic in turn, equation (\ref{HevoC-}) dictates that for inward-propagating trailing spirals (with $\mathcal{U}<0$), we have $\dd \mathcal{H}/\dd (-\mathcal{X}) < 0$, yet $\dd \left|\mathcal{U}\right|/\dd (-\mathcal{X}) > 0$ (since $\p_\Theta R_+ < 0$). That is, on inward-propagating $C_-$ characteristics, $\mathcal{H}$ decreases, whilst $\left|\mathcal{U}\right|$ increases.

Since $\mathcal{V}_m(\mathcal{U})$ is an increasing function of $\left|\mathcal{U}\right|$, and $C_-$ characteristics fill the plane (by assumption), it follows that $\mathcal{H} < \mathcal{V}_m(\mathcal{U})$ for all $\Theta$ and $\mathcal{X} < \mathcal{X}_i$, and so the solution remains smooth uniformly as the wave propagates inwards! This concludes the proof.

This argument then provides a non-linear generalisation to the linear analysis presented in section \ref{appxB21}, with the consequence that it's very natural to expect dispersion to prevent low-amplitude spiral waves from shocking. This indeed appears to be the case for planets of mass $M_p \lesssim 0.01 M_\text{th}$, as we saw in section \ref{SAS}.

A few final comments seem necessary. Firstly, the physical meaning of the condition $\p_\Theta R_+ < 0$ is slightly obscure. For small $\mathcal{H}$, it reads that $\mathcal{U}$ increases away from the nose. This is consistent with the wave trains observed in the simulations in section (\ref{SAS}), and the condition that rays are diverging in linear ray theory, meaning that the wave train is stable, and the tail propagates away from the nose: the requirement $\p_\Theta R_+ < 0$ might plausibly be viewed as a non-linear extension of this very reasonable condition.

Furthermore, the restriction to $\mathcal{F} \equiv \text{const}$ and $\tilde{\kappa}_0 \equiv 1$ seems particularly limiting. The choice $\tilde{\kappa}_0 \equiv 1$ restricts the analysis to waves well within their corotation radii in Keplerian discs. In practice, $\tilde{\kappa}_0 > 1$ in the inner disc, and these waves are correspondingly slightly more dispersive than the case we've considered. Far outside the corotation radius of a wave, $\tilde{\kappa}_0 \ll 1$, meaning that the wave increasingly resembles a (non-dispersive) acoustic wave. Relaxing the assumption of constant $\mathcal{F}$ means that the level of non-linearity of a wave is driven up/down as $\mathcal{F}$ decreases/increases inwards. Indeed, if $\mathcal{F}$ decreases indefinitely in the inner disc (as is the case in many power-law discs), the total radial wave action flux will eventually exceed the threshold $\mathcal{B}_\text{max} = 0.0217\mathcal{F}$ found in section \ref{SFPA}, beyond which no steady smooth waves are possible, and shock formation is guaranteed. Conversely, if $\mathcal{F}$ increases inwards, inward-propagating waves become less prone to shocking.

\section{2D shearing sheet WKB analysis}\label{appxc}

\renewcommand{\theequation}{C\arabic{equation}}
\setcounter{equation}{0}

In the following analysis we characterise the effect of linear dispersion on the amplitude of density waves in a local approximation for astrophysical discs, known as the shearing sheet (e.g. \citet{2017MNRAS.472.1432L}). We see that dispersive spreading of wave packets is too weak to prevent the waves from becoming non-linear, and so, in agreement with \citet{heinemann_weakly_2012}, we expect that dispersion is unable to prevent wave breaking in the shearing sheet. Consequently, we infer that the inner spiral wakes excited by embedded objects are more prone to shocking in thinner discs.

Adopting the notation of \citet{brown_horseshoes_2024} (for this appendix only), free waves in the shearing sheet satisfy \citep[equation (63a)]{brown_horseshoes_2024}:
\begin{equation}\label{FWESS}
\left[\mathcal{D}^2 + 1 \pm 3 \p_y - \nabla^2\right]J_{\pm} =  0,
\end{equation}
where $x$ and $y$ are non-dimensional local radial and azimuthal co-ordinates, $\mathcal{D} = -\frac{3}{2}x \p_y$, and $J_{\pm} = u \pm \chi$. Here $u$ is the non-dimensional radial velocity and $\chi$ the enthalpy. 

We now focus on $J_+$ in $x>0$ (which behaves equivalently to $-J_-$ in $x<0$). Let $\xi = \frac{3}{4}x^2$ and $k$ be the Fourier-conjugate variable of $y$. After a Fourier transform with respect to $y$, equation (\ref{FWESS}) becomes
\begin{equation}
    \left[-3\xi\left(k^2 + \p_\xi^2\right)+1+3\ii k-\tfrac{3}{2}\p_\xi + k^2\right]\tilde{J}_+ = 0.
\end{equation}
We seek a WKB solution for the $k$th mode. We have:
\begin{equation}
    \left[\p_\xi^2+\frac{1}{2\xi}\p_\xi+\left(k^2-\frac{1+3\ii k+k^2}{3\xi}\right)\right]\tilde{J}_+ = 0.
\end{equation}
Now let $\tilde{J}_+ = f \xi^{-1/4}$,

\[\implies f'' + \left(\frac{3}{16\xi^2} + k^2 - \frac{1+3\ii k+k^2}{3\xi}\right)f = 0,\]

\[\implies f \sim \xi^{1/2}\exp\left[\ii\left(k\xi - \frac{1}{6}\left(k + \frac{1}{k}\right)\ln \xi\right)\right] +\mathcal{O}\left(\xi^{-1/2}\right),\]

\begin{equation}
\implies J_+ \sim \int \tilde{A}(k) \xi^{1/4}e^{\ii k(y+\xi)-\tfrac{\ii}{6}\left(k+\tfrac{1}{k}\right)\ln\xi}\dd k,
\end{equation}
so for new `space' and `time' coordinates $X = y + \xi$, $T = \frac{1}{6}\ln \xi$, and effective frequency $\omega = k + 1/k$, we see (by the method of stationary phase):
\begin{multline}
    J_+ \sim \xi^{1/4}\tilde{A}(k_0)\sqrt{\frac{2\pi}{T\left|\omega''(k_0)\right|}}\cos\bigg(k_0X + \omega(k_0)T \\ + \arg\left[\tilde{A}(k_0)\ee^{-\ii\tfrac{\pi}{4}}\right]\bigg)
\end{multline}
on characteristics with $X = \omega'(k_0)T$ (a family of curves similar to $y = -\frac{3}{4}x^2$). That is, on such characteristics,
\begin{equation}
\left|J_+\right|\sim \sqrt{\frac{x}{\ln x}}.
\end{equation}

In contrast, suppose we were to artificially eliminate the dispersive inertial acceleration by setting the epicyclic frequency equal to zero in the above analysis. This leads to the corresponding toy problem\footnote{This model is derivable in the case $B = \Omega + \tfrac{r}{2}\frac{\dd \Omega}{\dd r} = 0$, corresponding to a disc with uniform specific angular momentum, using a slightly different non-dimensionalisation to that used for equation (\ref{FWESS}).}
\begin{equation}
\left[\mathcal{D}^2 \pm 3 \p_y - \nabla^2\right]J_{\pm} =  0.
\end{equation}
In this case we similarly obtain the asymptotic relation
\begin{equation}\label{Jpndint}
J_+ \sim \int \tilde{A}(k) \xi^{1/4}\exp{\left[\ii k\left(y+\xi - \tfrac{1}{6}\ln\xi\right)\right]}\dd k,
\end{equation}
in which we recognise the `space' and `time' coordinates $X = y + \xi$, $T = \frac{1}{6}\ln \xi$ from before, but now the effective frequency is $\omega=k$. The wave packet is correspondingly non-dispersive (and has constant group velocity $\omega'(k) \equiv 1$). Evaluating the expression (\ref{Jpndint}) gives
\begin{equation}
J_+ = \xi^{1/4}A(y+\xi - \tfrac{1}{6}\ln\xi) \sim \sqrt{x}.
\end{equation}
On characteristics with $y+\xi - \tfrac{1}{6}\ln\xi = \text{const}$, these artificially non-dispersive waves therefore have
\begin{equation}
\left|J_+\right|\sim \sqrt{x}.
\end{equation}
That is, linearly, in the shearing sheet, dispersion only reduces the amplitude of the density waves asymptotically by a factor of $\sqrt{\ln x}$, which is too little to overcome the shear amplification. Non-linearity, and presumably shock formation, is unavoidable for these waves.

\bsp	
\label{lastpage}
\end{document}